\documentclass[5p,times,final]{elsarticle}

\usepackage{graphicx,subcaption}
\usepackage{amssymb}
\usepackage{amsmath}
\usepackage{lineno}
\usepackage{tikz}
\usepackage[ruled, linesnumbered]{algorithm2e}

\SetCommentSty{mycommfont}

\usetikzlibrary{shapes.geometric, arrows}
\usepackage{nomencl} 
\usepackage{titlesec}
\usepackage{lipsum}

\usepackage{colordvi}
\usepackage{color, soul}

\usepackage{framed} %

\immediate\write18{%
makeindex -s nomencl.ist -o \jobname.nls -t \jobname.nlg \jobname.nlo%
}

\usepackage{multicol} %

\usetikzlibrary{patterns}
\usetikzlibrary{decorations.shapes}
\tikzset{decorate sep/.style 2 args={decorate, decoration={shape backgrounds, shape=circle, shape size=#1, shape sep=#2} }}

\tikzstyle{startstop} = [rectangle, rounded corners, minimum width=3cm, minimum height=1cm,text centered,text width = 6 cm, draw=black, fill=red!30]
\tikzstyle{io} = [trapezium, trapezium left angle=70, trapezium right angle=110, minimum width=3cm, minimum height=1cm, text centered, draw=black, fill=blue!30]
\tikzstyle{process} = [rectangle, minimum width=3cm, minimum height=1cm, text centered, text width = 6 cm ,draw=black, fill=orange!30]
\tikzstyle{decision} = [diamond, minimum width=3cm, minimum height=2cm, text centered, text width = 3.3cm, draw=black, fill=green!30]
\tikzstyle{arrow} = [thick,->,>=stealth]

\usepackage{comment}
\usepackage{bibentry}
\nobibliography{elsarticle_bib}

\titleformat{\paragraph}
{\normalfont\normalsize\it}{\theparagraph}{0.5em}{}
\titlespacing*{\paragraph}
{0pt}{1.0ex plus 0.5ex minus .2ex}{0.0ex plus .2ex}

\biboptions{sort&compress}

\newcommand{\angstrom}{\text{\normalfont \AA}}

\newcommand{\bc}{\mathbf{c}}
\newcommand{\bx}{\mathbf{x}}
\newcommand{\bn}{\mathbf{n}}
\newcommand{\bu}{\mathbf{u}}
\newcommand{\bq}{\mathbf{q}}
\newcommand{\rd}{\,\mathrm{d}}

\newcommand{\Kn}{\mathrm{Kn}}

\journal{Journal of Computational Physics [Accepted]}
\usepackage[]{xcolor}
\usepackage{booktabs}
\usepackage{multirow}

\begin{document}
\begin{frontmatter}

    \title{A Discontinuous Galerkin Fast Spectral Method for the Full Boltzmann Equation \\with General Collision Kernels}

\author[labelAero]{Shashank Jaiswal}
\ead{jaiswal0@purdue.edu}

\author[labelAero]{Alina A. Alexeenko}
\ead{alexeenk@purdue.edu}

\author[labelMath]{Jingwei Hu\corref{cor1}}
\ead{jingweihu@purdue.edu}

\cortext[cor1]{Corresponding author.}

\address[labelAero]{School of Aeronautics and Astronautics}
\address[labelMath]{Department of Mathematics}
\address{Purdue University, West Lafayette, IN 47907, USA}

\begin{abstract}
The Boltzmann equation, an integro-differential equation for the molecular distribution function in the physical and velocity phase space, governs the fluid flow behavior at a wide range of physical conditions, including compressible, turbulent, as well as flows involving further physics such as non-equilibrium internal energy exchange and chemical reactions. Despite its wide applicability, deterministic solution of the Boltzmann equation presents a huge computational challenge, and often the collision operator is simplified for practical reasons. In this work, we introduce a highly accurate deterministic method for the full Boltzmann equation which couples the Runge-Kutta discontinuous Galerkin (RKDG) discretization in time and physical space (\citeauthor{su2015parallel}, Comp. Fluids, 109 pp.~123-136, \citeyear{su2015parallel})  and the recently developed fast Fourier spectral method in velocity space  (\citeauthor{GHHH17}, SIAM J. Sci. Comput., 39 pp.~B658--B674, \citeyear{GHHH17}). The novelty of this approach encompasses three aspects: first, the fast spectral method for the collision operator applies to general collision kernels with little or no practical limitations, and in order to adapt to the spatial discretization, we propose here a singular-value-decomposition based algorithm to further reduce the cost in evaluating the collision term; second, the DG formulation employed has high order of accuracy at element-level, and has shown to be more efficient than the finite volume method; thirdly, the element-local compact nature of DG as well as our collision algorithm is amenable to effective parallelization on massively parallel architectures. The solver has been verified against analytical Bobylev-Krook-Wu solution. Further, the standard benchmark test cases of rarefied Fourier heat transfer, Couette flow, oscillatory Couette flow, normal shock wave, lid-driven cavity flow, and thermally driven cavity flow have been studied and their results are compared against direct simulation Monte Carlo (DSMC) solutions with equivalent molecular collision models or published deterministic solutions.
\end{abstract}


\end{frontmatter}

\section{Introduction}
In micro/rarefied gas flows, the gas  molecule wall-surface interactions lead to the formation of Knudsen layer (KL): a local thermodynamically non-equilibrium region extending $\sim$ $O(\lambda)$ from the surface, where $\lambda$ is the gas mean free path (MFP) \cite{sone2002kinetic}. The Knudsen number ($Kn$) is defined as $\lambda/H$, where $H$ is the characteristic length of the system. The classical constitutive relations of the Navier-Stokes-Fourier equations fail to predict nonlinear behavior in the KL and deviations are significant in the slip ($10^{-3} < Kn < 10^{-1}$) and transition flow regimes ($10^{-1} < Kn < 10$) \cite{sone2002kinetic, karniadakis2006microflows, Bird}. The Boltzmann equation, an integro-differential equation for the molecular distribution function in the physical and velocity phase space, governs the fluid flow behavior for a wide range of Knudsen numbers and physical conditions, including compressible, turbulent, as well as flows involving further physics such as non-equilibrium internal energy exchange and chemical reactions. Accurate physical models and efficient numerical methods are required for solving the Boltzmann equation so as to predict the non-equilibrium phenomenon encountered in such rarefied flows.

The approaches for numerical solution of the Boltzmann equation date back to as early as 1940s \cite{grad1949} using, for example, the now widely used direct simulation Monte Carlo (DSMC) method \cite{Bird1963,Bird2013}. The DSMC method, based on the kinetic theory of dilute gases, models the binary interactions between particles stochastically. However, it is this stochastic nature of the method that introduces high statistical noise in low-speed flows, and imposes strict constraints on cell-size and time-step. Moreover, the formal accuracy of particle time-stepping is linear. The stiffness properties of the Boltzmann equation further aggravates the time-step constraints. To overcome these limitations, improved particle-based approaches have been proposed \cite{gallis2009convergence}, including hybrid continuum/particle solvers \cite{sun2003hybrid,dimarco2008hybrid}, variance reduction methods \cite{baker2005variance}, and simplified Bernoulli trials \cite{stefanov2011dsmc}. 

It is to be noted that the assertion that DSMC solves the actual full Boltzmann equation is not strictly valid. Indeed, the DSMC method can be derived rigorously as the Monte Carlo solution of the $N$-particle master kinetic equation \cite{DSMC2016}. Wagner \cite{wagner1992convergence} established convergence proof for Bird's DSMC method for the Boltzmann equation in the limit of infinite number of particles, $N\to\infty$. Moreover, the proof has inherent assumptions on the boundedness of the collision operator which is clearly highlighted in Wagner's work (see section 5 in \cite{wagner1992convergence}). 

The deterministic solutions based on discretization of governing differential equations on representative grids is central to computational fluid dynamics (CFD). However, the multi-dimensional nature of the Boltzmann equation and the collision integral becomes a bottleneck resulting in excessive use of time and computing resources. To bypass this issue, simplified Boltzmann equation variants such as linearized Boltzmann (LB) \cite{gross1959kinetic}, Bhatnagar-Gross-Krook (BGK) \cite{bhatnagar1954model}, and ellipsoidal Bhatnagar-Gross-Krook (ES-BGK) \cite{holway1966new} equations are used. These simplified models perform better at low Knudsen number flows in slip and early transition regimes. Yet they often fail to capture the physics at high Knudsen numbers as well as for diffusion dominated flows at even low Knudsen numbers (see \cite{gallis2014direct,gallis2006normal}). Another way to reduce the dimensionality is to consider the moment closure of the Boltzmann equation. Introduced by Grad \cite{grad1949}, the moment method produces an evolution equation for the moments of the distribution function. Different level of approximations/closures lead to different hierarchies, e.g., \textit{Grad 13-moment} \cite{grad1949}, \textit{Levermore 14-moment} \cite{levermore1996moment}, and various regularized versions \cite{struchtrup2003regularization, gu2009high, CFL14}.

Over the past decades, the deterministic methods that solve the full Boltzmann equation have undergone considerable development. Without being exhaustive, we refer to \cite{Mieussens14, pareschi} for a comprehensive review. In this work, we employ the recently developed fast Fourier spectral method \cite{GHHH17} to solve the Boltzmann collision operator. Compared with other deterministic methods such as the discrete velocity models (DVM), the Fourier spectral method can provide significantly more accurate results with less numerical complexity; compared with DSMC, it produces smooth, noise-free solutions and can simulate low-speed flows such as those encountered often in micro-systems. On the other hand, the Fourier spectral method is still computationally demanding, as it requires $O(N^6)$ memory to store precomputed weights and has $O(N^6)$ numerical complexity \cite{PR00, GT09}, where $N$ is the number of discretization points in each velocity dimension. The main contribution in \cite{GHHH17} is a low-rank strategy to accelerate the direct Fourier spectral method so that it requires only $O(MN^4)$ memory to store precomputed weights (no precomputation is needed in certain cases) and has $O(MN^4\log N)$ complexity, where $M$ is the number of discretization points on the sphere and $M\ll N^2$. Furthermore, the fast method applies directly to arbitrary collision kernels and can be easily extended to general collision models including the multi-species and inelastic Boltzmann equations. We mention that there is another line of research that develops the fast Fourier spectral method based on Carleman representation of the collision operator \cite{MP06}. The complexity of the method is $O(MN^3\log N)$. However, its applicability is limited to hard sphere molecules. {The method has been extended to anisotropic scattering in \cite{wu2013deterministic}, but it assumes a special form of the kernel and requires recalibration of transport coefficients and parametric fitting therein. This methodology has been applied to Lennard-Jones potential and to many canonical flows in \cite{WRZ14},\cite{wu2015influence}.} {The method has been extended to anisotropic scattering in \cite{wu2013deterministic} and applied to many canonical flows in \cite{WRZ14} by assuming a special form of the kernel and performing a recalibration of transport coefficients and parametric fitting. In \cite{wu2015influence}, the Lennard-Jones potential was considered by fully resolving the kernel, resulting in the cost of $O(MN^4\log N)$.}

All of the former approaches have relied on low-order (up to second-order) finite volume (FV) or finite difference (FD) methods for spatial discretization of the Boltzmann equation. In this work, we employ the discontinuous Galerkin (DG) method for the spatial discretization, a class of high order method widely used for time dependent multi-dimensional hyperbolic equations \cite{cockburn1989tvb2,cockburn1989tvb3,cockburn1998runge,cockburn1990runge,hesthaven2007nodal}. Compared to high-order FV/FD methods, DG provides easy formulation on arbitrary meshes, high-order flux reconstruction, straightforward implementation of boundary conditions, high-order accuracy, as well as strong linear scaling on parallel processors due to the compactness of the scheme \cite{hesthaven2007nodal}. DG has been employed for solving the BGK and ES-BGK equations for 0D/1D \cite{alexeenko2008high}, and 2D \cite{su2015parallel,su2017stable} flow problems. It has also been used to approximate the moment systems of the Boltzmann equation in \cite{barth2006discontinuous,abdelmalik2016entropy}. To the best of our knowledge, DG discretization in the physical space hasn't been applied for solving the full Boltzmann equation till date.

To summarize, we present a 1D/2D-3V full Boltzmann equation solver by coupling the Runge-Kutta discontinuous Galerkin (RKDG) discretization in time and physical space \cite{su2015parallel} and the fast Fourier spectral method in velocity space \cite{GHHH17}. The method is high order in both physical space and time, and spectrally accurate in velocity space. There are no ad-hoc adjustments or parametric fitting involved in our present formulation for solving the collision operator. Moreover, our singular value decomposition (SVD) variant of the algorithm for evaluating weak form of the collision term is novel and unique to DG formulation. 

In the section that follows, we give a brief introduction of the Boltzmann equation and the collision kernel involved. Section 3 presents an overview of the DG method in general, and describes the weak DG formulation of the Boltzmann equation, including the direct and SVD variant of the algorithm for evaluating the collision term. Extensive numerical experiments and results are performed and discussed in Section 4. Concluding remarks are given in Section 5. A brief description of the fast Fourier spectral method is provided in the Appendix.

\section{The Boltzmann equation}

The Boltzmann equation for a single-species, monatomic gas without external forces can be written as (cf. \cite{Cercignani})
\begin{equation}
\frac{\partial  f}{\partial  t} + \bc\cdot \nabla_{\bx}f  =\mathcal{Q}(f,f), \quad t\geq 0,\,\, \bx\in \Omega_x, \, \,\bc\in \mathbb{R}^3,
\label{eq_bze}
\end{equation}
where $f=f(t,\bx,\bc)$ is the one-particle distribution function of time $t$, position $\bx$, and particle velocity $\bc$. $f\rd{\bx}\rd{\bc}$ gives the number of particles to be found in an infinitesimal volume $\rd{\bx}\rd{\bc}$ centered at the point $(\bx,\bc)$ of the phase space. $\mathcal{Q}(f,f)$ is the collision operator describing the binary collisions among particles, and acts only in the velocity space:
\begin{equation}
\mathcal{Q}(f,f)(\bc) =\int_{\mathbb{R}^3} \int_{\mathcal{S}^2} \mathcal{B} ( \bc -  \bc_*, \sigma) [  f(\bc')  f(\bc_*')-  f(\bc) f(\bc_*) ]\rd{\sigma} \rd{\bc_*},
\label{eq_dim_Q_full}
\end{equation}
where $(\bc, \bc_*)$ and $(\bc', \bc'_*)$ denote the pre- and post- collision velocity pairs, which are related through momentum and energy conservation as
\begin{equation} \label{cc}
\bc'=\frac{\bc+\bc_*}{2}+\frac{|\bc-\bc_*|}{2}\sigma, \quad \bc_*'=\frac{\bc+\bc_*}{2}-\frac{|\bc-\bc_*|}{2}\sigma,
\end{equation}
with the vector $\sigma$ varying over the unit sphere $\mathcal{S}^2$. The quantity $\mathcal{B}$ ($\geq 0$) is the collision kernel depending only on $| \bc - \bc_*|$ and the scattering angle $\chi$ (angle between $\bc-\bc_*$ and $\bc'-\bc'_*$), and can be expressed as
\begin{equation}
	\mathcal{B}(\bc - \bc_*, \sigma) = B(|\bc - \bc_*|, \cos\chi), \quad \cos \chi = \frac{\sigma \cdot (\bc - \bc_*)}{| \bc - \bc_*|}.
	\label{eq_dim_B}
\end{equation}

Given the interaction potential between particles, the specific form of $B$ can be determined using the classical scattering theory (cf. \cite{FW}):
\begin{equation}
B(|\bc - \bc_*|, \cos\chi)=|\bc-\bc_*|\Sigma(|\bc-\bc_*|,\chi),
\label{eq_dim_B1}
\end{equation}
where $\Sigma$ is the differential cross-section given by
\begin{equation}
\Sigma(|\bc-\bc_*|,\chi)=\frac{b}{\sin \chi}\left | \frac{\rd{b}}{\rd{ \chi}} \right|,
\label{eq_differentialCrossSection}
\end{equation}
with $b$ being the impact parameter. 

With a few exceptions (e.g. hard sphere molecules), the explicit form of $\Sigma$ can be hard to obtain since $b$ is related to $\chi$ implicitly. To avoid this complexity, phenomenological collision kernels are often used in practice with the aim to reproduce the correct transport coefficients. Koura et al. \cite{koura1991variable} introduced a scattering model so called as variable soft sphere (VSS) by assuming an \textit{explicit} cosine dependence between the scattering angle and impact parameter:
\begin{equation}
	\chi = 2 \cos^{-1} \{(b/d)^{1/\alpha}\},
	\label{eq_VSSmodel}
\end{equation}
where $\alpha$ is the scattering parameter, and $d$ is the diameter borrowed from Bird's \cite{Bird} variable hard sphere (VHS) model:
\begin{equation}
	d = d_\mathrm{ref} \Bigg[ \Bigg(\frac{4 R T_\mathrm{ref}}{|\bc-\bc_*|^2}\Bigg)^{\omega-0.5} \frac{1}{\Gamma(2.5-\omega)} \Bigg]^{1/2}.
	\label{eq_dVSS}
\end{equation}
Here $R=k_B/m$ is the gas constant ($k_B$ is the Boltzmann's constant and $m$ is the single particle mass), $\Gamma$ denotes the usual Gamma function, $d_\mathrm{ref}$, $T_\mathrm{ref}$, and $\omega$ are, respectively, the reference diameter, reference temperature, and viscosity index. The diameter $d$ and exponent $\alpha$ are determined so that the transport (viscosity and diffusion) coefficients of VSS are consistent with experimental data \cite{Weaver,Stephani}. 

Substituting (\ref{eq_VSSmodel}), (\ref{eq_dVSS}) into (\ref{eq_differentialCrossSection}) and (\ref{eq_dim_B1}), we obtain the general form of $B$ as
\begin{equation}
	B = b_{\omega,\,\alpha} \; |\bc - \bc_*|^{2(1 - \omega)} \; (1 + \cos \chi)^{\alpha-1},
	\label{eq_B_gen}
\end{equation}
where $b_{\omega,\,\alpha}$ is a constant given by
\begin{equation}
	b_{\omega,\,\alpha} = \frac{d_\mathrm{ref}^2}{4} \left(4 R T_\mathrm{ref}\right)^{\omega - 0.5} \frac{1}{\Gamma(2.5 - \omega)} \;\frac{\alpha}{2^{\alpha-1}}.
	\label{eq_b_vhs_dimensional}
\end{equation}
In particular, the VHS kernel is obtained when $\omega \in [0.5, 1]$ and $\alpha = 1$ ($\omega=\alpha=1$ corresponds to the Maxwell molecules, and $\omega=0.5$, $\alpha=1$ to the hard spheres); and the VSS kernel is obtained when $\omega \in [0.5, 1]$ and $\alpha \in (1, 2]$.

It is worth emphasizing that although the collision kernel (\ref{eq_B_gen}) is adopted in the present work for easy comparison with DSMC solutions, the fast spectral method we use for the collision operator applies straightforwardly to any kernel of the form (\ref{eq_dim_B}), i.e., $B$ can be any function of the relative velocity and scattering angle as long as the collision integral makes sense (see Appendix). This generality allows us to treat many well studied/calibrated collision models in the existing literature, for example, for Lennard-Jones interactions, one can use the tabulated kernel $B$ as obtained in \cite{VenkattramanLJ}.

Given the distribution function $f$, the macroscopic quantities can be obtained via its moments:
\begin{equation} \label{moments}
\begin{split}
&n=\int_{\mathbb{R}^3}f \rd{\bc}, \quad \bu=\frac{1}{n}\int_{\mathbb{R}^3}f\bc \rd{\bc}, \quad T=\frac{1}{3Rn}\int_{\mathbb{R}^3}f|\bc-\bu|^2\rd{\bc},\\
&\mathbb{P}=m\int_{\mathbb{R}^3}f(\bc-\bu)\otimes(\bc-\bu)\rd{\bc},\quad \bq=\frac{1}{2}m\int_{\mathbb{R}^3}f(\bc-\bu)|\bc-\bu|^2\rd{\bc},
\end{split}
\end{equation}
where $n$, $\bu$, $T$, $\mathbb{P}$, and $\bq$ are, respectively, the number density, bulk velocity, temperature, stress tensor, and heat flux vector.

\subsection{Non-dimensionalization}

To reduce the parameters, it is convenient to non-dimensionalize all variables and functions. 

We first choose the characteristic length $H_0$, characteristic temperature $T_0$, and characteristic number density $n_0$, and then define the characteristic velocity $u_0=\sqrt{2RT_0}$ and characteristic time $t_0=H_0/u_0$.

Now we rescale $t$, $\bx$, $\bc$, and $f$ as follows
\begin{equation}
\tilde{t}=\frac{t}{t_0}, \quad \tilde{x}=\frac{x}{H_0}, \quad \tilde{\bc}=\frac{\bc}{u_0}, \quad \tilde{f}=\frac{f}{n_0/u_0^3},
\end{equation}
the macroscopic quantities as
\begin{equation}
\tilde{n}=\frac{n}{n_0}, \quad \tilde{\bu}=\frac{\bu}{u_0}, \quad \tilde{T}=\frac{T}{T_0}, \quad \tilde{\mathbb{P}}=\frac{\mathbb{P}}{mn_0RT_0}, \quad \tilde{\bq}=\frac{\bq}{mn_0RT_0u_0},
\end{equation}
and the collision kernel $B$ as
\begin{equation}
\tilde{B}=\frac{B}{2^{1-\omega}\pi d_{\text{ref}}^2(4RT_{\text{ref}})^{\omega-0.5}u_0^{2(1 - \omega)}},
\end{equation}
then the equation (\ref{eq_bze}) becomes
\begin{equation} \label{eq_nbze}
\frac{\partial  \tilde{f}}{\partial  \tilde{t}} + \tilde{\bc}\cdot \nabla_{\tilde{\bx}}\tilde{f}  =\frac{1}{Kn}\tilde{\mathcal{Q}}(\tilde{f},\tilde{f}),
\end{equation}
with the collision operator 
\begin{equation} \label{nCO}
\begin{split}
	\mathcal{Q}(\tilde{f},\tilde{f})(\tilde{\bc}) = \int_{\mathbb{R}^3} \int_{\mathcal{S}^2} &\tilde{B} (|\tilde{ \bc} -  \tilde{\bc}_*|, \cos\chi) [  \tilde{f}(\tilde{\bc}')  \tilde{f}(\tilde{\bc}_*') \\& -\tilde{f}(\tilde{\bc}) \tilde{f}(\tilde{\bc}_*) ]\rd{\sigma} \rd{\tilde{\bc}_*},
\end{split}
\end{equation}
where
\begin{equation}
\tilde{B} (|\tilde{ \bc} -  \tilde{\bc}_*|, \cos\chi)=\frac{\alpha}{2^{2-\omega+\alpha}\Gamma(2.5-\omega)\pi}|\tilde{\bc} - \tilde{\bc}_*|^{2(1 - \omega)} \; (1 + \cos \chi)^{\alpha-1},
\end{equation}
The Knudsen number $Kn$ is given by
\begin{equation} 
	Kn = \frac{1}{\sqrt{2} \pi\; n_0\; d^2_\text{ref}\; (T_\text{ref}/T_0)^{\omega-0.5}\; H_0},
	\label{eq_Kn}
\end{equation}
which is the ratio between the MFP and characteristic length (consistent to equation~(4.65) in \cite{Bird,Bird2013}). Finally, the definition (\ref{moments}) in rescaled variables reduces to
\begin{equation} \label{moments1}
\begin{split}
&\tilde{n}=\int_{\mathbb{R}^3}\tilde{f} \rd{\tilde{\bc}}, \quad \tilde{\bu}=\frac{1}{\tilde{n}}\int_{\mathbb{R}^3}\tilde{f}\tilde{\bc} \rd{\tilde{\bc}}, \quad \tilde{T}=\frac{2}{3\tilde{n}}\int_{\mathbb{R}^3}\tilde{f}|\tilde{\bc}-\tilde{\bu}|^2\rd{\tilde{\bc}},\\
&\tilde{\mathbb{P}}=2\int_{\mathbb{R}^3}\tilde{f}(\tilde{\bc}-\tilde{\bu})\otimes(\tilde{\bc}-\tilde{\bu})\rd{\tilde{\bc}},\quad \tilde{\bq}=\int_{\mathbb{R}^3}\tilde{f}(\tilde{\bc}-\tilde{\bu})|\tilde{\bc}-\tilde{\bu}|^2\rd{\tilde{\bc}}.
\end{split}
\end{equation}

Henceforth, we will always refer to the non-dimensionalized equations (\ref{eq_nbze})-(\ref{moments1}) in our presentation, and $\sim$ will be dropped for simplicity. %

\section{Discontinuous Galerkin formulation}

\subsection{Brief overview}

The Runge Kutta discontinuous Galerkin (RKDG) method \cite{cockburn1989tvb2,cockburn1989tvb3,cockburn1998runge,cockburn1990runge,hesthaven2007nodal} is a class of finite element methods coupling RK discretization in time and DG discretization in space which provides high-order numerically accurate solutions to governing partial differential equations. Higher order accuracy is desirable for simulating flows with strong gradients, droplet collisions as in multi-phase flows, combustion-modeling, reactors, and micro-mechanical systems. RKDG can recover flow properties at the domain boundaries with the same high-order accuracy as in the interior of the domain.

In the Boltzmann equation simulations, the computational domain consists of physical and velocity domains. We propose to use the RKDG method in time and physical space and the Fourier spectral method in the velocity space. Hence the velocity space is partitioned using the Cartesian type grid point (with reasons to be explained in section 3.3), and the physical space is split up into a set of line segments (in 1D), triangles/quadrilaterals (in 2D), and tetrahedrals/prisms/hexahedrals (in 3D) for instance. In particular for 2D grids of quadrilateral cells, each cell in the physical space has four faces. The cell connectivity is such that a cell face is either internal and intersects two cells only, or comprises part of an external boundary and belongs to single cell only. 

In such a grid system, the DG method is developed to solve the Boltzmann equation at each velocity grid point $\bc^j$. Within a given spatial element $i$, the distribution function $f$ is approximated as a linear combination of orthogonal basis functions $\phi_l^i(\mathbf{x})$ as
\begin{equation}
	f_j^i = \sum_{l = 1}^{K} \mathcal{F}_l^{i,j} \phi_l^i(\mathbf{x}),
	\label{eq_fp}
\end{equation}
where $K$ is the number of unknowns in the element also known as local degree of freedom. The task is to determine the coefficients $\mathcal{F}_l^{i,j}$ of the expansion for all elements. Therefore, the complexity of the problem is proportional to the number of velocity nodes, the number of spatial elements, the order of basis functions, and the number of time integration steps. Due to the multi-dimensionality of the problem, and the typical size of phase space considered in the current work (order of millions), parallel computation is highly desirable. 

In finite element setting, the information is exchanged between two-adjacent elements using the shared nodes between them. The DG method, in contrast to the classical finite element method that relies on global stiffness matrices, duplicates the values that are shared between the elements. To connect the elements at the shared nodes, DG introduces monotone interface flux (as in finite volume method). It is this flux that allows element-to-element decoupling, recovers a meaningful global solution, and allows for explicit time stepping (see \cite{hesthaven2007nodal}). It is this element-to-element decoupling and element local-nature of the DG method that makes it amenable to strong scaling on parallel processors, and therefore our choice of spatial discretization scheme.

\subsection{Discretization in the physical space}

Assume that the Boltzmann equation~(\ref{eq_nbze}) is posed in the domain $\Omega_x$ with boundary $\partial \Omega_{x}$ in the physical space. We decompose $\Omega_x$ into $I$ variable sized elements $D^i_x$:
\begin{equation}
	\Omega_{x} \approx \bigcup\limits_{i=1}^{I} D^i_x.
\end{equation}
In each element $D_x^i$, we approximate the function $f(t,\bx,\bc)$ by a polynomial of order $N_p$:
\begin{equation}
	\mathbf{x} \in D_x^{i}: \quad f^{i}(t,\bx,\bc) = \sum_{l = 1}^{K} \mathcal{F}_l^{i}(t,\bc) \; \phi_l^i(\mathbf{x}),
	\label{eq_DG_local_approx}
\end{equation}
where $\phi_l^i(\mathbf{x})$ is the basis function supported in $D_x^{i}$, $K$ is the total number of terms in the local expansion, and $\mathcal{F}_l^{i}(t,\bc)$ is the elemental degree of freedom. In general $K$ depends on elemental-shape. In 1D, $K = N_p+1$. In 2D, $K=(N_p+1)^2$ for quadrilateral elements, and $K=(N_p+1)(N_p+2)/2$ for triangular elements. In 3D, $K=(N_p+1)^3$ for hexahedral elements, and $K=(N_p+1)(N_p+2)(N_p+3)/6$ for tetrahedral elements.

We now present a general 3D spatial weak DG formulation for the Boltzmann equation. Reduction to the 2D case can be achieved by choosing a 2D basis, and ignoring the $z$-axis dependence. Similarly for the 1D case. Time and velocity space are left as continuous at the moment. 

We first form the residual by substituting the expansion (\ref{eq_DG_local_approx}) into the equation (\ref{eq_nbze}):
\begin{equation}
\begin{split}
\mathcal{R}^i=	\sum_{l=1}^{K} \phi_l^i \frac{\partial}{\partial t} \mathcal{F}_l^{i}+ \sum_{l=1}^{K} \mathcal{F}_l^{i} \bc \cdot \nabla_{\bx} \phi_l^i -\frac{1}{Kn}\sum_{l_1=1}^{K} \sum_{l_2=1}^{K} \mathcal{Q} \left(\mathcal{F}_{l_1}^{i}, \mathcal{F}_{l_2}^{i}\right) \phi_{l_1}^i \phi_{l_2}^i,
	\label{eq_bze_weak_multiply}
\end{split}
\end{equation}
where we used the quadratic nonlinearity of the collision operator.

We then require that the residual is orthogonal to all test functions. In the Galerkin formulation, the test function is the same as the basis function, thus
\begin{equation}
	\int_{D_x^i} \mathcal{R}^i \, \phi_m^i  \rd{\bx} = 0, \quad 1\leq m \leq K,
	\label{eq_innerProduct}
\end{equation}
in each element $D_x^i$.

Substituting (\ref{eq_bze_weak_multiply}) into (\ref{eq_innerProduct}) and applying the divergence theorem, we obtain
\begin{equation}
\begin{split}
	&\sum_{l=1}^{K} \left(\int_{D_x^i} \phi_m^i\, \phi_l^i \rd{\bx}\right)\frac{\partial}{\partial t} \mathcal{F}_l^{i} -\sum_{l=1}^K \mathcal{F}_l^{i} \int_{D_x^i}  \phi_l^i \,\nabla_{\bx}\cdot (\bc \, \phi_m^i) \rd{\bx}\\ 
	=&- \int_{\partial D_x^i}  \phi_m^i  \left( {\bf F}^*\cdot \hat{\bn}^i \right)\rd{\bx} \\ &+ \frac{1}{Kn}\sum_{l_1=1}^{K} \sum_{l_2=1}^{K} \mathcal{Q} (\mathcal{F}_{l_1}^{i}, \mathcal{F}_{l_2}^{i}) \left(\int_{D_x^i} \phi_m^i \,\phi_{l_1}^i\, \phi_{l_2}^i \rd{\bx}\right),
\end{split}
\label{eq_dg_innerProd}
\end{equation}
where $\hat{\bn}^i$ is the local outward pointing normal and $\bf F^*$ denotes the numerical flux. Specifically, the surface integral in the above equation is defined as follows
\begin{equation}
\int_{\partial D_x^i}  \phi_m^i  \left( {\bf F}^*\cdot \hat{\bn}^i \right)\rd{\bx} =\sum_{e\in \partial D_x^i} \int_{e}  \phi_m^i  \left( {\bf F}^*_e\cdot \hat{\bn}^i_e \right) \rd{\bx},
\end{equation}
with $\hat{\bn}^i_e$ and ${\bf F}^*_e$ being the outward normal and numerical flux along the face $e$. In our implementation, we choose the upwind flux:
\begin{equation}
{\bf F}^*_e = 
\begin{cases}
\bc \,f^i(t, \mathbf{x}_{e,\;int(D_x^i)},\bc), \quad \bc \cdot \hat{\bn}_e^i \geq 0
\\
\bc \, f^i(t, \mathbf{x}_{e,\;ext(D_x^i)},\bc), \quad \bc \cdot \hat{\bn}_e^i< 0
\end{cases}
\label{eq_numericalFlux_upwind}
\end{equation}
where \textit{int} and \textit{ext} denote interior and exterior of the face $e$ respectively.

Note that the second term in equation (\ref{eq_dg_innerProd}) can be expanded as
\begin{equation}
\begin{split}
	\int_{D_x^i} \phi_l^i \,\nabla_{\bx}\cdot (\bc \, \phi_m^i) \rd{\bx}&=c_1 \int_{D_x^i} \phi_l^i\,\frac{\partial \phi_m^i}{\partial x} \rd{\bx}+c_2  \int_{D_x^i} \phi_l^i\,\frac{\partial \phi_m^i}{\partial y} \rd{\bx}\\&+c_3 \int_{D_x^i} \phi_l^i\,\frac{\partial \phi_m^i}{\partial z} \rd{\bx},
	\end{split}
\end{equation}
where $c_1$, $c_2$, $c_3$ are the three components of $\bc$.

Finally, let us define the mass matrix $\mathcal{M}_{ml}$, stiffness matrices $\mathcal{S}^{x}_{ml}$, $\mathcal{S}^{y}_{ml}$,$\mathcal{S}^{z}_{ml}$, and the tensor $\mathcal{H}_{m l_1 l_2}$ as
\begin{equation}
\mathcal{M}_{ml} = \int_{D^i_x} \phi_m^{i}(\mathbf{x}) \, \phi_l^{i}(\mathbf{x}) \rd{\bx},
\label{eq_massMatrix}
\end{equation}
\begin{equation}
\mathcal{S}^{x}_{ml} = \int_{D^i_x} \phi_l^{i}(\mathbf{x}) \, \frac{\partial}{\partial x} \phi_m^{i}(\mathbf{x}) \rd{\bx},
\label{eq_stiffMatrix1}
\end{equation}
\begin{equation}
\mathcal{S}^{y}_{ml} = \int_{D^i_x} \phi_l^{i}(\mathbf{x}) \, \frac{\partial}{\partial y} \phi_m^{i}(\mathbf{x}) \rd{\bx},
\label{eq_stiffMatrix2}
\end{equation}
\begin{equation}
\mathcal{S}^{z}_{ml} = \int_{D^i_x} \phi_l^{i}(\mathbf{x}) \, \frac{\partial}{\partial z} \phi_m^{i}(\mathbf{x}) \rd{\bx},
\label{eq_stiffMatrix_Sz}
\end{equation}
\begin{equation}
\mathcal{H}_{m\, l_1 l_2} = \int_{D^i_x} \phi_m^{i}(\mathbf{x}) \, \phi_{l_1}^{i}(\mathbf{x}) \,\phi_{l_2}^{i}(\mathbf{x}) \rd{\bx}.
\label{eq_stiffMatrix3}
\end{equation}
Then the equation (\ref{eq_dg_innerProd}) can be recast as
\begin{equation}
\begin{split}
&\sum_{l=1}^{K} \mathcal{M}_{ml} \frac{\partial}{\partial t} \mathcal{F}_l^i - c_1 \sum_{l=1}^{K}\mathcal{S}^{x}_{ml} \mathcal{F}_l^i  - c_2 \sum_{l=1}^{K}\mathcal{S}^{y}_{ml} \mathcal{F}_l^i  - c_3 \sum_{l=1}^{K}\mathcal{S}^{z}_{ml} \mathcal{F}_l^i \\
=&-\sum_{e\in \partial D_x^i}  \int_{e}  \phi_m^i  \left( {\bf F}^*_e\cdot \hat{\bn}^i_e \right) \rd{\bx}+ \frac{1}{Kn}\sum_{l_1,l_2=1}^{K} \mathcal{H}_{m \, l_1  l_2}  \mathcal{Q} \left(\mathcal{F}_{l_1}^i,  \mathcal{F}_{l_2}^i\right),
\label{eq_weakForm}
\end{split}
\end{equation}
for $1\leq m \leq K$. Equation (\ref{eq_weakForm}) is the DG system we are going to solve in each element $D_x^i$ of the physical space.

\subsection{Discretization in the velocity space}
\label{subsec:vel}

To further discretize the system (\ref{eq_weakForm}) in the velocity space, we employ a finite difference (or discrete velocity) discretization. Each velocity component $c_i$ ($i\,  \in \, \{1, 2, 3\}$) is discretized uniformly with $N$ points in the interval $[-L,L]$. The grid points are chosen as $-L+ (j-1/2) \Delta c$, with $j=1,\dots, N$ and $\Delta c = 2L/N$ (the choice of $L$ is given below). For brevity we will use $\bc^j$ to denote the 3D velocity grid point.

The reason of using the uniform velocity grid is because our fast algorithm for the collision operator is based on Fourier transform, which is naturally done on a uniform mesh (see Appendix for details). Simply speaking, it takes the function values at the grid points as input, does the calculation (including forward and backward FFTs) in a black box solver, and outputs the values of the collision operator at the same grid points. Inside the solver, it assumes the distribution function has a compact support, and chooses a relatively large computational domain enclosing this support, then periodically extends the function to the whole space $\mathbb{R}^3$. As such, the method can achieve spectral accuracy (subject to domain truncation error which is usually very small); furthermore, the simple mid-point rule would also allow one to construct the moments with spectral accuracy.

To determine the domain size $L$, we first choose the maximum temperature $T_{\text{max}}$ and velocity ${\bf u}_{\text{max}}$ specified at all boundaries, and estimate $\mu$ such that the interval $[c_{\text{min}},c_{\text{max}}]$ defined as
\begin{equation}
c_{\text{max}}, \, c_{\text{min}} = |{\bf u}_\text{max}| \pm \mu \sqrt{T_{\text{max}}},
\end{equation}
can produce the correct values of $T_{\text{max}}$ and ${\bf u}_{\text{max}}$ (i.e., it is large enough that the tail truncation effects of the Gaussian characterized by $T_{\text{max}}$ and ${\bf u}_{\text{max}}$ are negligible). Finally, $L$ is chosen as
\begin{equation}
L= 2.2 \max (|c_{\text{max}}|,|c_{\text{min}}|),
\end{equation}
which is a relatively safe choice to avoid aliasing effect (\cite{PR00}). In general, the parameter $\mu$ ranges between 1 to 3.

With the above setup, we just need to solve the system (\ref{eq_weakForm}) at each velocity grid $\bc^j$ and in each spatial element $D_x^i$.

The macroscopic quantities defined in (\ref{moments1}): density, bulk velocity, temperature, stress tensor, and heat flux in the spatial element $D_x^i$ can be recovered using numerical integration (mid-point rule) of the distribution function over the entire velocity grid:
\begin{equation}
\begin{split}
&	n^i(t,\bx) = \sum_j f^{i}(t,\bx,\bc^j)\, \Delta \bc,\\
&	\bu^i(t,\bx) = \frac{1}{n^i} \sum_j f^{i}(t,\bx,\bc^j) \bc^j\,  \Delta \bc,\\
&	T^i(t,\bx)= \frac{2}{3n^i} \sum_j f^{i}(t,\bx,\bc^j) |\bc^j - \bu^i|^2  \, \Delta \bc,\\
&	\mathbb{P}^i(t,\bx) = 2\sum_j f^{i}(t,\bx,\bc^j) (\mathbf{c}^j - \bu^i) \otimes (\mathbf{c}^j - \bu^i) \,\Delta \bc,\\
&	\mathbf{q}^i(t,\bx) = \sum_j f^{i}(t,\bx,\bc^j) (\mathbf{c}^j - \bu^i) |\mathbf{c}^j - \bu^i|^2 \,\Delta \bc,
\end{split}
\end{equation}
where $\Delta \bc=\Delta c^3$. Note that $n^i$, $\bu^i$, $T^i$, $\mathbb{P}^i$, $\mathbf{q}^i$ are polynomials defined in each element since $f^{i}(t,\bx,\bc^j)$ are polynomials.

\subsubsection{Evaluation of the collision term}
\label{subsec:collision}

We are now left with the issue of evaluating the collision term in (\ref{eq_weakForm}):
\begin{equation}
\sum_{l_1,l_2=1}^{K} \mathcal{H}_{m \, l_1  l_2}  \mathcal{Q} \left(\mathcal{F}_{l_1}^i,  \mathcal{F}_{l_2}^i\right)(\bc).
\label{HH}
\end{equation}
Note here that the collision operator $\mathcal{Q}$ acts only in the velocity variable $\bc$.

The method we use was proposed in \cite{GHHH17}. Given a function $f$ at $N^3$ velocity grid, it produces $\mathcal{Q}(f,f)$ at the same grid with $O(MN^4\log N)$ complexity, where $M$ is the number of quadrature points on the sphere and $M\ll N^2$. In the Appendix, we give a brief description of this method for evaluating the operator of the form $\mathcal{Q}(f,g)$ with general collision kernel~(\ref{eq_dim_B}). Compared to the original method in \cite{GHHH17}, we improve the accuracy and efficiency by using a different quadrature on the half sphere.

Equipped with the fast collision solver, the complexity of evaluating (\ref{HH}) would be $O(K^2MN^4\log N)+O(K^3N^3)$ for all $m$, where the first term is to generate $\mathcal{Q} \left(\mathcal{F}_{l_1}^i,  \mathcal{F}_{l_2}^i\right)(\bc)$ for all $l_1$ and $l_2$, and the second term is to evaluate the outer double summation. For (relatively) high-order polynomial approximations, $K$ can be large. To further reduce the cost, here we propose a simple approach based on singular value decomposition (SVD).

For each fixed m ($1\leq m \leq K$), we precompute the SVD of the matrix $(\mathcal{H}_{m \, l_1  l_2})_{K\times K}$ as
\begin{equation}
	\mathcal{H}_{m \, l_1  l_2}= \sum\limits_{r=1}^{R_m} U_{l_1,r}^m V_{r,l_2}^m,
\label{eq_coln_derivation_SVD_3}
\end{equation}
where $R_m$ is the rank of the matrix and $R_m\leq K$ (the diagonal matrix in the usual SVD has been absorbed in the term $V$ in the above notation). Substituting (\ref{eq_coln_derivation_SVD_3}) into (\ref{HH}) yields
\begin{equation}
\begin{split}
\sum_{l_1,l_2=1}^{K}	\sum\limits_{r=1}^{R_m} U_{l_1,r}^m V_{r,l_2}^m \mathcal{Q} (\mathcal{F}_{l_1}^i,  \mathcal{F}_{l_2}^i)(\bc) = \sum\limits_{r=1}^{R_m} \mathcal{Q} \left( f_r^{i,m},  g_r^{i,m} \right)(\bc),\\
	\text{with} \quad  f_r^{i,m}:=\sum_{l_1=1}^{K} U_{l_1,r}^m \mathcal{F}_{l_1}^i(\bc), \quad g_r^{i,m}:=\sum_{l_2=1}^{K} V_{r,l_2}^m \mathcal{F}_{l_2}^i(\bc).
	\end{split}
\label{eq_coln_derivation_SVD_4}
\end{equation}
Note that the functions $f_r^{i,m}$ and $g_r^{i,m}$ can be computed in a different loop. Therefore, the complexity of evaluating (\ref{HH}) becomes $O(\sum_{m=1}^K R_m MN^4\log N)+O(\sum_{m=1}^K R_mKN^3)$. For the conventional nodal DG basis \cite{karniadakis1999spectral,hesthaven2007nodal} used in the current work, we found that for many $m$, $R_m$ can be much smaller than $K$, thus $\sum_{m=1}^K R_m$ is strictly less than $K^2$. Comparing with the aforementioned direct method, we can see that the SVD approach always saves. Considering that the evaluation of the collision operator always constitutes the main bottleneck in the computation, this saving, may not be in the order of magnitude, is still appreciable.

We mention that the rank $R_m$ of the matrix $(\mathcal{H}_{m \, l_1  l_2})_{K\times K}$ strongly depends on the underlying DG basis. The structure of $\mathcal{H}_{m \, l_1  l_2}$ for various element shapes is currently under study and will be reported in future work.

\subsection{Discretization in time}

Once the spatial and velocity discretization is done, the time discretization can be performed by simply applying an explicit Runge-Kutta method to the system (\ref{eq_weakForm}). Here we adopt the widely used strong-stability-preserving (SSP) RK schemes \cite{GKS11}.

For notational simplicity, we rewrite the system (\ref{eq_weakForm}) as 
\begin{equation}
	\frac{\partial}{\partial t} \mathcal{F}^i = \mathcal{L} \;(\mathcal{F}^i),
\end{equation}
and use $\mathcal{F}^i$ to denote the solution vector with components $\mathcal{F}^i_m$, $1\leq m\leq K$.

Then the 2nd order SSP-RK scheme is given by
\begin{align} \label{SSPRK2}
\left\{
\begin{array}{l}
\displaystyle v^{(1)}= \mathcal{F}^i+\Delta t \mathcal{L} \;(\mathcal{F}^i), \\
\displaystyle \mathcal{F}^{i,\text{new}} = \frac{1}{2}\mathcal{F}^i+\frac{1}{2}v^{(1)}+\frac{1}{2}\Delta t \mathcal{L} \;(v^{(1)});
\end{array} \right.
\end{align} 
and the 3rd order SSP-RK scheme is given by
\begin{align} \label{SSPRK3}
\left\{
\begin{array}{l}
\displaystyle v^{(1)}= \mathcal{F}^i+\Delta t \mathcal{L} \;(\mathcal{F}^i), \\
\displaystyle v^{(2)}= \frac{3}{4}\mathcal{F}^i+\frac{1}{4}v^{(1)}+\frac{1}{4}\Delta t \mathcal{L} \;(v^{(1)}), \\
\displaystyle \mathcal{F}^{i,\text{new}} = \frac{1}{3}\mathcal{F}^i+\frac{2}{3}v^{(2)}+\frac{2}{3}\Delta t \mathcal{L} \;(v^{(2)}).
\end{array} \right.
\end{align}

\subsection{Initial and boundary conditions}
The initial value of the distribution function is set to Maxwellian at given initial macroscopic conditions $n_{ini}(\bx)$, $T_{ini}(\bx)$, and $\mathbf{u}_{ini}(\bx)$:
\begin{equation}
	f_{ini}(\bx,\bc) = \frac{n_{ini}}{(\pi T_{ini})^{3/2}} \exp \Bigg[ - \frac{(\mathbf{c} - \mathbf{u}_{ini})^2}{T_{ini}}  \Bigg].
\end{equation} 

For the test cases considered in the current work, the fully diffusive Maxwell boundary condition is assumed at the wall \cite{mieussens2000discrete} except the normal shock wave example in Section~\ref{sec:normal}. Consider a wall moving with velocity $\mathbf{u}_w(t,\bx)$, and is at temperature $T_w(t,\bx)$, the inflow boundary condition at $\bx \in \partial \Omega_x$ with the local outward pointing normal ${\bf \hat{n}}$ is given by
\begin{equation}
	f(t,\bx,\bc)=n_w f_w, \quad (\bc-{\bf u}_w)\cdot {\bf \hat{n}}<0,
\end{equation} 
with
\begin{equation}
f_w(t,\bx,\bc)=  \exp \Big[ - \frac{(\mathbf{c} - \mathbf{u}_w)^2}{T_w}  \Big],
\end{equation}
and $n_w$ is determined from conservation of mass as
\begin{equation}
	n_w = - \frac{\int_{(\mathbf{c} - \mathbf{u}_w)\cdot {\bf \hat{n}} \geq 0}(\mathbf{c} - \mathbf{u}_w)\cdot {\bf \hat{n}} f \rd{\bc}}
	{ \int_{(\mathbf{c} - \mathbf{u}_w)\cdot {\bf \hat{n}} < 0} (\mathbf{c} - \mathbf{u}_w)\cdot {\bf \hat{n}}  f_w \rd{\bc}}.
\end{equation}
{For the normal shock wave example, we use the inflow boundary condition at $\bx \in \partial \Omega_x$:}
\begin{equation}
	f_{in}(t,\bx,\bc)=  \frac{n_{in}}{(\pi T_{in})^{3/2}} \exp \Big( - \frac{(\bc - \mathbf{u}_{in})^2}{T_{in}} \Big), \quad (\bc)\cdot {\bf \hat{n}}<0,
\end{equation}
{where $n_{in}(t,\bx)$, $\mathbf{u}_{in}(t,\bx)$, $T_{in}(t,\bx)$ are the prescribed inlet conditions.}
Details about other boundary conditions can be found in \cite{Bird,Cercignani,su2015parallel}.

\section{Numerical experiments and results}
In this section, we evaluate the accuracy of the proposed discontinuous Galerkin fast spectral method, which we shall denote by the acronym DGFS in the following. A nodal DG basis has been used similar to the ones described in \cite{karniadakis1999spectral,hesthaven2007nodal}.

Standard benchmark cases of Bobylev-Krook-Wu (BKW) solution \cite{bobylev1975exact,krook1977exact}, planar Fourier heat transfer, Couette flow, oscillatory Couette flow, normal shock, lid driven cavity flow, and thermally driven cavity flow have been considered in the present work. The results are compared with those obtained from the DSMC method \cite{Bird} with equivalent molecular collision models, analytical solution, or published deterministic solutions, wherever applicable. 

\subsection{Solver configurations}
\textit{SPARTA} \cite{gallis2014direct} has been employed for carrying out DSMC verifications in the present work. It implements the DSMC method as proposed by Bird \cite{Bird}. The solver has been benchmarked \cite{gallis2014direct} and widely used for studying hypersonic, subsonic and thermal \cite{gallis2017molecular,gallis2016direct,pekardan2016rarefaction,sebastiao2018direct} gas flow problems. In this work, cell size less than $\lambda/3$ has been ensured in all the test cases. A minimum of 30 DSMC simulator particles per cell are used in conjunction with the no-time collision (NTC) algorithm. Each \textit{steady-state} simulation has been averaged for a minimum 100,000 steps so as to minimize the statistical noise.

Our numerical tests in this work are restricted to monatomic gases. Argon gas with mass $m=6.63 \times 10^{-26} $ kg, reference viscosity of $2.117 \times 10^{-5}$ N/m$\cdot$s at reference temperature $T_\mathrm{ref}$ of $273 K$ is selected. The molecular diameters are selected so as to maintain the reference viscosity: $d_\mathrm{ref}=4.59 \angstrom$, $\omega=1.0$ for the Maxwell
collision model, and $d_{ref}=4.17 \angstrom$, $\omega=0.81$ for the VHS collision model. These values are consistent for both DSMC and DGFS in all test cases unless otherwise explicitly stated. 

In rarefied gas dynamics, two widely used definitions of Knudsen number exist. The first definition is by Cercignani \cite{Cercignani}, the second definition is by Bird \cite{Bird} (i.e. the equation~(\ref{eq_Kn}) in the present work). Here we want to compare our results with DSMC results published in the literature, for instance, Fourier heat transfer in Gallis at al. \cite{gallis2002calculations}, Couette flow in Gu et al. \cite{gu2009high}. These works use Cercignani's definition. Therefore, for consistency, the Knudsen number defined in this section follows Cercignani's definition, i.e., $Kn=2\mu/n_0 m\bar{c}H_0$, where $\bar{c}=(8 k_B T_0/\pi m )^{1/2}$ and $\mu$ is the dynamic viscosity. Using Bird's power law for viscosity, the ratio of these two Knudsen numbers is simply a constant given by
\begin{align}
\frac{\Kn_\text{Cercignani}}{\Kn_\text{Bird}} = \pi \frac{5(\alpha+1)(\alpha+2)}{4 \alpha (5-2\omega) (7-2\omega)}.
\end{align}

\subsection{Hardware configurations}
MPI-parallel implementation of DSMC solver (SPARTA) is run on Intel E5-2680 Xeon(TM) Processor v2 2.80 GHz (Conte cluster at Purdue). The operating system used is 64-bit RHEL 6.7. The solver has been written in C++ and is compiled using OpenMPI mpic++ 1.8.1, g++ 5.2.0 with OpenMP-4.0 support, and third level optimization flags.GPU-parallelized implementations of DGFS solver are run on Intel Xeon E5 2623 v4 2.60 GHz CPU with NVIDIA Titan-X (Pascal) GPU accompanying CUDA driver 8.0 and CUDA runtime 8.0. The operating system used is 64-bit Red Hat 6.9 (Santiago). The GPU has 5376 CUDA cores, 12GB device memory, and compute capability of 6.1. The solver has been written in C++/CUDA and is compiled using g++ 5.3.0, and  nvcc 8.0.44 compiler with third level optimization flag. All the simulations are done with double precision floating point values.

\subsection{0D case: BKW solution}

For constant collision kernel $B = 1/(4 \pi)$, an analytical solution to the spatially homogeneous Boltzmann equation 
\begin{equation}
	\frac{\partial f}{\partial t} = \mathcal{Q}(f,f)
\label{eq_bkw_pde}
\end{equation}
can be constructed as (see \cite{bobylev1975exact,krook1977exact})
\begin{equation}
f(t,\bc) = \frac{1}{2(2\pi K(t))^{3/2}} \exp \left(-\frac{\bc^2}{2K(t)}\right) \left(\frac{5 K(t) - 3}{K(t)} + \frac{1 - K(t)}{K^2(t)} \bc^2\right),
\label{eq_bkw_f}
\end{equation}
where $K(t) = 1 - \exp(-t/6)$. 
{Upon differentiation, one recovers the exact $\mathcal{Q}$ as}
\begin{align}
\mathcal{Q}(f,f) = \frac{\partial f}{\partial t} = K'(t) \left( -\frac{3}{2K(t)} + \frac{\bc^2}{2K(t)^2} \right) f \; +\nonumber \\ \left[\frac{1}{2(2\pi K(t))^{3/2}} \exp\left(-\frac{\bc^2}{2K(t)}\right) \left( \frac{3}{K(t)^2} + \frac{K-2}{K^3} \bc^2\right) \right] K'(t),
\label{eq_bkw_Q}
\end{align}
{where $K'(t)=\exp(-t/6)/6$}. The initial time $t_0$ must be greater than $6 \ln(2.5) \approx 5.498$ for $f$ to be positive. An arbitrary time of $t_0=5.5$ has been picked in the present work. The 3rd order
SSP-RK scheme (\ref{SSPRK3}) with $\Delta t = 0.01$ is employed for time integration. Velocity domain size $[-6.62,\;6.62]^3$ has been used for the present case.

\subsubsection{Error in evaluation of the collision operator}
\label{sec_bkw_error_Nr}
{
Using~(\ref{eq_bkw_Q}), one can verify the accuracy of the proposed method without introducing additional time discretization error. Table~\ref{tab_bkw_error} shows the error in evaluating the collision operator i.e., $\|Q_\text{numerical}-Q_\text{exact}\|_{L^\infty}$. As noted in the Appendix, the total number of Gauss-Legendre quadrature points $N_r$ in the radial direction $c_r$ should be on order of $O(N)$. As per \cite{GHHH17}, a more precise estimate is $\approx 0.8\,N$. However, there is no good rule to select optimal $N_r$. From Table~\ref{tab_bkw_error}, we observe that the error is relatively unaffected upon reducing $N_r$ from $N$ to $N/2$. However, we note that $N$ is a safer choice. For all cases considered henceforth, $N_r=N$, unless otherwise explicitly stated.
}
\begin{table}[!ht]
\caption{{$\|\mathcal{Q}_\text{numerical}-\mathcal{Q}_\text{exact}\|_{L^\infty}$ evaluated at different time instants. $M = 6$ points are used on the half-sphere for all cases. $N$ discretization points in the velocity space, and $N_r$ Gauss-Legendre quadrature points are used in the radial direction.}}
\centering
\begin{tabular}{@{}lc|ccc@{}}
\toprule
$N$ & $N_r$ & \multicolumn{3}{c}{$\|\mathcal{Q}_\text{numerical}-\mathcal{Q}_\text{exact}\|_{L^\infty}$} \\ 
    &       & $t_0=5.5$ & $t_0=6.5$ & $t_0=10$\\ 
\midrule
 8 &  2 & 4.06e-03 & 2.26e-03 & 6.04e-04 \\
   &  4 & 1.51e-03 & 5.58e-04 & 1.00e-03 \\
   &  8 & 1.26e-03 & 6.92e-04 & 1.16e-03 \\
16 &  4 & 1.89e-03 & 7.42e-04 & 9.06e-05 \\
   &  8 & 1.65e-04 & 8.22e-05 & 7.39e-06 \\
   & 16 & 1.71e-04 & 8.38e-05 & 6.56e-06 \\
24 &  6 & 7.72e-04 & 5.18e-04 & 6.70e-05 \\
   & 12 & 2.41e-05 & 4.22e-06 & 9.05e-08 \\
   & 24 & 2.42e-05 & 4.19e-06 & 9.07e-08 \\
32 &  8 & 2.10e-04 & 5.77e-05 & 2.39e-06 \\
   & 16 & 5.22e-08 & 3.90e-08 & 7.04e-08 \\
   & 32 & 5.23e-08 & 3.90e-08 & 7.04e-08 \\
48 & 12 & 7.40e-07 & 1.26e-07 & 7.02e-08 \\
   & 24 & 1.88e-08 & 3.81e-08 & 7.04e-08 \\
   & 48 & 1.88e-08 & 3.81e-08 & 7.04e-08 \\
64 & 16 & 1.88e-08 & 3.81e-08 & 7.05e-08 \\
   & 32 & 1.88e-08 & 3.81e-08 & 7.05e-08 \\
   & 64 & 1.88e-08 & 3.81e-08 & 7.05e-08 \\
\bottomrule
\end{tabular}
\label{tab_bkw_error}
\end{table}
\subsubsection{Normalized error}
\begin{figure}[!ht]
	\centering
    \includegraphics[width=80mm]{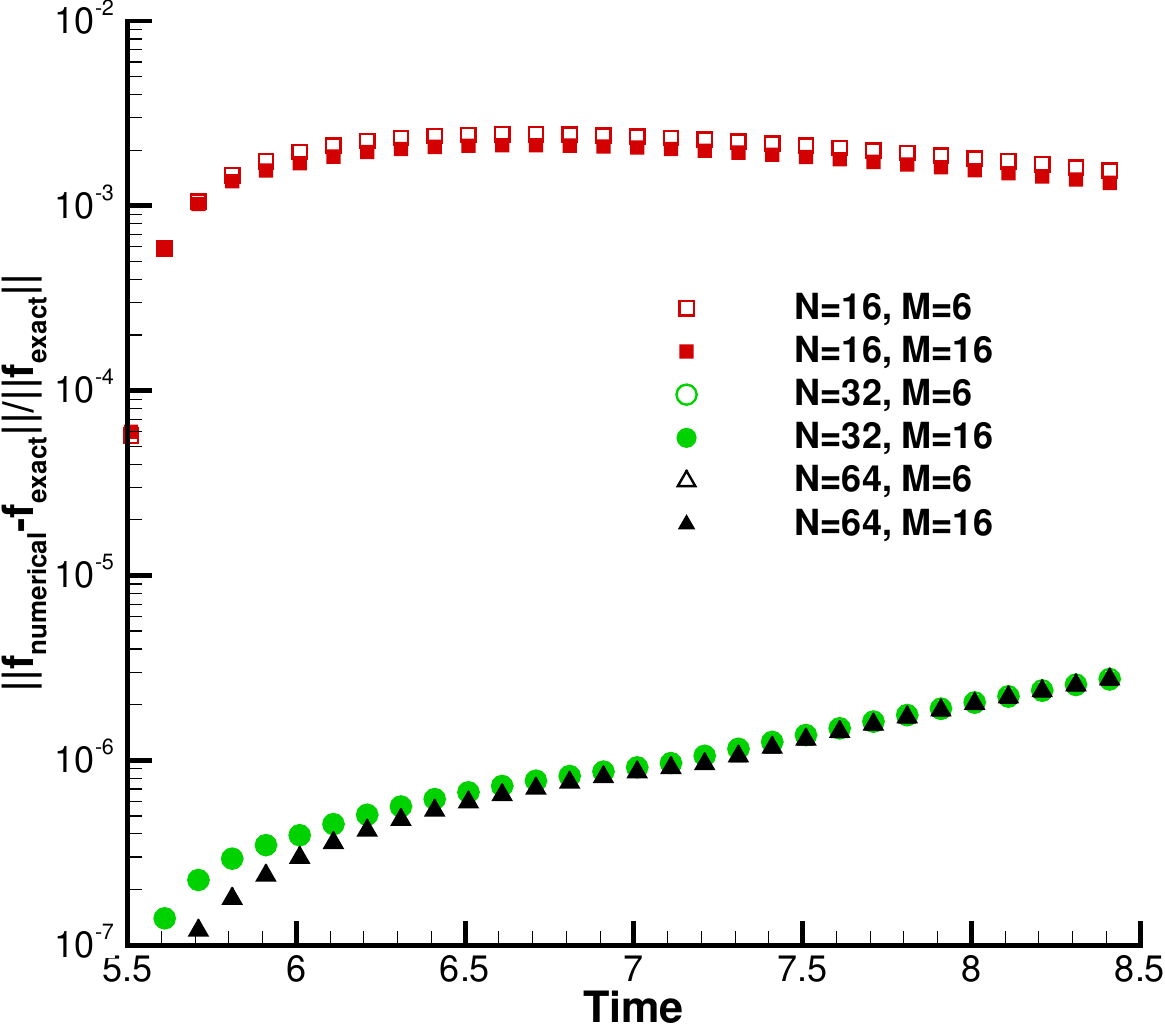}
	\caption{Comparison of BKW analytical and numerical solutions over time with logarithmic $y$-axis.}
	\label{fig_bkw_comparison}
\end{figure}

Figure \ref{fig_bkw_comparison} shows the time evolution of normalized error in $L^\infty$ norm between the numerical and analytical solutions with logarithmic $y$-axis. We have considered the cases $N= 16$, 32, and 64 points in each velocity dimension; and $M=6$, 16 spherical design quadrature points on the half sphere. A good agreement between analytical and numerical solutions is clearly evident from the figure. It is also observed that the differences between $M=6$ and $M=16$ solutions, i.e. $\|\;f_{numerical}|_{M=16}-f_{numerical}|_{M=6}\;\|_{L^\infty}$ are small (quantitatively on the order of $10^{-5}$). The slight increase of the error in the cases of $N=32$ and $64$ is due to the aliasing effect of the spectral method as discussed in \cite{PR00}.

\subsubsection{Time evolution of the distribution function}

\begin{figure}[!ht]
	\centering
     \includegraphics[width=85mm]{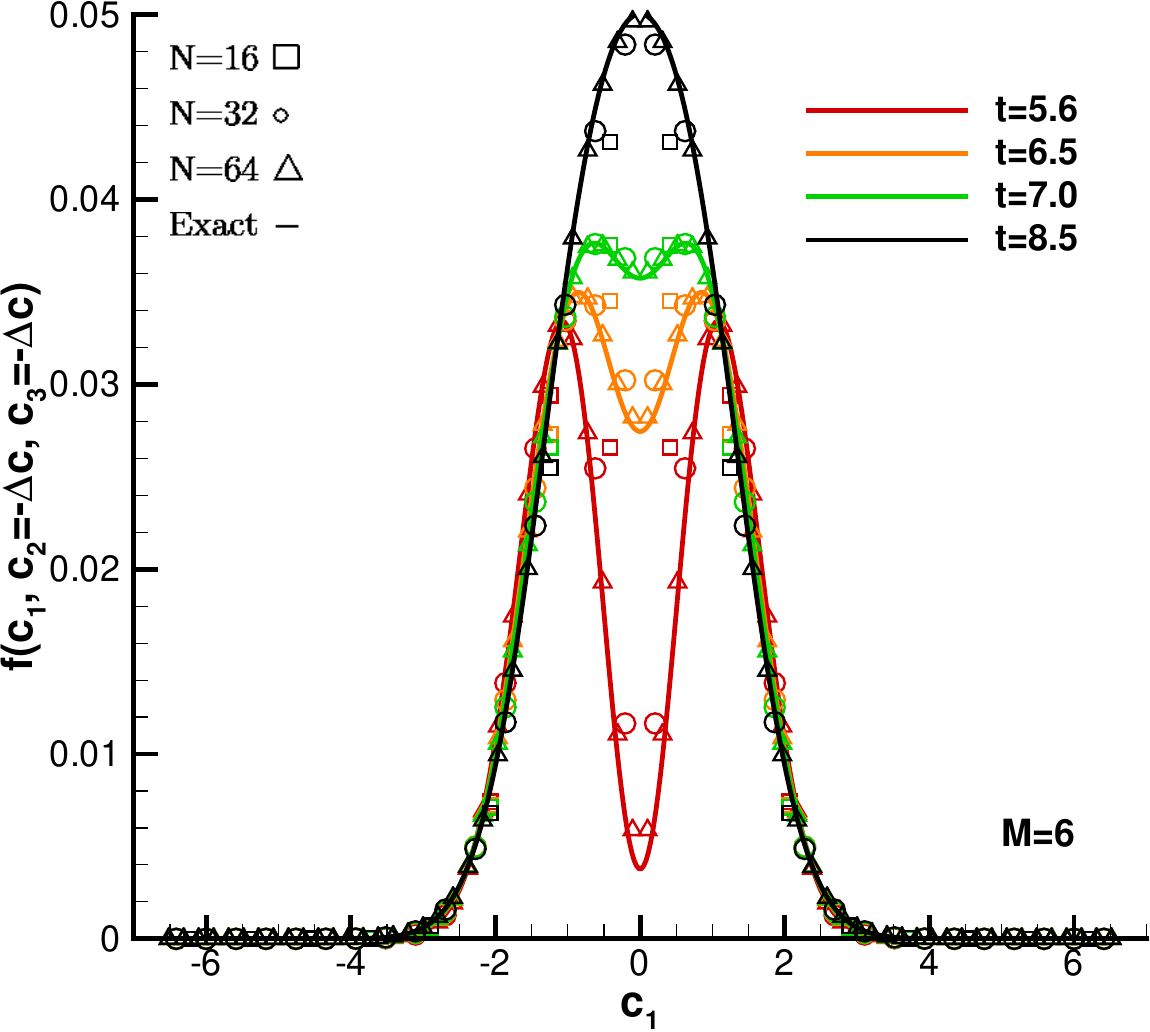}
	\caption{Comparison of BKW analytical and numerical solutions over time. The analytical solution is plotted by discretizing the velocity space with $N=256$ points. The numerical solution is evaluated by discretizing the velocity space with $N=16$, $32$ and $64$ as indicated in the plot. $M=6$ is used on the half sphere in all cases.}
	\label{fig_bkw_distEvolution_M6}
\end{figure}

Figure \ref{fig_bkw_distEvolution_M6} illustrates the time evolution of the distribution function sliced along the velocity domain centerline, i.e., $f(:, N/2, N/2)$. The \textit{smooth} analytical solution is plotted by discretizing the velocity space with $N=256$ points. The numerical solution is evaluated by discretizing the velocity space with $N=16$, $32$ and $64$ respectively. $M=6$ spherical design quadrature points is used on the half sphere in all cases. It is observed that: a) as $N$ increases, the numerical solution moves closer to the \textit{smooth} analytical solution at different time instants; b) as time goes by, the distribution function tends toward the Maxwellian. 

\subsubsection{Time evolution of the entropy}
\begin{figure*}[!ht]
\begin{subfigure}[t]{0.50\textwidth}
  \centering
  \includegraphics[width=75mm]{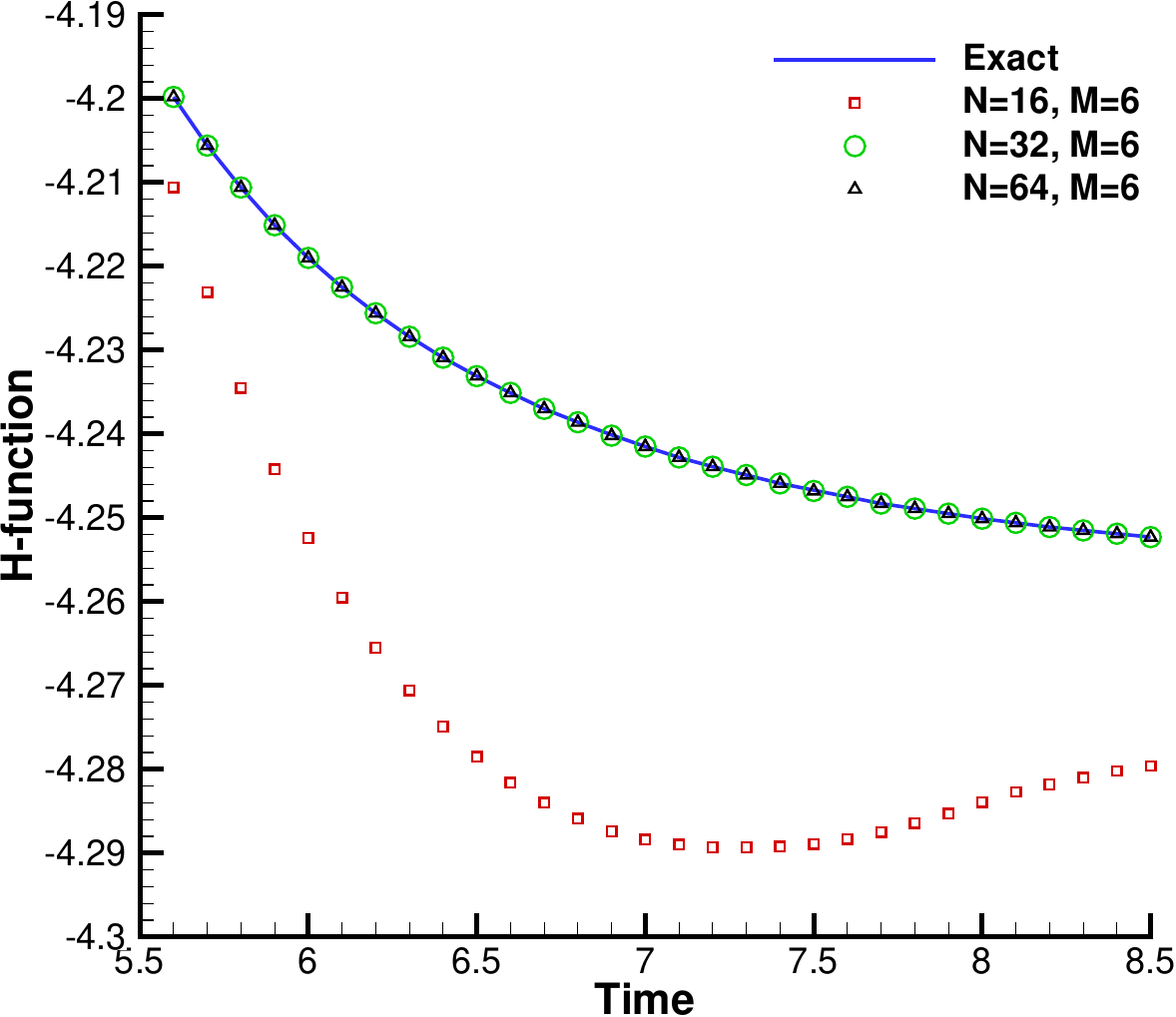}
  \caption{}
  \label{fig_res_hFunc}
\end{subfigure}%
~
\begin{subfigure}[t]{0.50\textwidth}
  \centering
    \includegraphics[width=75mm]{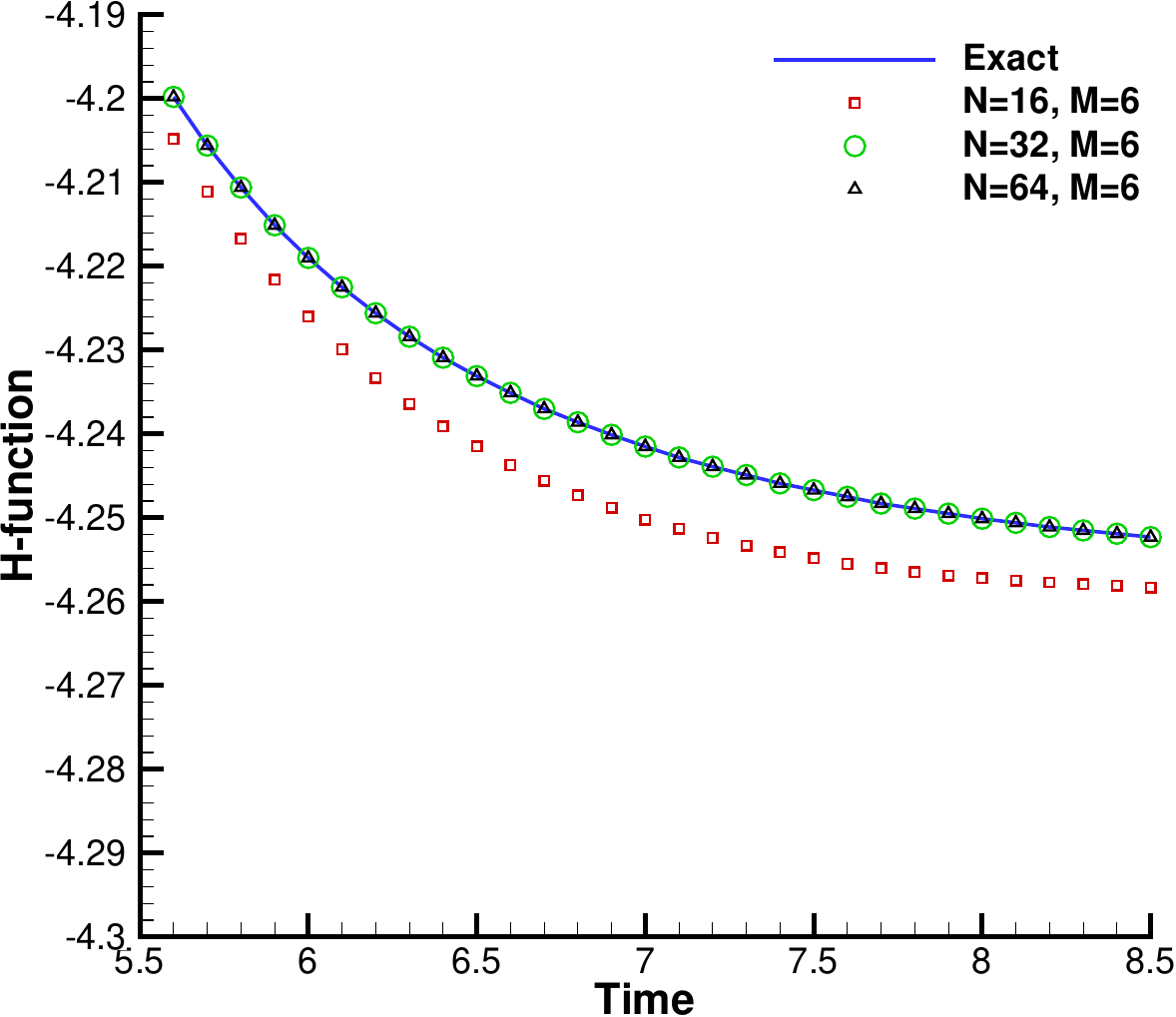}
    \caption{}
  \label{fig_res_hFunc_2}
\end{subfigure}
\caption{Time evolution of BKW analytical and numerical entropy. The analytical entropy is evaluated by discretizing the velocity space with $N=256$ points. The numerical entropy is evaluated by discretizing the velocity space with $N=16$, $32$ and $64$ as indicated in the plot. $M=6$ is used on the half sphere in all cases. At points where $f$ is negative, the left figure evaluates the entropy using $|f|$, whereas the right figure evaluates the entropy by ignoring the negative values.}
\label{fig_res_hFunc_all}
\end{figure*}

The H-theorem states that the entropy is always decreasing (the physical entropy is increasing), which can be expressed mathematically as
\begin{equation}
    \frac{\partial}{\partial t}\int_{\mathbb{R}^3} f\ln f\rd{\bc} = \int_{\mathbb{R}^3} \mathcal{Q}(f,f) \ln f\rd{\bc}\leq 0,
\end{equation}
where $\int_{\mathbb{R}^3} f\ln f\rd{\bc}$ is the so-called H-function or entropy. The entropy can be a powerful quantity for verification of numerical solutions in rarefied flows. Using the mid-point rule, the entropy can be approximated as
\begin{equation}
\int_{\mathbb{R}^3} f\ln f\rd{\bc} \approx \sum_{j}^{} f_j \ln{f_j} \Delta ~c.
\end{equation}The Fourier spectral approximations do not necessarily maintain the positivity of the distribution function. At points where $f_j$ becomes negative, we consider two approaches: (a) evaluate the entropy using the absolute value $|f_j|$, or (b) ignore the contribution of these points in the entropy. Figure \ref{fig_res_hFunc} illustrates the time evolution of the analytical entropy and numerical entropy using the approach (a); and Figure \ref{fig_res_hFunc_2} illustrates the same quantity evaluated using the approach (b). We observe that, in particular, for $N=16$ velocity grid, although the relative error of the distribution function is on the order of $10^{-3}$ as shown in Figure \ref{fig_bkw_comparison}, the entropy in this case evaluated using approach (a) is significantly lower than the analytical one, and also qualitatively violates the second law of thermodynamics; the entropy evaluated using approach (b) is not very accurate as well, however, it does predict the correct trend. Hence the second approach is preferable. We believe that this comparison of analytical/numerical entropy is fairly significant for establishing the importance of the fast spectral method.

\subsection{1D case: Fourier heat transfer}
\begin{figure}[!ht]
	\centering

\begin{tikzpicture}
		\def\pr{1.23};
		\foreach \x in {1,...,5}
			\fill ({-\pr + (\pr)^\x},0cm) circle (0.05cm);
			
		\foreach \x in {1,...,5}
				\fill ({ 4 + \pr - (\pr)^(\x)},0cm) circle (0.05cm);

		\draw(0,0) -- (4,0) ;
		\draw[blue] (0.0,0.0) circle (0.1cm);
		\draw[red] (4.0,0.0) circle (0.1cm);
		
		\draw[-latex] (0.0,-0.5) -- (0.0, 0.0) node[below, yshift=-0.5cm] {$\mathbf{u}_l,\;T_l$};
		\draw[-latex] (4.0,-0.5) -- (4.0, 0.0) node[below, yshift=-0.5cm] {$\mathbf{u}_r,\;T_r$};
		
		\draw[-latex] (-2.0,-1) -- (-1.0, -1) node[anchor=north] {x};
		\draw[-latex] (-2.0,-1) -- (-2.0, 0) node[anchor=south] {y};
\end{tikzpicture} 	\caption{Numerical setup for 1D Fourier/Couette/oscillatory-Couette flow. Distance between the walls is fixed as $H$. Note that the cells are finer in the near-wall region. The domain size $H$ is fixed to $10^{-3}$ meter.}
	\label{fig_couetteFlowSchematic}
\end{figure}
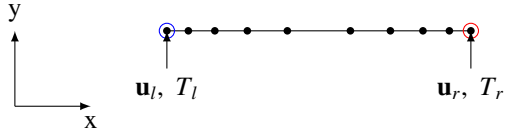
For the spatially inhomogeneous case, analytical solutions for the Boltzmann equation do not exist. Therefore, we compare our results with DSMC which solves the Boltzmann equation stochastically. In the current test, the coordinates are chosen such that the walls are parallel to the $y$ direction and $x$ is the direction perpendicular to the walls. The geometry as well as boundary conditions are shown in Figure \ref{fig_couetteFlowSchematic}. The two parallel walls, at rest, are set $H$ distance apart. The reference ($T_\mathrm{ref}$), left-wall ($T_l$), and right-wall ($T_r$) temperatures are 273K, 263K, and 283K, respectively. The simulation is carried out at three different Knudsen numbers namely $Kn=0.4745$, $Kn=1.582$, and $Kn=4.745$ by varying the density while keeping the $H$ fixed. The 2nd order SSP-RK scheme (\ref{SSPRK2}) is used for time evolution. Argon with $Maxwell$ collision model is taken as the working gas (see \cite{gallis2002calculations} for additional DSMC conditions). 

\subsubsection{Validation: SVD v.s. direct algorithm}
\begin{figure}[!ht]
	\centering
	\includegraphics[width=80mm]{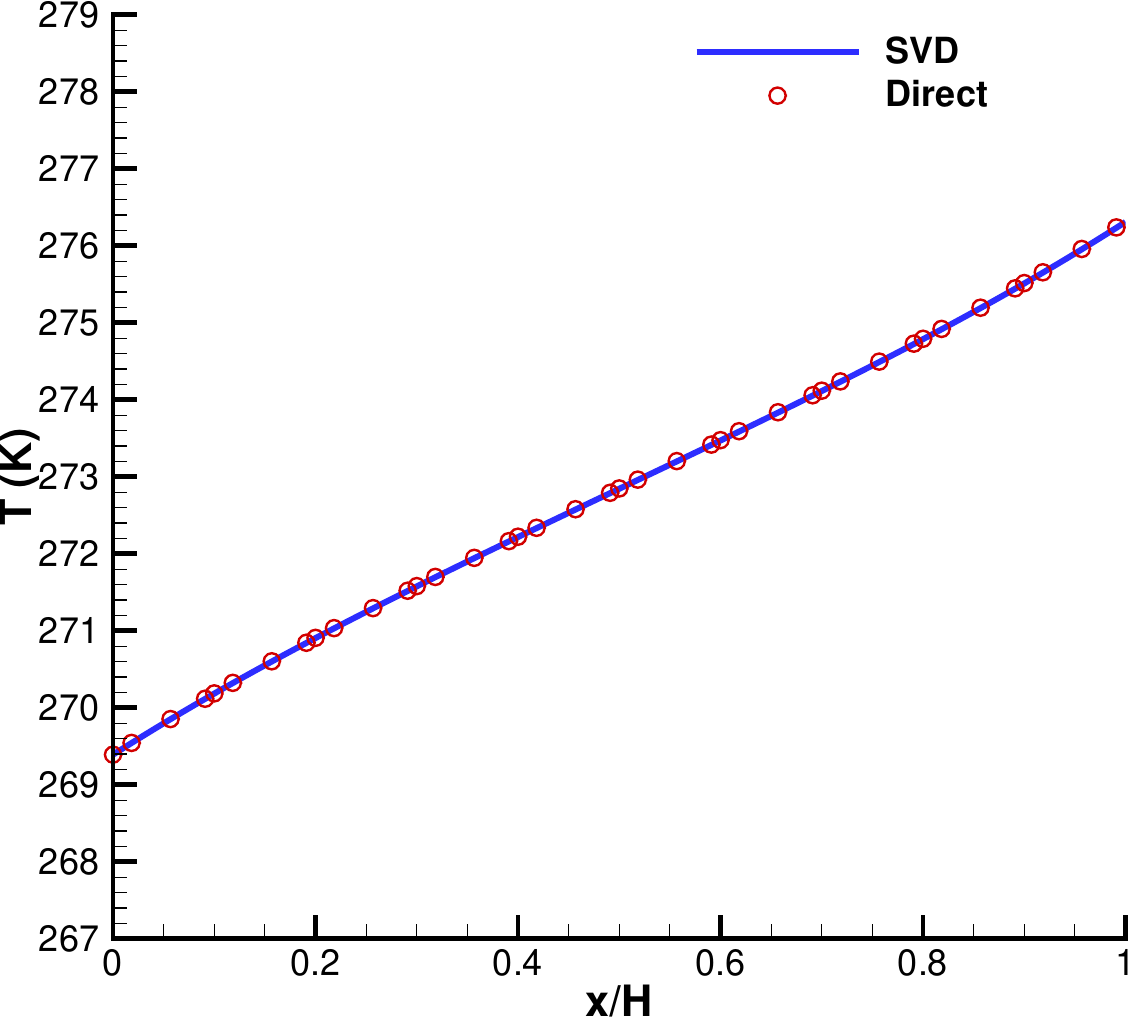}    
	\caption{Variation of temperature along the domain length, obtained using SVD and direct algorithm variants of DGFS at $Kn=1.582$ using Maxwell collision model for Argon molecules. The walls are kept at the temperature difference of 20K. The physical space is discretized using 10 cells and polynomial order of 2, while the velocity space $[-5.09,\;5.09]^3$ is discretized using $N^3=24^3$ points. $M=6$ is used on the half sphere in all cases.}
	\label{fig_fc_res_SVDvsDirect}
\end{figure}

Figure \ref{fig_fc_res_SVDvsDirect} illustrates the temperature profile along the domain length obtained using the SVD and direct variants of the collision algorithm. It is observed that the corresponding two curves are inextricable which verifies that both algorithms evaluate the same Boltzmann collision operator. 

\subsubsection{Temperature at different Knudsen numbers}

\begin{figure}[!ht]
	\centering
	\includegraphics[width=80mm]{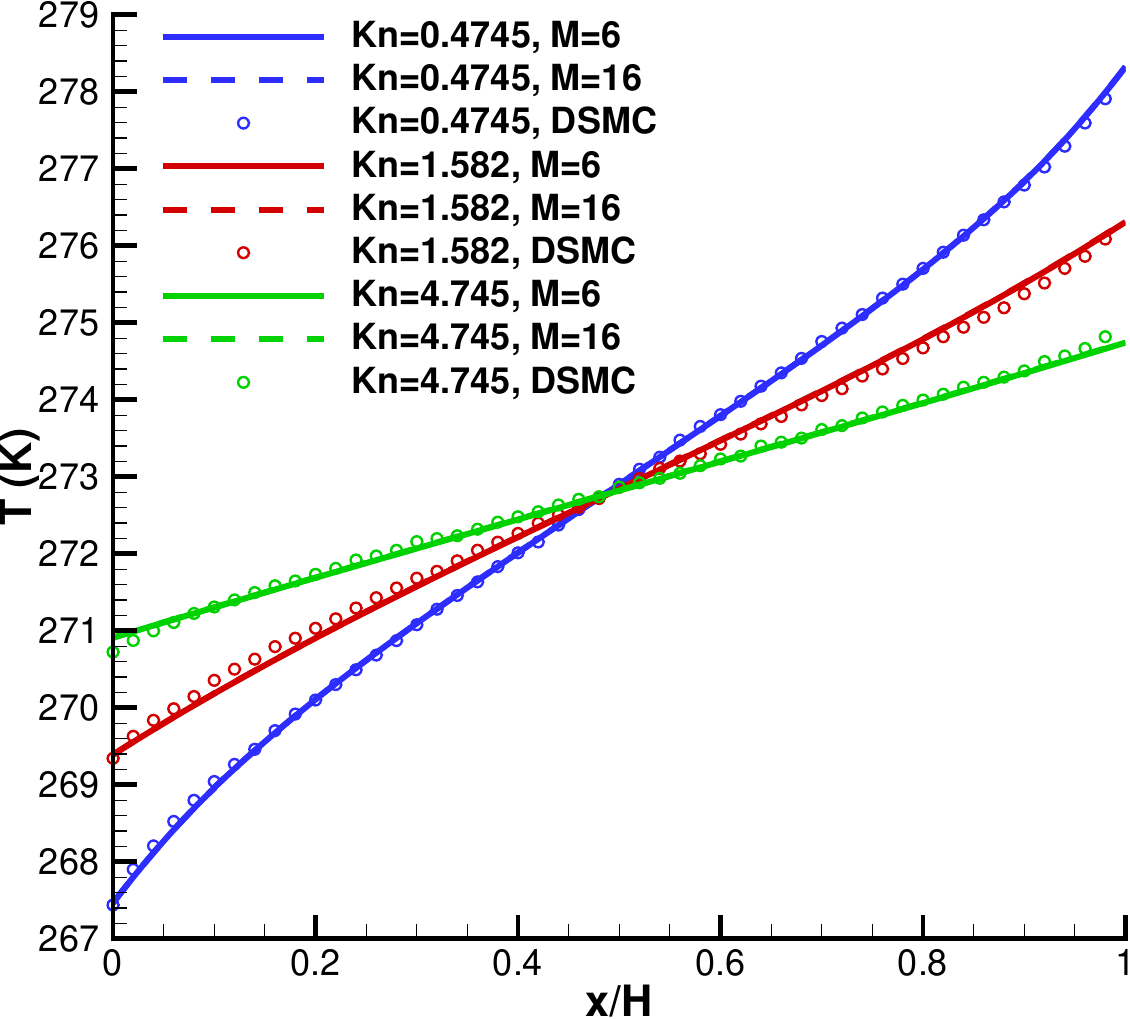}
	\caption{Variation of temperature along the domain length for $Kn=0.4745$, $1.582$, and $4.745$ using Maxwell collision model for Argon molecules obtained with DSMC and DGFS. The walls are kept at the temperature difference of 20K. The physical space is discretized using 10 cells and polynomial order of 2, while the velocity space $[-5.09,\;5.09]^3$ is discretized using $N^3=24^3$ points. $M=6$ and $M=16$ are used on the half sphere.}
	\label{fig_fc_compare_all_T}
\end{figure}

Figure \ref{fig_fc_compare_all_T} illustrate the temperature profile along the domain length for different Knudsen numbers obtained using the SVD variant of the algorithm. The results are compared against the DSMC data \cite{gallis2014direct}, where our DGFS implementation captures the \textit{nonlinear} \cite{lilley2007velocity} nature of temperature profiles in the near wall region, i.e., the Knudsen layer. 

\begin{figure}[!ht]
	\centering
	\includegraphics[width=80mm]{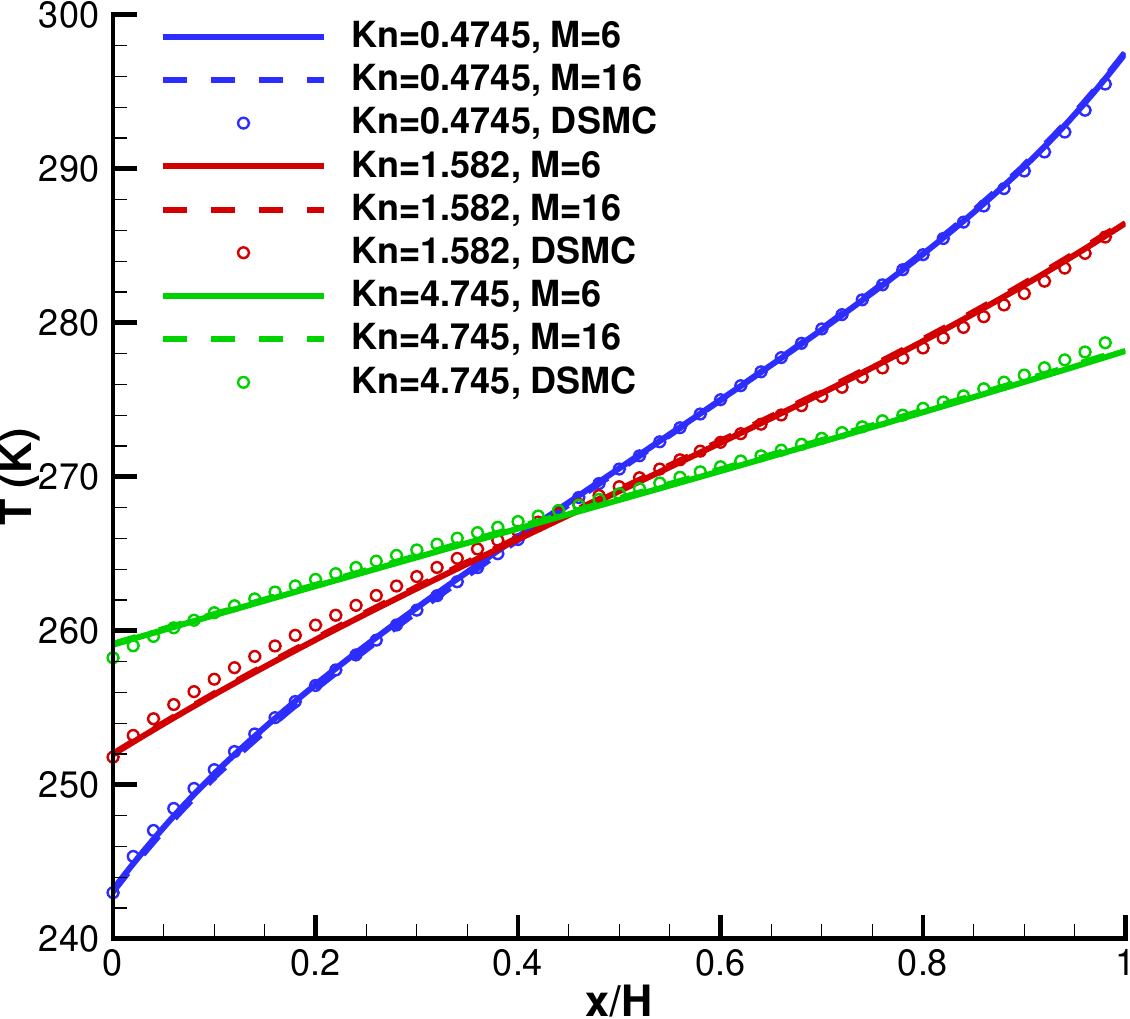}
	\caption{Variation of temperature along the domain length for $\Kn=0.4745$, $1.582$, and $4.745$ using Maxwell collision model for Argon molecules obtained with DSMC and DGFS. The walls are kept at the temperature difference of 100K. The physical space is discretized using 10 cells and polynomial order of 2, while the velocity space $[-5.09,\;5.09]^3$ is discretized using $N^3=24^3$ points. $M=6$ and $M=16$ are used on the half sphere.}
	\label{fig_fc_compare_all_T_100}
\end{figure}

To further highlight the nature of DGFS, we increase the temperature difference between the two walls to $100K$ i.e., $T_\mathrm{l}=223K$ and $T_\mathrm{r}=323K$. Figure \ref{fig_fc_compare_all_T_100} illustrates the results for this case. The results for Fourier heat transfer cases suggest that the combination of $M=6$, $N^3=24^3$, and velocity domain size of $[-5.09,\;5.09]^3$ suffices. In particular, the use of $M=16$ does not change the result significantly. 

From a computation viewpoint, DSMC-SPARTA simulations at $Kn=0.4745$, $\Delta T=100K$ with 500 cells, 30 particles per cell, a time-step of 2e-9 sec, 1 million unsteady time-steps, and 100 million steady time-steps, on 32 CPU processors took 11321.6 sec. These DSMC parameters have been taken from the Gallis et al. \cite{gallis2002calculations}. The parameters have been selected partially to minimize the statistical fluctuations, and avoid linear time-stepping errors inherent to DSMC simulations. On the other hand, DGFS simulations on a single GPU at $Kn=0.4745$, $\Delta T=100K$, with 10 elements, 2nd order polynomial, $N^3=24^3$, $M=6$ took 4456.54 sec to achieve $(\|f^{n+1}-f^{n}\|/\|f^{n}\|_{L_2})/(\|f^{2}-f^{1}\|/\|f^{1}\|_{L_2}) < 5\times10^{-5}$, where $f^{n}$ is the distribution function at $n^\text{th}$ timestep. Note that these are representative simulation times for indicating the computational efforts required in DGFS and DSMC for 1-D simulations. A detailed comparison between CPU and GPU performance is subject of future study.

\subsection{1D case: steady Couette flow}

We now consider the effect of velocity gradient on the solution. The geometry remains the same as in previous case. The left and right parallel walls move with a velocity of $\mathbf{u}_w=(0, \mp50, 0) \;m/s$, and the reference ($T_\mathrm{ref}$), left-wall ($T_\mathrm{l}$), and right-wall ($T_\mathrm{r}$) temperatures are  set to a constant value of 273K. The simulation is carried out at three different Knudsen numbers namely $Kn=0.5$, $Kn=1.0$, and $Kn=5.0$ by varying the density while keeping the $H$ fixed. The 2nd order SSP-RK scheme (\ref{SSPRK2}) is used for time evolution. Argon with $VHS$ collision model is taken as the working gas (see \cite{gu2009high} for additional DSMC conditions).

\begin{figure}[!ht]
	\centering
    \includegraphics[width=80mm]{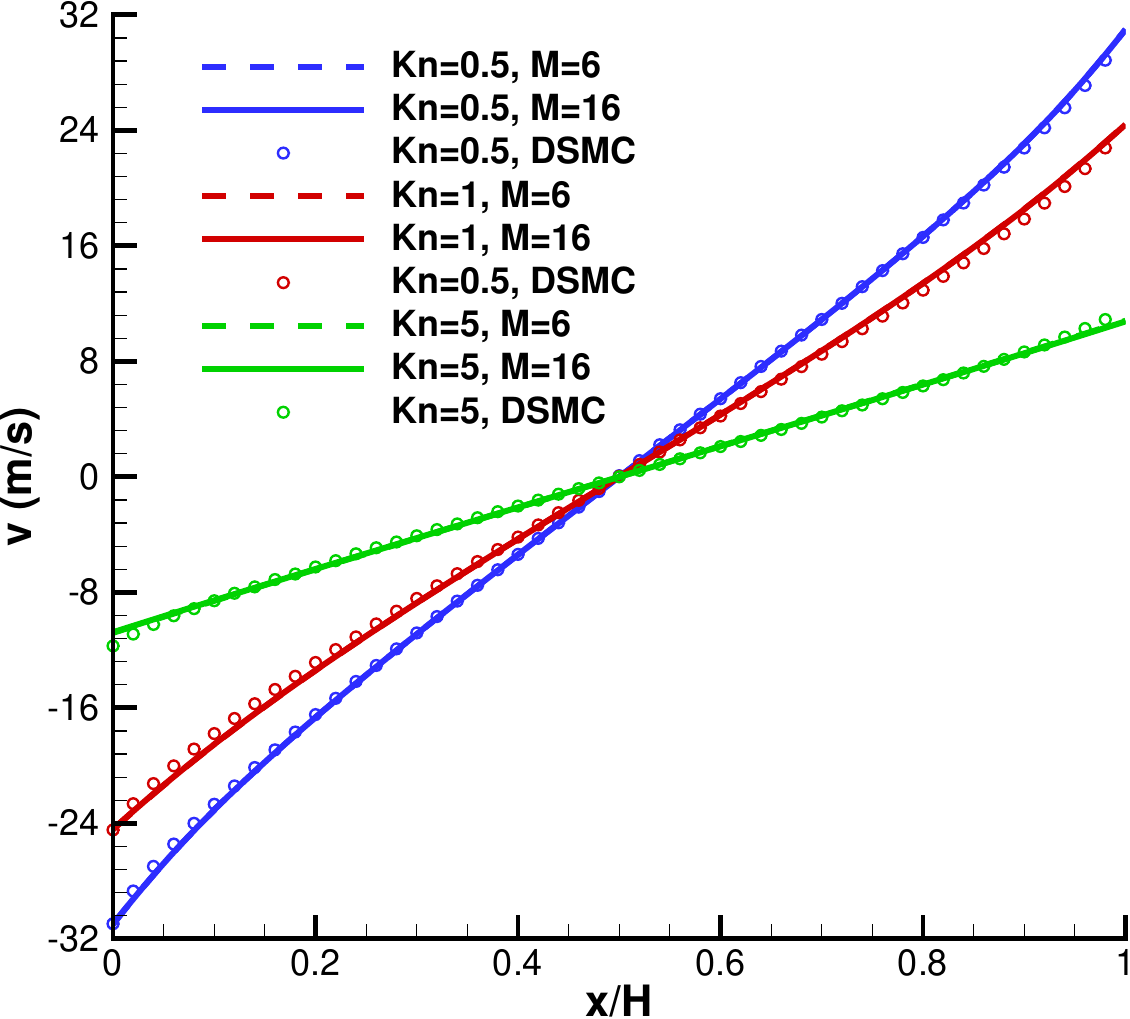}
	\caption{Variation of velocity along the domain length for $\Kn=0.5$, $1.0$, and $5.0$ obtained with DSMC and DGFS using VHS collision model for Argon molecules. The walls move with velocity of $(0,\mp 50,0) m/s$. The physical space is discretized using 10 cells and polynomial order of 2, while the velocity space $[-5.14,\;5.14]^3$ is discretized using $N^3=24^3$ points. $M=6$ and $M=16$ are used on the half sphere.}
	\label{fig_couette_compare_all_U}
\end{figure}

Figure \ref{fig_couette_compare_all_U} illustrates the velocity along the domain length. The deterministic solution is in excellent agreement with the DSMC solution \cite{gu2009high}, and again our model captures the nonlinearity in the near-wall region. 

\begin{figure}[!ht]
	\centering
	\includegraphics[width=80mm]{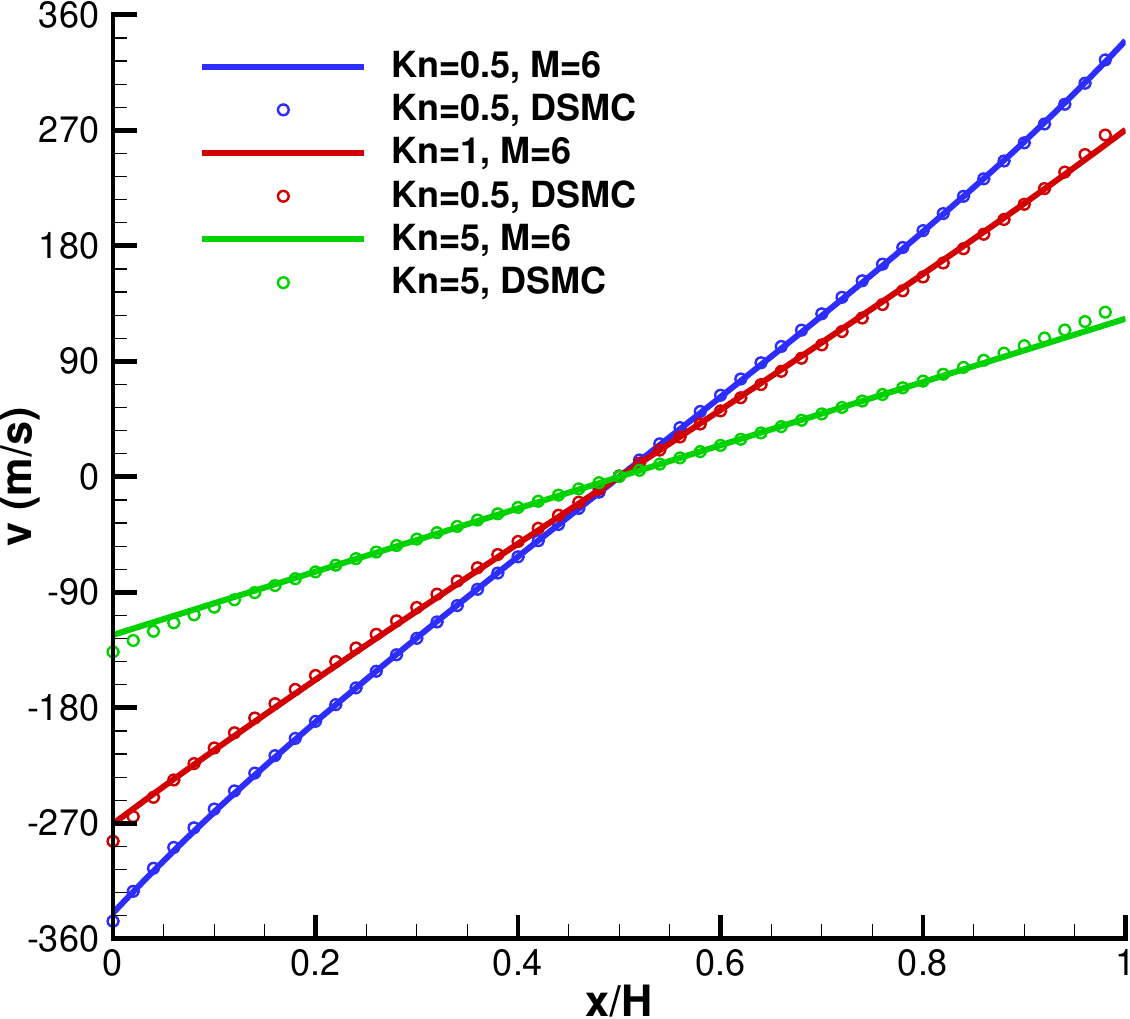}
	\caption{Variation of velocity along the domain length for $\Kn=0.5$, $1.0$, and $5.0$ using VHS collision model for Argon molecules obtained with DSMC and DGFS. The walls move with a relative velocity of $(0,\mp 500,0) m/s$. The physical space is discretized using 10 cells and polynomial order of 2, while the velocity space $[-6.14,\;6.14]^3$ is discretized using $N^3=24^3$ points. $M=6$ is used on the half sphere.}
	\label{fig_couette_compare_all_U_500}
\end{figure}

To further highlight the nature of DGFS, we increase the velocity difference between the two walls to 1000 $m/s$ $\sim$ Mach=3 i.e., $\mathbf{u}_w=(0,\mp500,0)\;m/s$ at the left and the right walls respectively. Figure \ref{fig_couette_compare_all_U_500} illustrates the results for this case. The results for Couette flow cases suggest that the combination of $M=6$, $N^3=24^3$, and velocity domain size of $[-5.14,\;5.14]^3$ suffices for subsonic flows. However, one needs larger $[-6.14,\;6.14]^3$ velocity domain for supersonic flow problems. We want to emphasize that these parameters are rather derived from heuristics, and it is certainly possible that one can obtain good results with other combinations of $M$, $N$, and velocity-space size. There's a trade-off between the accuracy and computational cost.

From a computation viewpoint, DSMC-SPARTA simulations at $Kn=0.5$, $\bu_w=(0,\,\mp500,\,0)$ with 500 cells, 30 particles per cell, a time-step of 2e-9 sec, 1 million unsteady time-steps, and 100 million steady time-steps, on 32 processors took 11206.6 sec. The parameters have been again selected to minimize the statistical fluctuations, and avoid linear time-stepping errors inherent to DSMC simulations. On the other hand, DGFS simulations on a single GPU at $Kn=0.5$, $\mathbf{u}_w=(0,\,\mp500,\,0)$ with 10 elements, 2nd order polynomial, $N^3=24^3$, $M=6$ took 4541.98 sec to achieve  $(\|f^{n+1}-f^{n}\|/\|f^{n}\|_{L_2})/(\|f^{2}-f^{1}\|/\|f^{1}\|_{L_2}) < 2\times10^{-5}$. 

\subsection{1D unsteady case: oscillatory Couette flow}

To demonstrate the time accuracy of the DGFS, we consider the effect of time varying velocity gradient on the solution. The geometry and flow parameters remain the same as in previous case, except that the left wall is at rest, and the right wall moves with a velocity of $\mathbf{u}_w=(0, 50, 0) \sin(\zeta t) \;m/s$, where $\zeta = 2\pi/5e-5 \approx 125663.71 \;s^{-1}$. The simulation is carried out at $Kn=1.0$. The 2nd order SSP-RK scheme (\ref{SSPRK2}) with $\Delta t = 2\times 10^{-8}$ is employed for time integration. Specifically for DSMC simulations, the domain is discretized into 50 cells with 100000 particles per cell (PPC) and the results are averaged for every 1000 ($N_\text{avg}$) time steps. 

\begin{figure}[!ht]
	\centering
    \includegraphics[width=80mm]{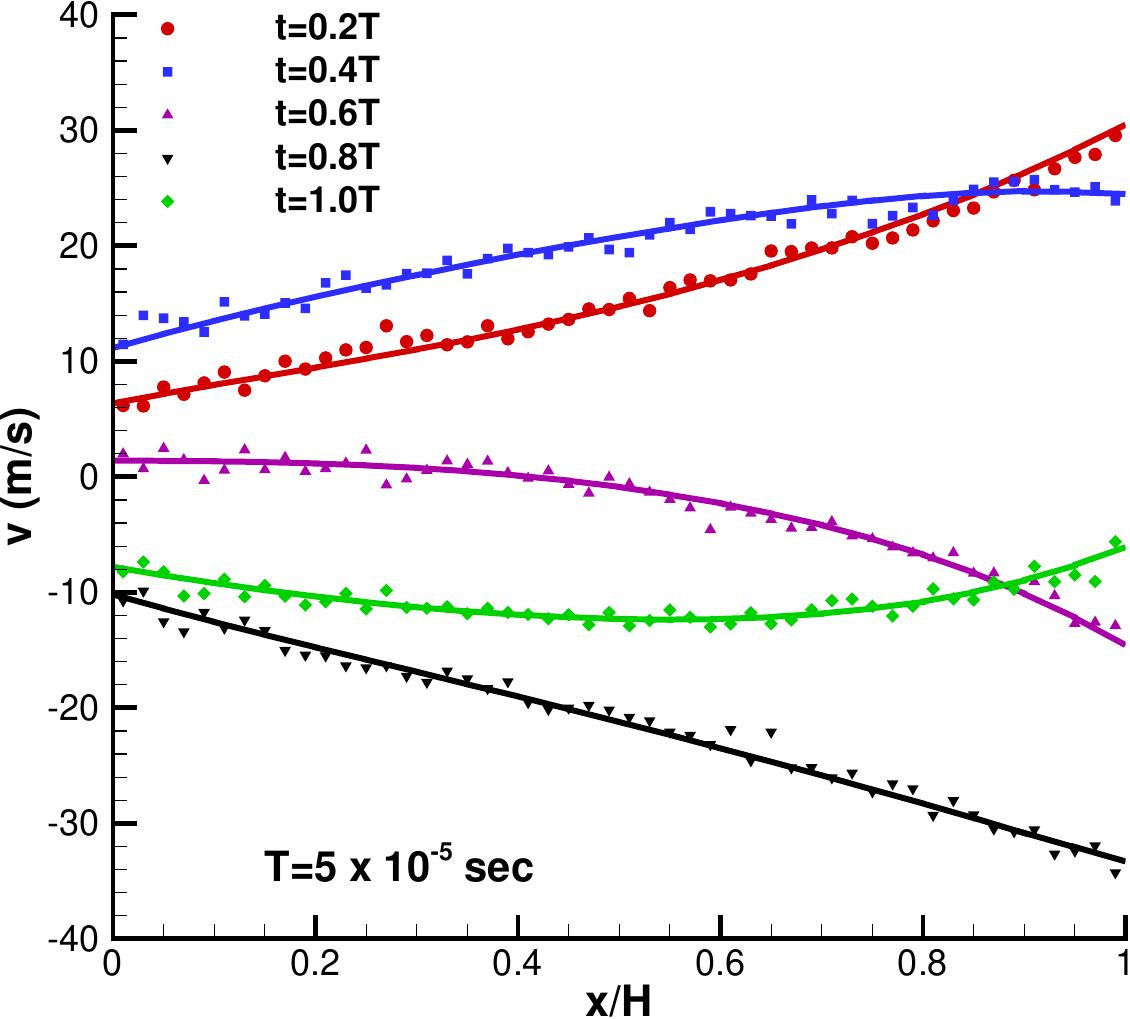}
	\caption{Time evolution of velocity along the domain length for oscillatory Couette flow at $Kn=1.0$ using VHS collision model for Argon molecules. We use 50 cells, 100000 PPC, and 1000 $N_\text{avg}$. Symbols and lines denote DSMC and DGFS results respectively. The physical space is discretized using 20 cells and polynomial order of 2, while the velocity space $[-5,\;5]^3$ is discretized using $N^3=24^3$ points. $M=6$ is used on the half sphere in all cases.}
	\label{fig_oscCouette_compare_all_U}
\end{figure}

Figure \ref{fig_oscCouette_compare_all_U} depicts the time evolution of velocity along the domain for both DSMC and DGFS results. Since the present case is unsteady, high statistical noise is observed in DSMC solutions. In contrast, DGFS produces a sufficiently smooth solution. Nevertheless, both results are in fair agreement with each other. Further, we observe a high amount of slip ($\approx 20\%$) at the left wall since the flow is in transition regime. 

\begin{figure}[!ht]
	\centering
    \includegraphics[width=80mm]{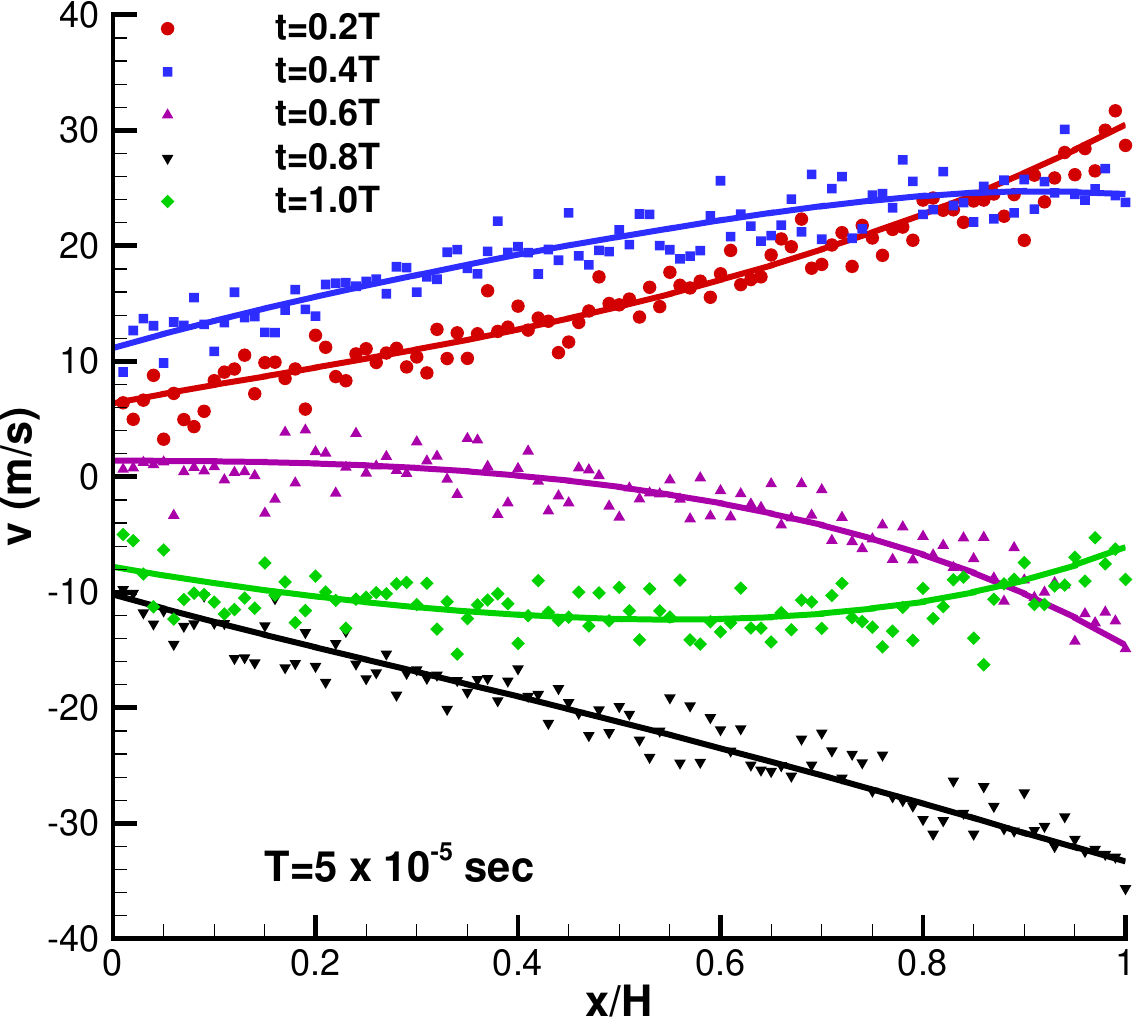}
	\caption{Time evolution of velocity along the domain length for oscillatory Couette flow at $Kn=1.0$ using VHS collision model for Argon molecules. We use 500 cells, 10000 PPC, and 1000 $N_\text{avg}$. Symbols and lines denote DSMC and DGFS results respectively. The physical space is discretized using 20 cells and polynomial order of 2, while the velocity space $[-5,\;5]^3$ is discretized using $N^3=24^3$ points. $M=6$ is used on the half sphere in all cases.}
	\label{fig_oscCouette_compare_all_U_s500p10000Navg1000}
\end{figure}

\begin{figure}[!ht]
	\centering
    \includegraphics[width=80mm]{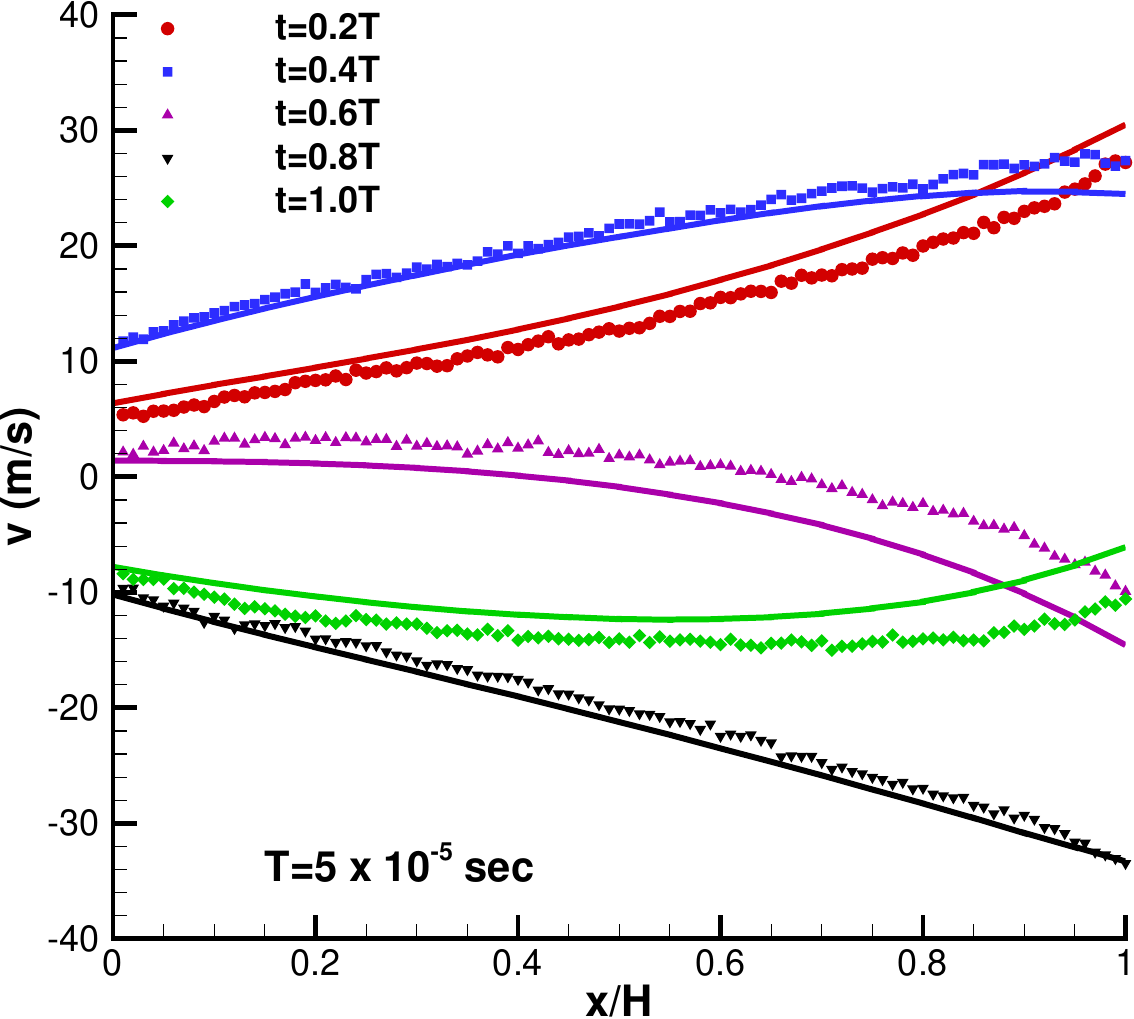}
	\caption{Time evolution of velocity along the domain length for oscillatory Couette flow at $Kn=1.0$ using VHS collision model for Argon molecules. We use 500 cells, 10000 PPC, and 100000 $N_\text{avg}$. Symbols and lines denote DSMC and DGFS results respectively. The physical space is discretized using 20 cells and polynomial order of 2, while the velocity space $[-5,\;5]^3$ is discretized using $N^3=24^3$ points. $M=6$ is used on the half sphere in all cases.}
	\label{fig_oscCouette_compare_all_U_s500p10000Navg100000}
\end{figure}

\begin{figure}[!ht]
	\centering
    \includegraphics[width=80mm]{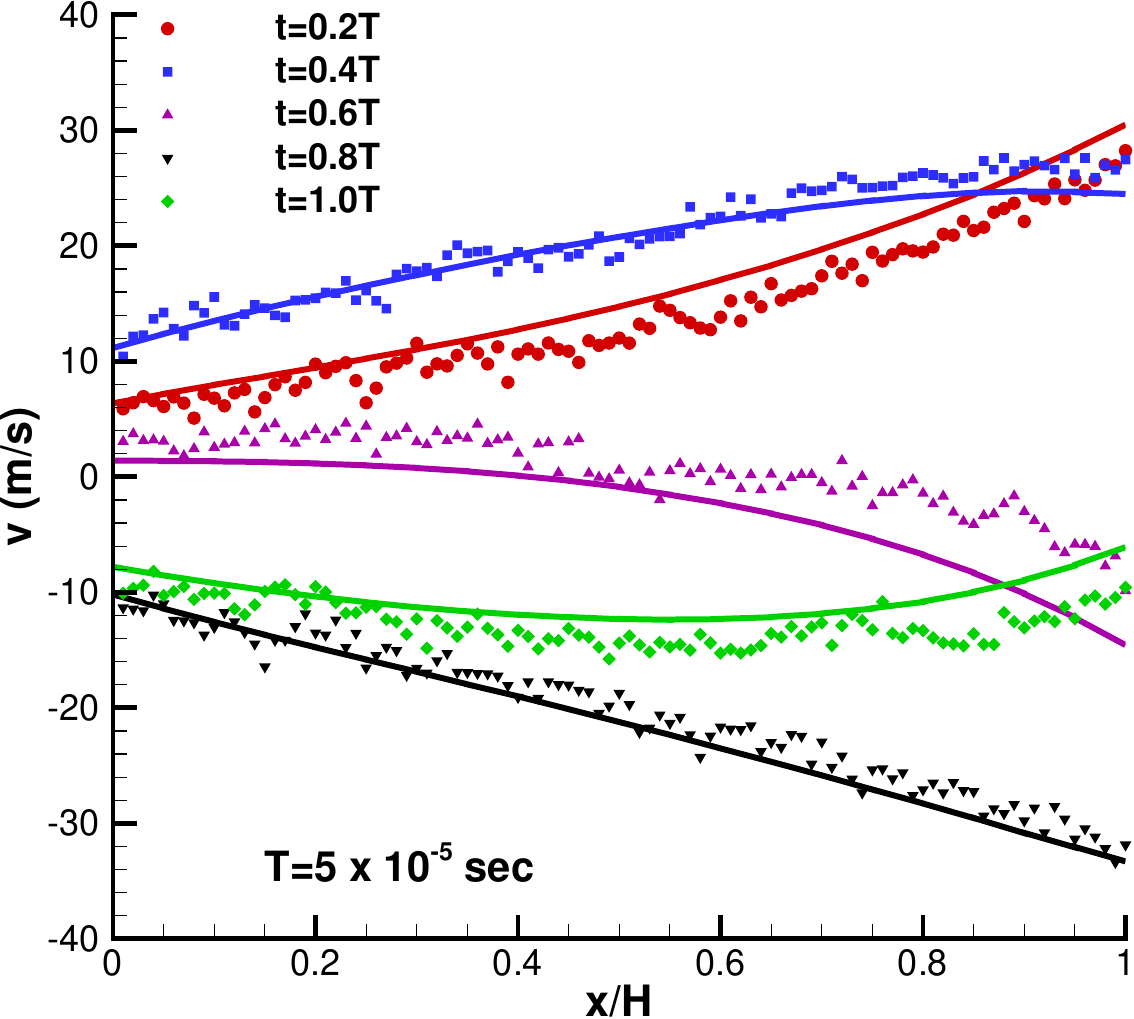}
	\caption{Time evolution of velocity along the domain length for oscillatory Couette flow at $Kn=1.0$ using VHS collision model for Argon molecules. We use 500 cells, 1000 PPC, and 100000 $N_\text{avg}$. Symbols and lines denote DSMC and DGFS results respectively. The physical space is discretized using 20 cells and polynomial order of 2, while the velocity space $[-5,\;5]^3$ is discretized using $N^3=24^3$ points. $M=6$ is used on the half sphere in all cases.}
	\label{fig_oscCouette_compare_all_U_s500p1000Navg100000}
\end{figure}

An accurate unsteady DSMC result is inherently tricky. We carried out set of simulations by varying PPC, cell-count, and $N_\text{avg}$. It is observed that keeping $N_\text{avg}$ fixed, with decrease in PPC, the sample size decreases and consequently the statistical noise increases as illustrated in Figures \ref{fig_oscCouette_compare_all_U}, \ref{fig_oscCouette_compare_all_U_s500p10000Navg1000} and \ref{fig_oscCouette_compare_all_U_s500p10000Navg100000}, \ref{fig_oscCouette_compare_all_U_s500p1000Navg100000}. Keeping PPC fixed, with increase in $N_\text{avg}$, the sample size increases and consequently the statistical noise decreases, but the simulation lags behind in time as a result of high $N_\text{avg}$. These observations are depicted in Figures \ref{fig_oscCouette_compare_all_U_s500p10000Navg100000} and \ref{fig_oscCouette_compare_all_U_s500p1000Navg100000}.

Through Figures \ref{fig_oscCouette_compare_all_U_s500p10000Navg1000}, \ref{fig_oscCouette_compare_all_U_s500p10000Navg100000}, \ref{fig_oscCouette_compare_all_U_s500p1000Navg100000}, we want to emphasize the smooth time accurate results obtained from DGFS, and the well-known stochastic nature of DSMC solutions. In present case, we used as large as 10000 particles per cell for obtaining time accurate results. In large scale simulations, 10000 particles per cell might not be feasible computationally, and hence the results from DSMC would always be inaccurate in those cases. 

\subsection{1D steady case: normal shock wave}
\label{sec:normal}
\begin{figure*}[!ht]
\begin{subfigure}[t]{0.5\textwidth}
  \centering
  \includegraphics[width=75mm]{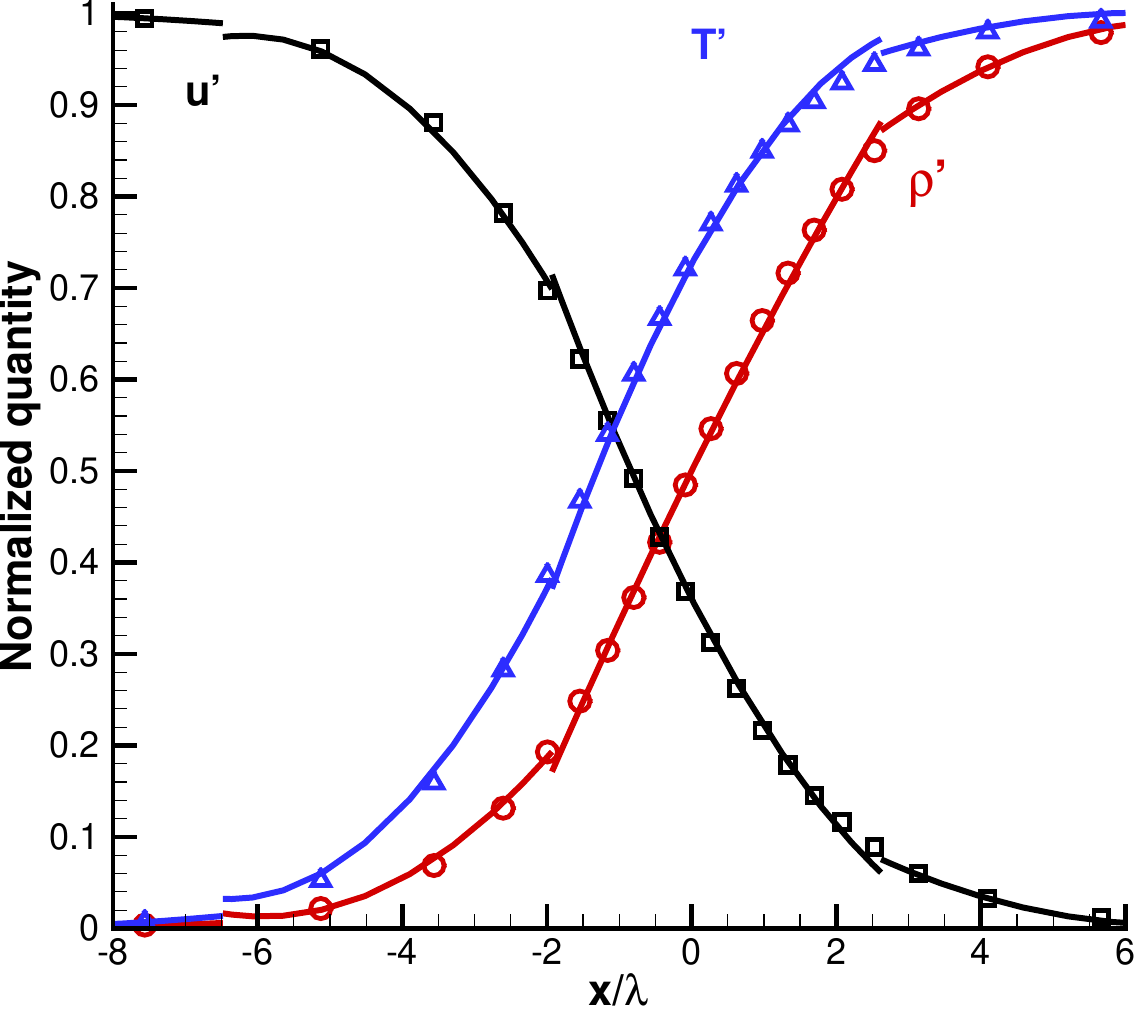}
  \caption{}
  \label{fig_shockHeOhwada_Ma_1_59_s4k3v32}
\end{subfigure}
\begin{subfigure}[t]{0.5\textwidth}
  \centering
  \includegraphics[width=75mm]{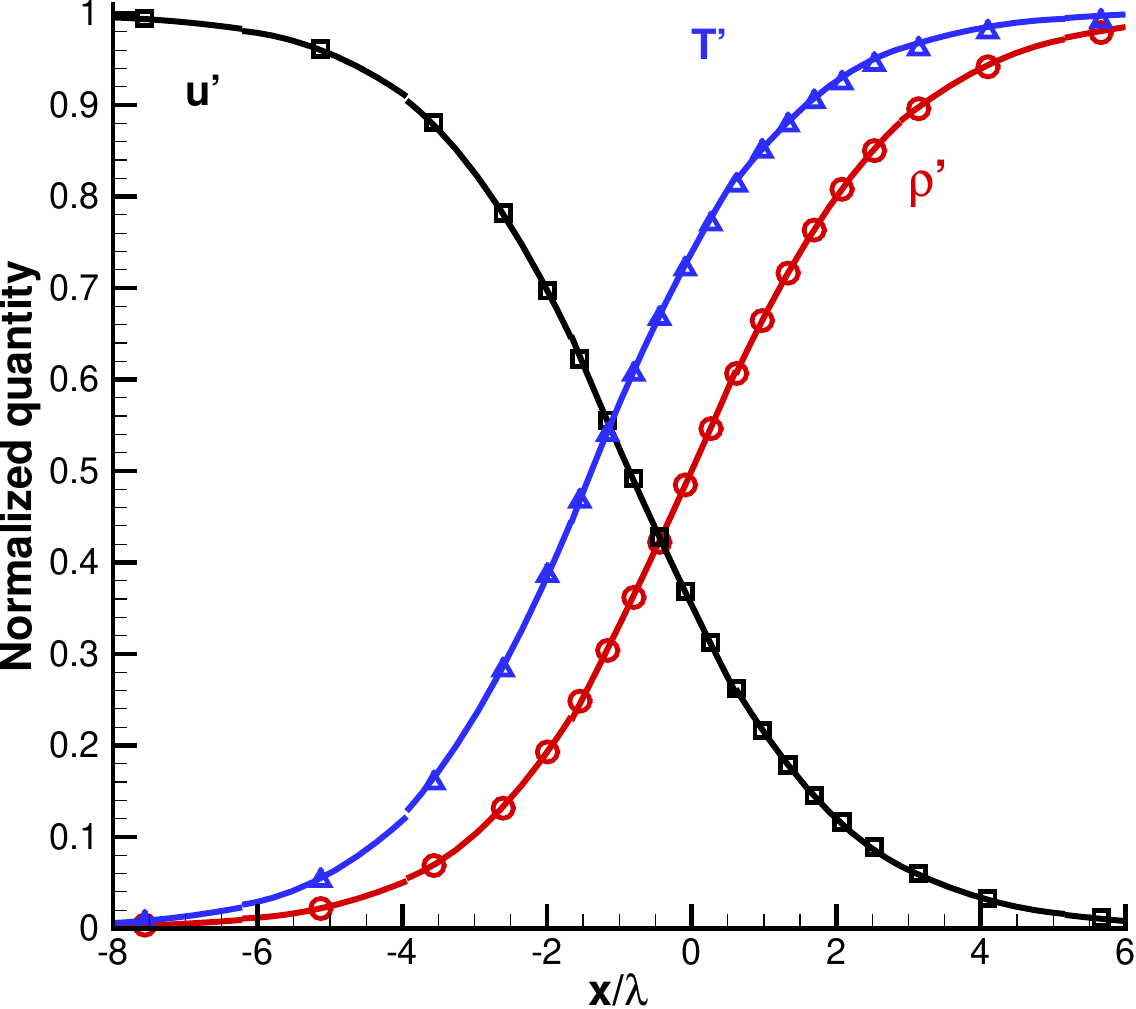}
  \caption{} 
  \label{fig_shockHeOhwada_Ma_1_59_s8k3v32}
\end{subfigure}
\caption{Variation of normalized flow properties along the domain for Mach~1.59 Helium normal shock. Symbols denote results from Ohwada et al.~\cite{ohwada1993structure}, and lines denote DGFS solutions. Note that the position of the shock wave has been adjusted to the location with the average density $(\rho_1+\rho_2)/2$ as per \cite{ohwada1993structure}. The normalized quantities are defined using: $\rho'=(\rho-\rho_1)/(\rho_2-\rho_1)$, $T'=(T-T_1)/(T_2-T_1)$, and $u'=(u-u_2)/(u_1-u_2)$. Here subscript $1$ and $2$ denote upstream and downstream conditions respectively. While the velocity space $[-7,\;7]^3$ is discretized using $N^3=32^3,\,M=6$ points, the physical space $[-15 \times 10^{-3},\;15 \times 10^{-3}]$ is discretized using 3rd order DGFS employing: a) 4 elements, and b) 8 elements.}
\label{fig_shockHeOhwada_Ma_1_59}
\end{figure*}

\begin{figure*}[!ht]
\begin{subfigure}[t]{0.5\textwidth}
  \centering
  \includegraphics[width=75mm]{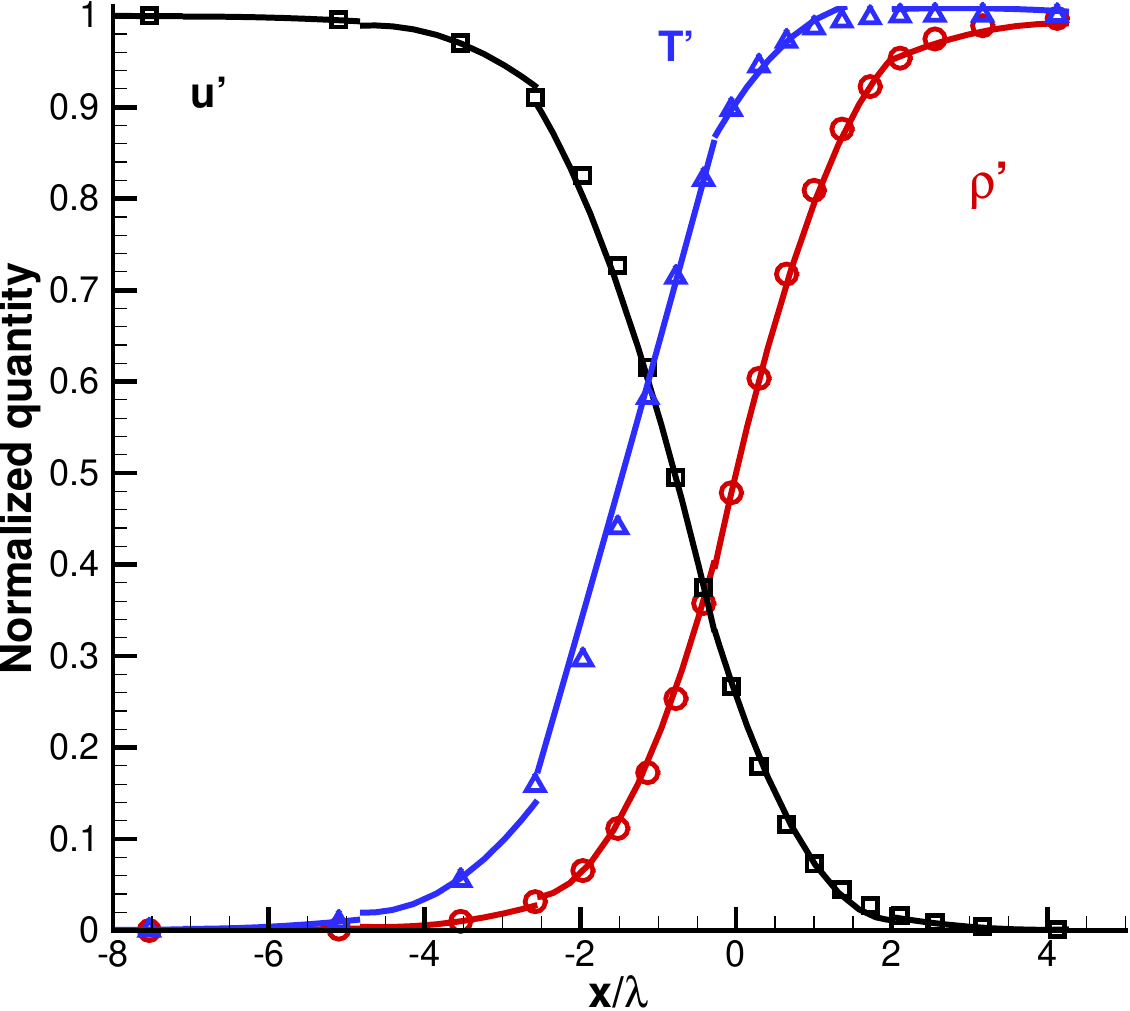}
  \caption{}
  \label{fig_shockHeOhwada_Ma_3_s8k3v48}
\end{subfigure}
\begin{subfigure}[t]{0.5\textwidth}
  \centering
  \includegraphics[width=75mm]{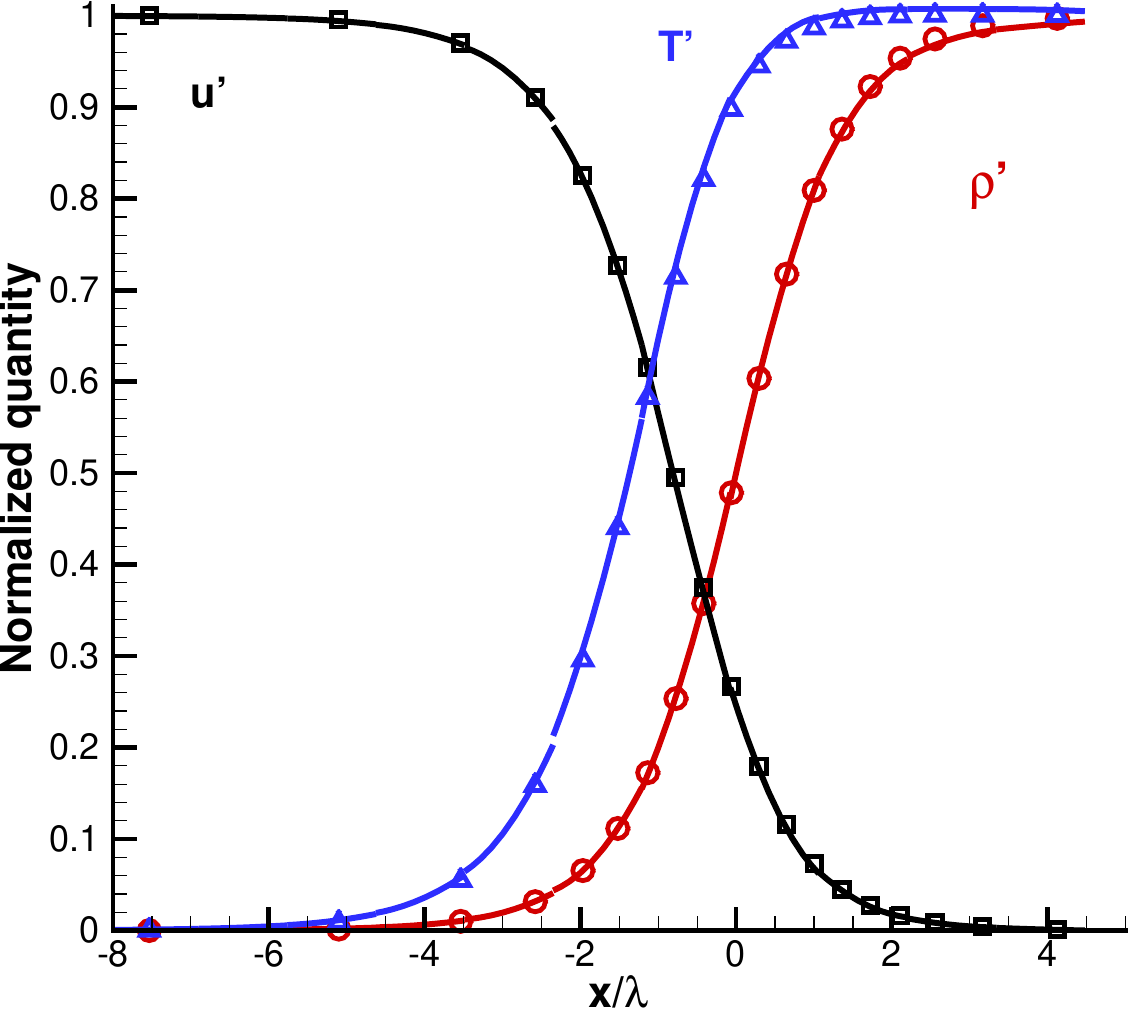}
  \caption{}
  \label{fig_shockHeOhwada_Ma_3_s16k3v48}
\end{subfigure}
\caption{Variation of normalized flow properties along the domain for Mach~3.0 Helium normal shock. Symbols denote results from Ohwada et al.~\cite{ohwada1993structure}, and lines denote DGFS solutions. Note that the position of the shock wave has been adjusted to the location with the average density $(\rho_1+\rho_2)/2$ as per \cite{ohwada1993structure}. The normalized quantities are defined using: $\rho'=(\rho-\rho_1)/(\rho_2-\rho_1)$, $T'=(T-T_1)/(T_2-T_1)$, and $u'=(u-u_2)/(u_1-u_2)$. Here subscript $1$ and $2$ denote upstream and downstream conditions respectively. While the velocity space $[-11,\;11]^3$ is discretized using $N^3=48^3,\,M=6$ points, the physical space $[-15 \times 10^{-3},\;15 \times 10^{-3}]$ is discretized using 3rd order DGFS employing: a) 8 elements, and b) 16 elements.}
\label{fig_shockHeOhwada_Ma_3}
\end{figure*}

To demonstrate the advantage of high order DGFS approximations, we consider the normal shock wave and compare our solutions with the finite-difference solutions reported in \cite{ohwada1993structure}. The numerical parameters are listed in Table~\ref{tab_normalShock_conditions}. Specifically for these cases, since the flow is in early slip regime hence the collision term is stiffer, the method acquires steady-state slowly. A convergence criterion of $(\|f^{n+1}-f^{n}\|/\|f^{n}\|_{L_2})$ /$(\|f^{2}-f^{1}\|/\|f^{1}\|_{L_2}) < 2\times10^{-5}$ has been used. Note in particular that the spatial domain has been discretized with just 4 elements, and 3rd order DG for Mach~1.59 case. Limiters have \textit{not} been used in the present study. Figure~\ref{fig_shockHeOhwada_Ma_1_59} illustrates the variation of normalized density, temperature, and velocity for Ma~1.59 normal shock. Note that the position of the shock wave has been adjusted to the location with the average density $(\rho_u+\rho_d)/2$ as per \cite{ohwada1993structure}. Based upon these results, one can infer that DGFS is able to resolve the normal shock with just 4 elements within \textit{engineering} ($\pm 5\%$) accuracy. Note that the discontinuity in the flow profile is the characteristic of the DG method. On increasing the number of elements to 8, the results from \cite{ohwada1993structure} match fairly well with DGFS. Similarly, Figure \ref{fig_shockHeOhwada_Ma_3} depicts the variation of normalized density, temperature, and velocity for Mach~3 normal shock with 8 elements and 3rd order DG. Again, the Mach~3 shock is captured well using just 8 elements. 

\begin{table}[!ht]
\caption{\normalfont Numerical parameters for the normal shock wave \cite{ohwada1993structure}.}
\centering
\begin{tabular}{@{}lcc@{}}
\toprule
Parameter & Case 01 & Case 02\\ 
\midrule
Working Gas & Helium & Helium \\ 
Mach number & 1.59 & 3.0 \\
Physical space ($mm$) & $[-15,\,15]$ & $[-15,\,15]$ \\ 
Velocity space & $[-7,\,7]^3$ & $[-11,\,11]^3$  \\ 
$N^3$ & $32^3$ & $48^3$ \\
$M$ & $6$ & $6$  \\
Spatial elements & 4 & 8 \\
DG order & 3 & 3 \\
Viscosity index: $\omega$ & 0.5 & 0.5 \\
Ref. diameter: $d_\text{ref}$ ($m$) & $2.17 \times 10^{-10}$ & $2.17 \times 10^{-10}$  \\
Ref. Temperature: $T_\text{ref}$ ($K$) & 273 & 273 \\
\midrule
Upstream conditions: \\
Velocity: $u_1$ ($m/s$) & 1398.771 & 2639.19 \\
Temperature: $T_1$ ($K$) & 223 & 223  \\
Density: $\rho_1$ ($kg/m^3$) & $1.916 \times 10^{-5}$ & $1.916 \times 10^{-5}$  \\
Mean free path: $\lambda$ (m) & 0.001648 & 0.001648 \\
\midrule
Downstream conditions: \\
Velocity: $u_2$ ($m/s$) & 764.659 & 879.73   \\
Temperature: $T_2$ ($K$) & 354.762 & 817.67  \\
Density: $\rho_2$ ($kg/m^3$) & $3.505 \times 10^{-5}$ & $5.748 \times 10^{-5}$   \\
\bottomrule
\end{tabular}
\label{tab_normalShock_conditions}
\end{table}

\begin{figure*}[!ht]
\begin{subfigure}[t]{0.5\textwidth}
  \centering
  \includegraphics[width=75mm]{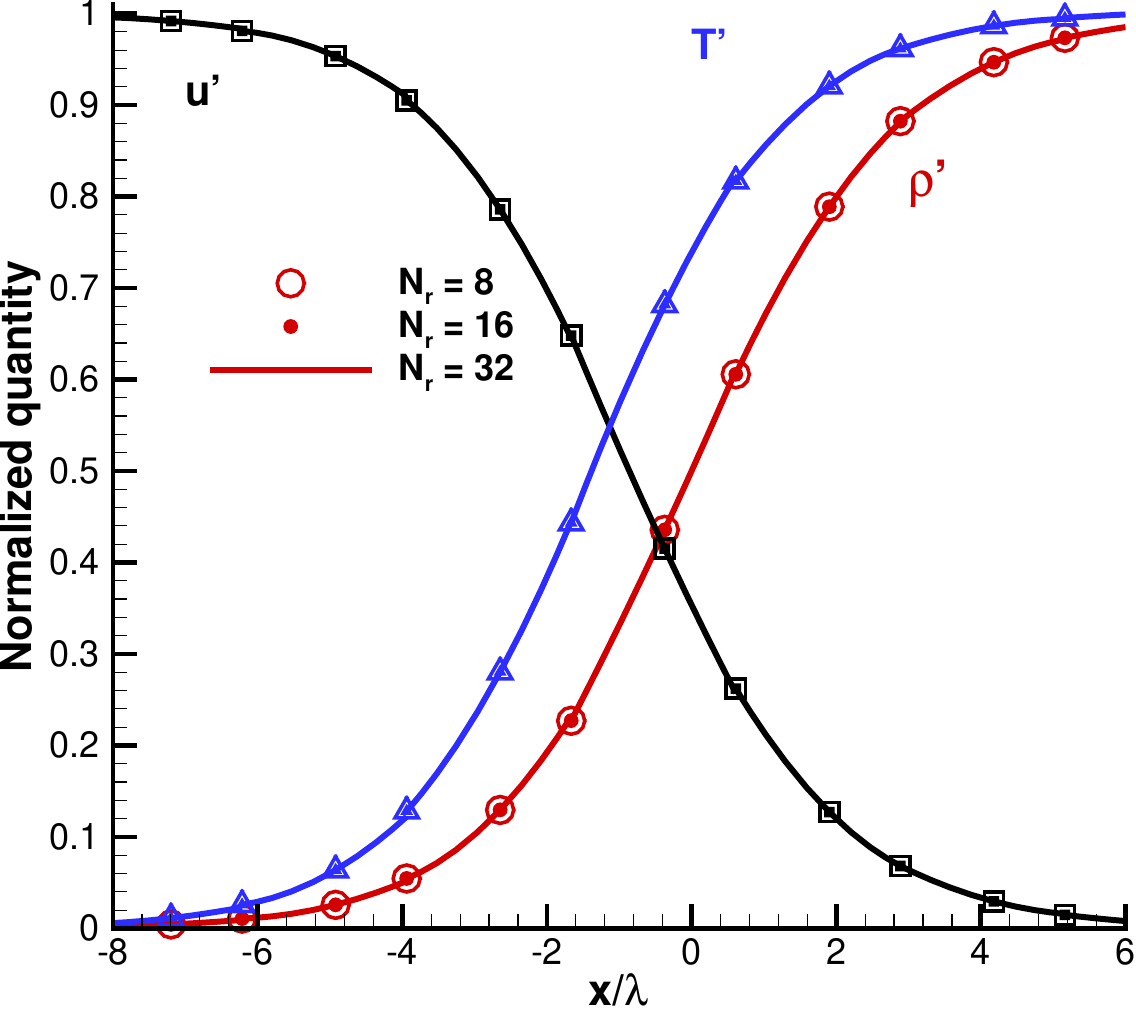}
  \caption{}
  \label{fig_shockHeOhwada_Ma_3_s8k3v32_Nrho}
\end{subfigure}
\begin{subfigure}[t]{0.5\textwidth}
  \centering
  \includegraphics[width=75mm]{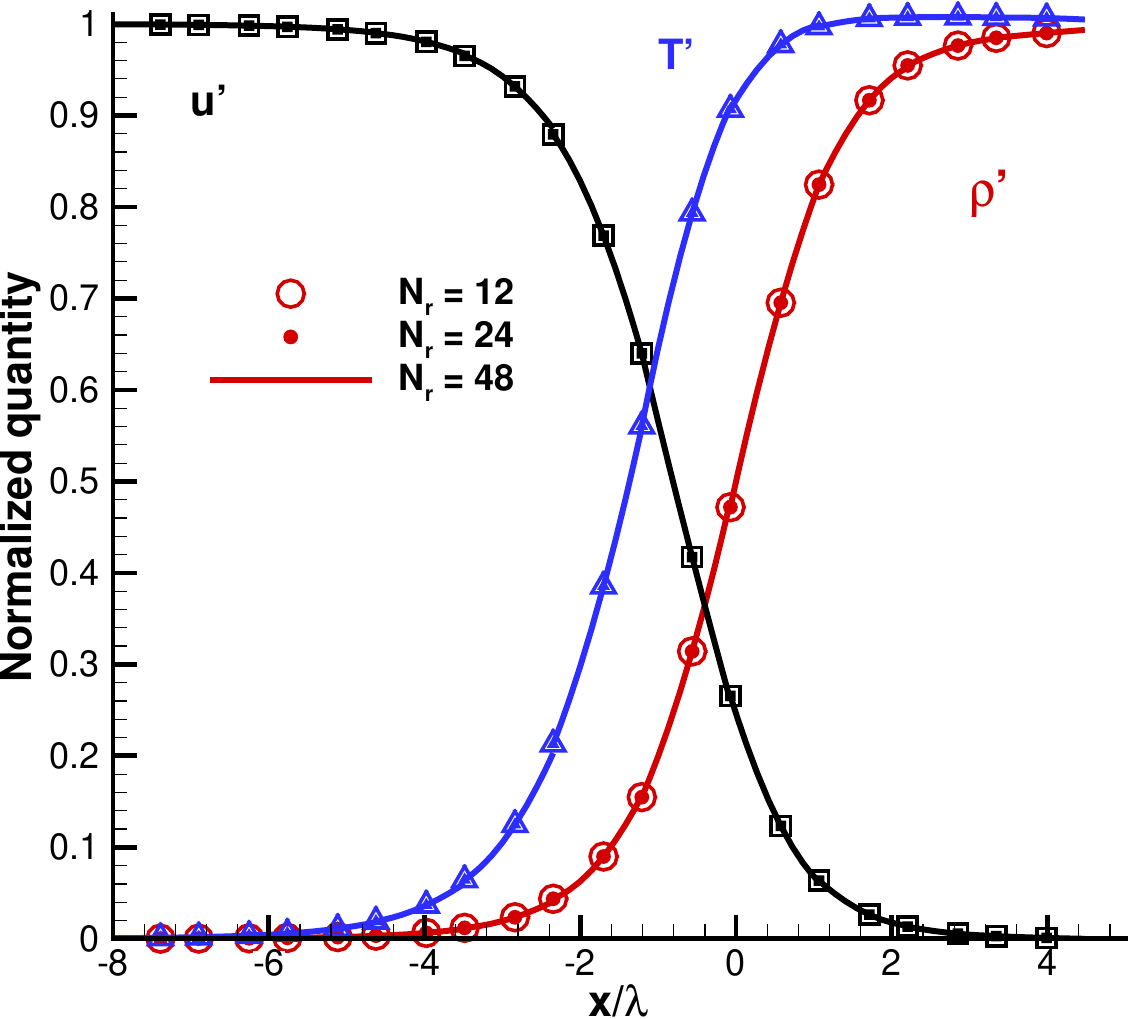}
  \caption{}
  \label{fig_shockHeOhwada_Ma_3_s16k3v32_Nrho}
\end{subfigure}
\caption{{Variation of normalized flow properties along the domain: a) Mach~1.59 Helium normal shock as in Fig.~\ref{fig_shockHeOhwada_Ma_1_59_s8k3v32}, and b) Mach~3.0 Helium normal shock as in Fig.~\ref{fig_shockHeOhwada_Ma_3_s16k3v48}. For both cases, $N_r$, the number of quadrature points in the radial direction -- an important component in Fourier spectral approximation (see Appendix) -- is varied. Symbols and lines denote the DGFS solutions with different $N_r$.}}
\label{fig_shockHeOhwada_Ma_3_Nrho}
\end{figure*}

{
Next, similar to the BKW solution in section~\ref{sec_bkw_error_Nr}, we quantify the effect of $N_r$ (number of quadrature points used in the radial direction) on the recovered bulk properties. Figure~\ref{fig_shockHeOhwada_Ma_3_Nrho} shows the normalized flow properties. We observe that the bulk properties are relatively unaffected upon reducing $N_r$ to $N/2$ and $N/4$.
}

\begin{figure}[!tbp]
\begin{subfigure}[b]{0.5\textwidth}
  \centering
  \includegraphics[width=75mm]{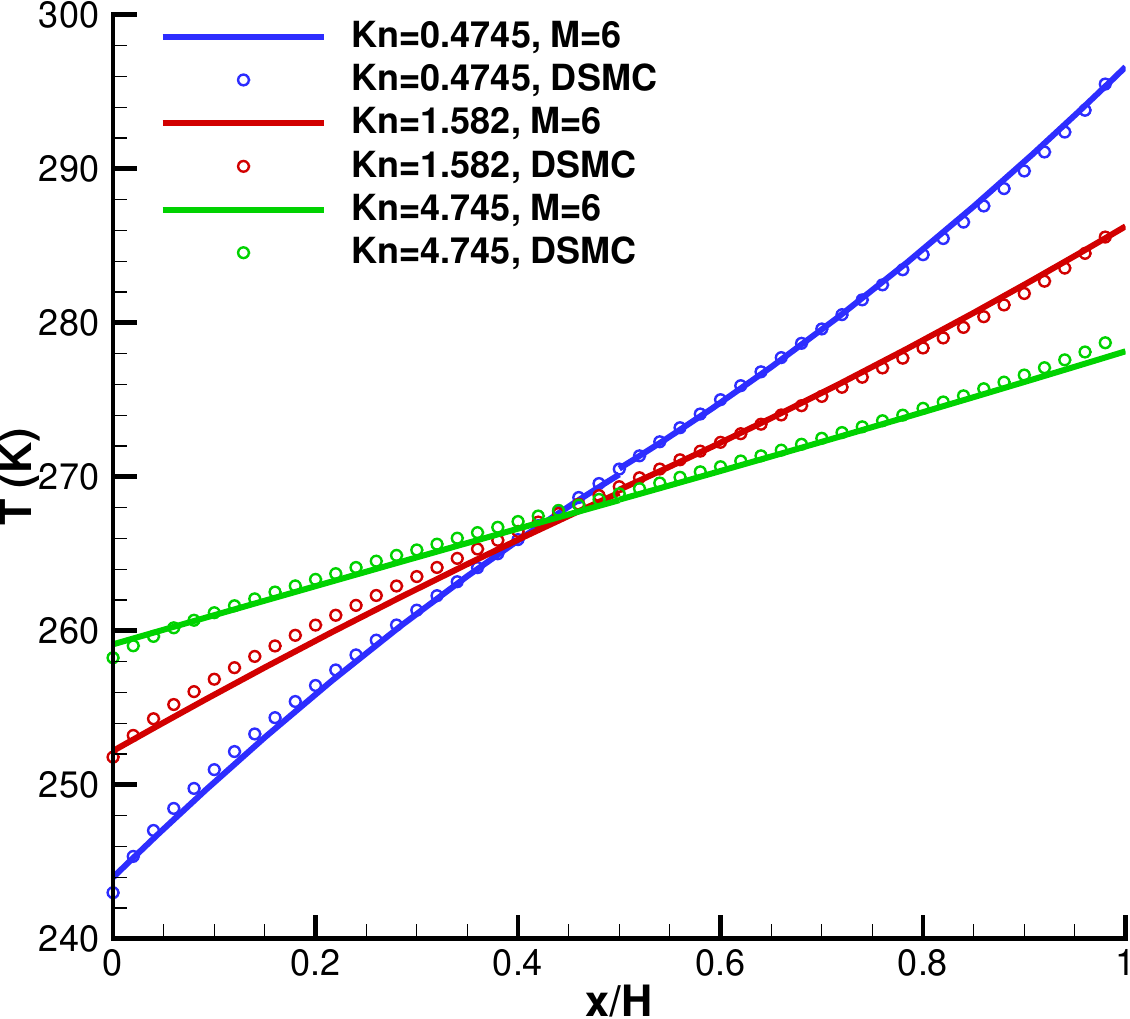}
  \caption{}
  \label{fig_couette_fourier_T100_s2k3v24}
\end{subfigure}

\begin{subfigure}[b]{0.5\textwidth}
  \centering
  \includegraphics[width=75mm]{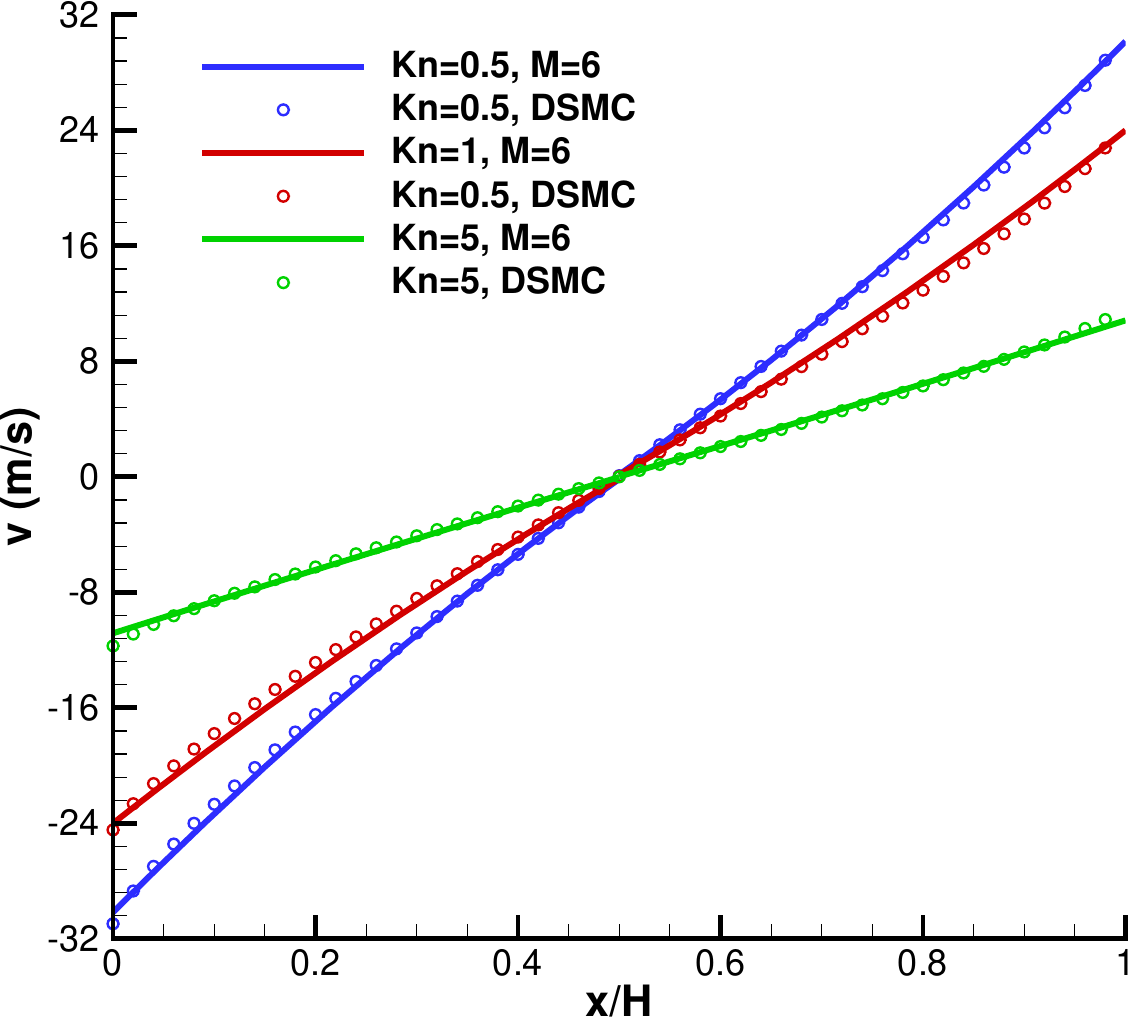}
  \caption{}
  \label{fig_couette_U50_s2k3v24}
\end{subfigure}

\begin{subfigure}[b]{0.5\textwidth}
  \centering
  \includegraphics[width=75mm]{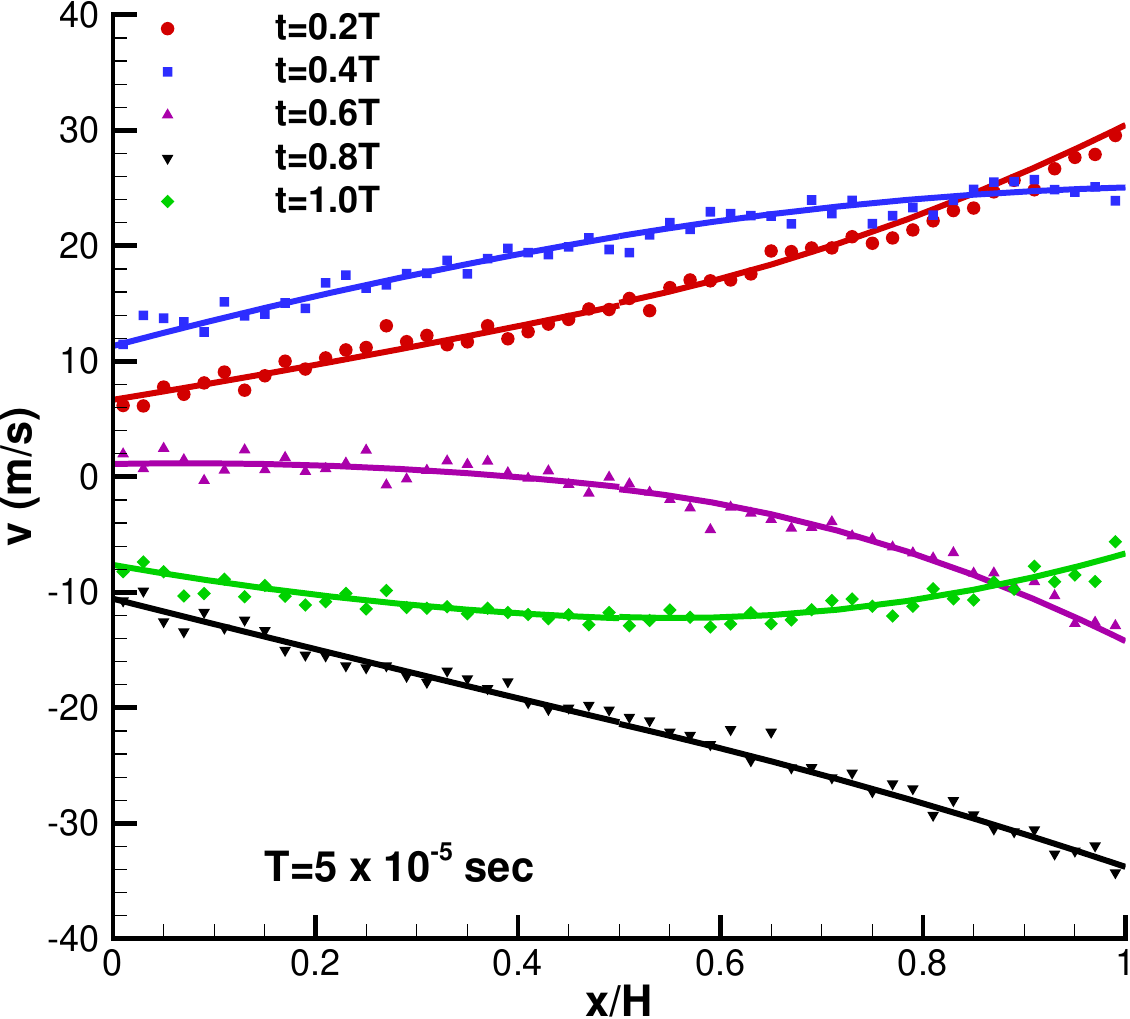}
  \caption{}
  \label{fig_oscCouette_U50_s2k3v24}
\end{subfigure}
\caption{Variation of flow properties along the domain for various cases using 2 elements and 3rd order DG: a) Temperature for Fourier heat transfer with $\Delta T=100K$ similar to Fig~\ref{fig_fc_compare_all_T_100}, b) y-velocity for Couette flow with $\mathbf{u}_w=(0,\,\mp 50,\,0)$ similar to Fig~\ref{fig_couette_compare_all_U}, and c) y-velocity for oscillatory Couette flow similar to Fig~\ref{fig_oscCouette_compare_all_U}. Note the small discontinuity at x/H=0.5 which marks the shared boundary of the two elements.} 
\label{fig_fourier_couette_oscCouette_s2k3v24}
\end{figure}

Having established that one can recover the shock profile reasonably using 8 elements, we can now hypothesize that one can capture the rarefied Couette flow, Fourier heat transfer, and oscillatory Couette flow with just 2 elements and the 3rd order DG. Figure~\ref{fig_fourier_couette_oscCouette_s2k3v24} serves as a proof of this hypothesis. This is precisely why the high order accurate methods such as Discontinuous Galerkin and Fast Spectral are useful. However, it is imperative that one would need more number of elements if the flow gradients are made stronger as in hypersonic cases.

\subsection{2D case: lid driven cavity flow}
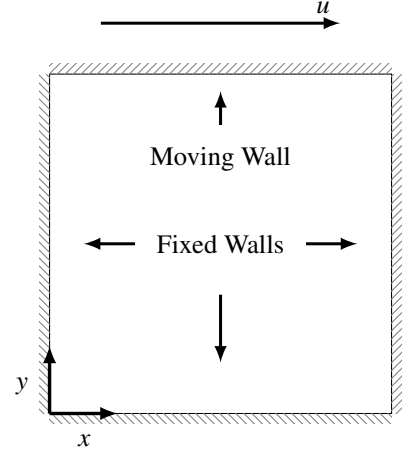
\begin{figure}[ht]
	\centering
\begin{tikzpicture}[scale=0.9]		
	\def\W{5};
	\def\H{5};
	\def\pW{0.15};	%
	\def\aW{0.75};
	\def\aH{0.75};
	
	\draw (0,0) -- (\W,0) -- (\W,\H) -- (0,\H) -- (0,0);		%

	\fill[pattern=north west lines, pattern color=gray, line width = 0.1mm, thin] (0,0) rectangle ({-\pW},\H);
	\fill[pattern=north east lines, pattern color=gray, line width = 0.1mm, thin] (\W,0) rectangle ({\W+\pW},\H);
		
	\fill[pattern=north west lines, pattern color=gray, line width = 0.1mm, thin] (0,0) rectangle (\W,{-\pW});
	\fill[pattern=north east lines, pattern color=gray, line width = 0.1mm, thin] (0,\H) rectangle ({\W},{\H+\pW});
		
	\draw[thick,-latex, line width=0.40mm] ({\W*0.75},{\H*0.5}) -- ({\W*0.75+\aW},{\H*0.5});	%
	\draw[thick,-latex, line width=0.40mm] ({\W*0.25},{\H*0.5}) -- ({\W*0.25-\aW},{\H*0.5});	%
	\draw[thick,-latex, line width=0.40mm] ({\W*0.5},{\H*0.5-\aH}) -- ({\W*0.5},{\aH});	%
	\node[] at ({\W*0.5},{\H*0.5}) {Fixed Walls};					%

	\node[] at ({\W*0.5},{\H*0.75}) {Moving Wall};					%
	\draw[thick,-latex, line width=0.40mm] ({\W*0.5},{\H*0.85}) -- ({\W*0.5},{\H*1.1-\aH});	%
	\draw[thick,-latex, line width=0.40mm] ({\aW},{\H+\aH}) -- ({\W-\aW},{\H+\aH}) node[anchor=south east] {$u$};	%

	\draw[thick,-latex, line width=0.45mm, black] (0,0) -- (0,1);				%
	\draw[thick,-latex, line width=0.45mm, black] (0,0) -- (1,0);				%
		
	\node[] at (0.5,-0.4) {$x$};
	\node[] at (-0.4,0.4) {$y$};	
\end{tikzpicture} 	\caption{Numerical setup for lid driven cavity flow.}
	\label{fig_lidCavitySchematic}
\end{figure}

As the first 2D example, we consider the standard lid driven cavity flow. We consider a square box of length $H=1\times10^{-3}$ meters. All the walls are kept at temperature of $T=273K$. At the top wall, a velocity of $u_w=50~m/s$ is introduced. The setup of the problem is given in Figure \ref{fig_lidCavitySchematic}. The Knudsen number is fixed at $Kn=1$ \cite{john2010investigation}. The 2nd order SSP-RK scheme is used for time evolution. {The velocity space $[-5,\,5]^3$ is discretized using $N^3=24^3$ points. $M=6$ is used on the half sphere for all the cases.} A convergence criterion of $(\|f^{n+1}-f^{n}\|/\|f^{n}\|_{L_2})$ /$(\|f^{2}-f^{1}\|/\|f^{1}\|_{L_2}) < 9\times10^{-6}$ has been used.

\begin{figure*}[tbp]
\centering
\begin{subfigure}{.5\textwidth}
  \centering
  \includegraphics[width=75mm,trim={0cm 0cm 0cm 0cm},clip]{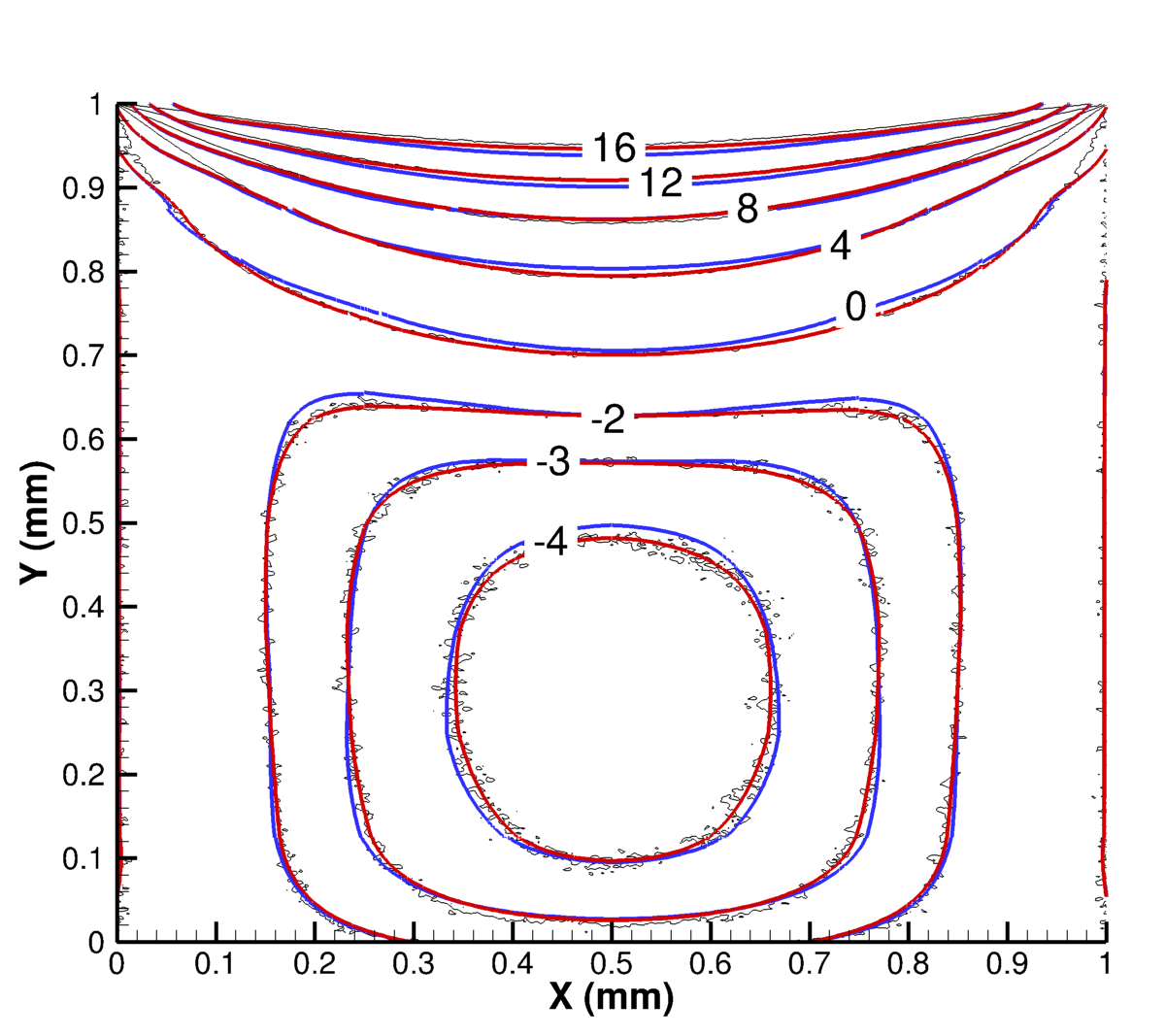}
  \caption{$x$-component of velocity}
  \label{fig_lidCavity_Ux_overlay}
\end{subfigure}%
\begin{subfigure}{.5\textwidth}
  \centering
  \includegraphics[width=75mm,trim={0cm 0cm 0cm 0cm},clip]{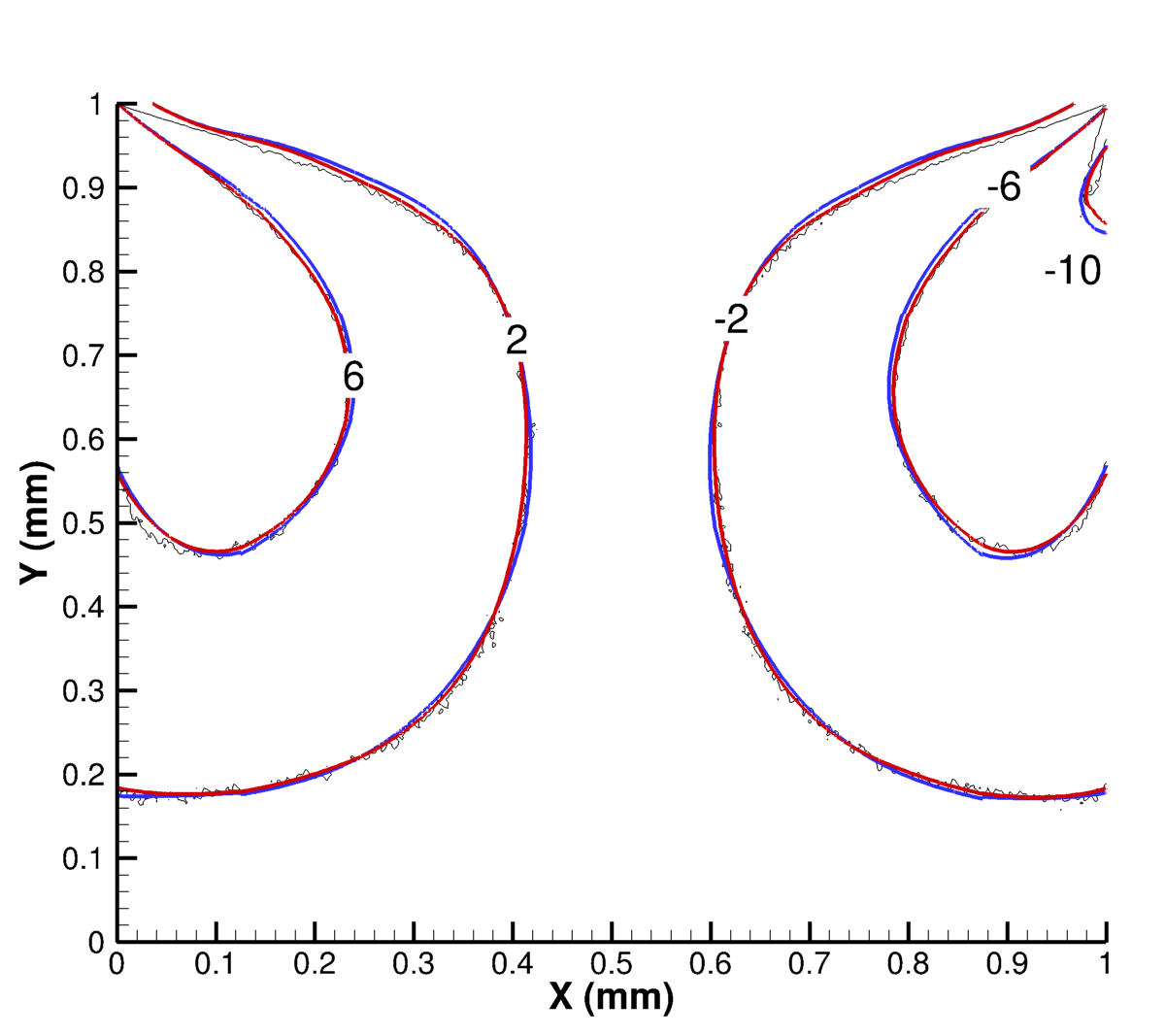}
  \caption{$y$-component of velocity}
  \label{fig_lidCavity_Ux_ver}
\end{subfigure}

\begin{subfigure}{.5\textwidth}
  \centering
  \includegraphics[width=75mm,trim={0cm 0cm 0cm 0cm},clip]{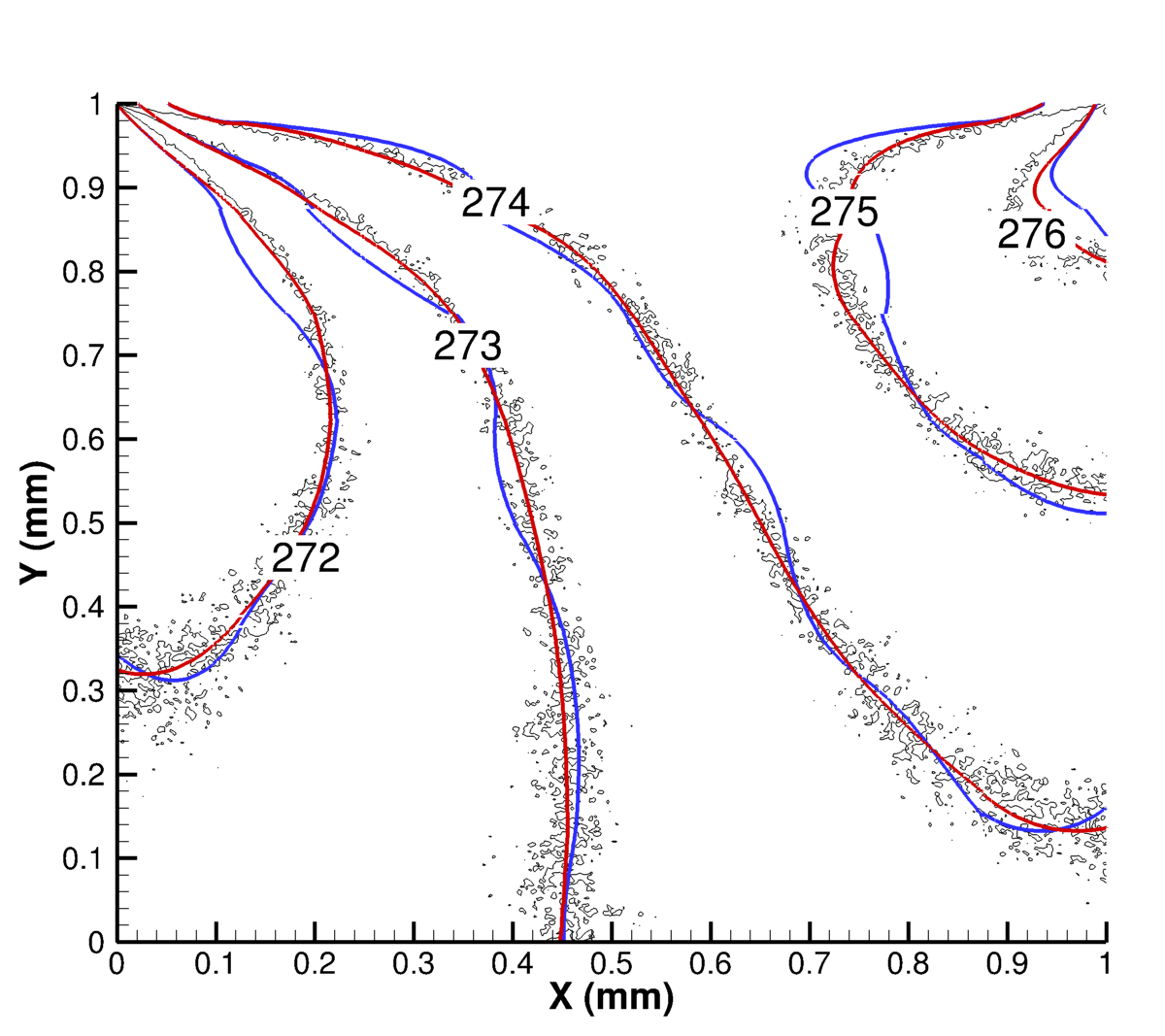}
  \caption{temperature}
  \label{fig_lidCavity_Uy_hor}
\end{subfigure}%
\begin{subfigure}{.5\textwidth}
  \centering
  \includegraphics[width=75mm,trim={0cm 0cm 0cm 0cm},clip]{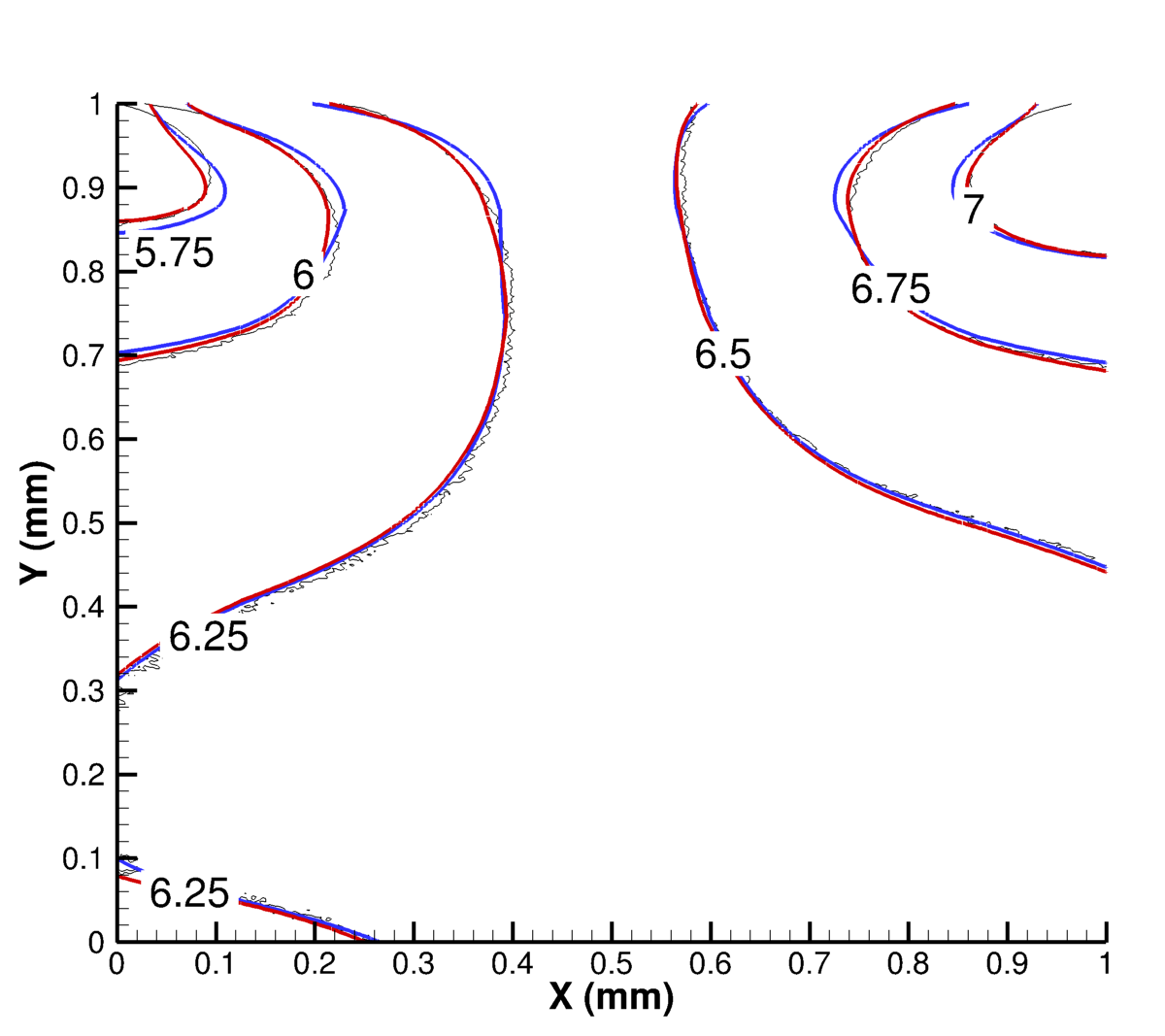}
  \caption{$xx$-component of stress}
  \label{fig_lidCavity_Uy_ver}
\end{subfigure}

\begin{subfigure}{.5\textwidth}
  \centering
  \includegraphics[width=75mm,trim={0cm 0cm 0cm 0cm},clip]{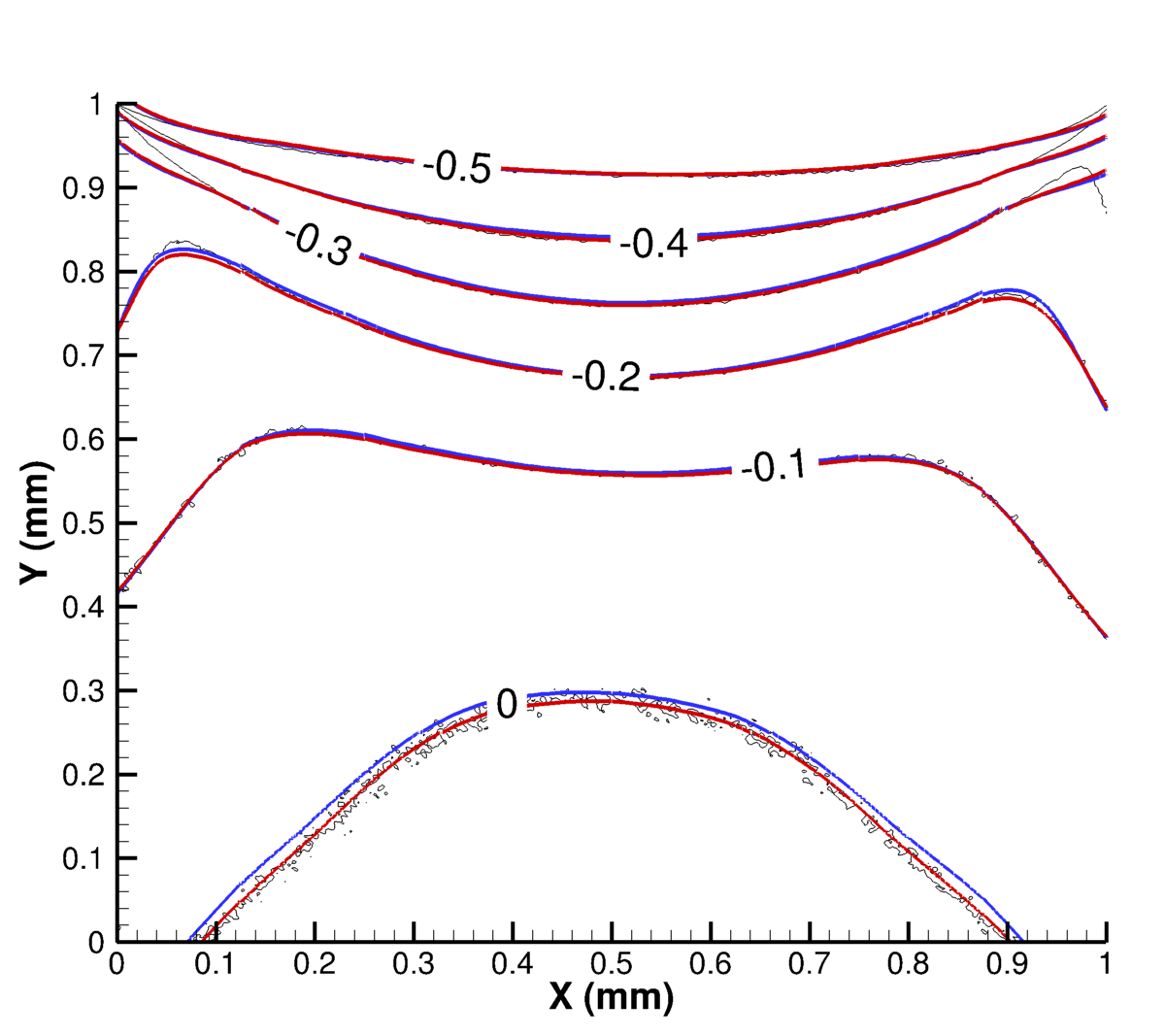}
  \caption{$xy$-component of stress}
  \label{fig_lidCavity_T_hor}
\end{subfigure}%
\begin{subfigure}{.5\textwidth}
  \centering
  \includegraphics[width=75mm,trim={0cm 0cm 0cm 0cm},clip]{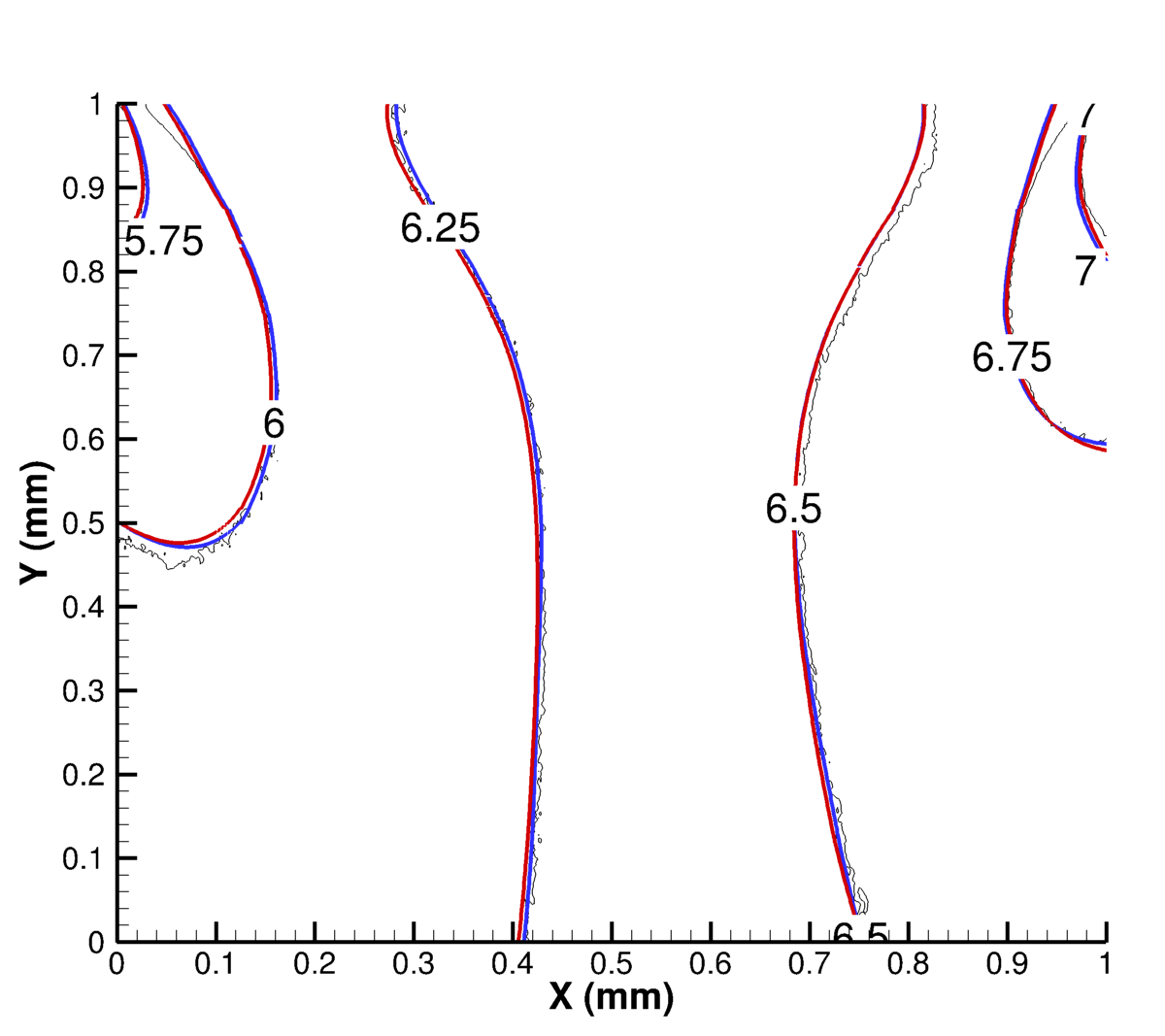}
  \caption{$yy$-component of stress}
  \label{fig_lidCavity_T_ver}
\end{subfigure}

\caption{{Contours of various flow properties for lid-driven cavity flow at $\Kn=1$ obtained with DSMC (thin black lines), DGFS employing velocity space $[-5,\,5]^3$ discretized with $N^3=24^3$ points (solid blue lines), and DGFS employing velocity space $[-6,\,6]^3$ discretized with $N^3=48^3$ points and $N_r=12$ (solid red lines). For DGFS, the physical space is discretized using $8 \times 8$ cells and DG order of 3. $M=6$ is used on the half sphere in all cases.}}
\label{fig_lidCavity_contour}
\end{figure*}

{Figure~\ref{fig_lidCavity_contour} shows the contour plot of various flow properties.} Figures~\ref{fig_lidCavity_UxUyT},~\ref{fig_lidCavity_PxxPxyPyy} illustrate the comparison of flow properties on vertical and horizontal lines along and across the square cavity. From these results, ignoring the statistical fluctuations, one can infer that DGFS results match well with DSMC. Additionally in Fig.~(\ref{fig_lidCavity_Uy_hor}), the y-velocity profile along central horizontal axis ($y/H=0.5$) is compared with \cite{john2010investigation}, and is again found to be in fair agreement. 

{
In the present case, the flow is driven by a velocity gradient in the x-direction, while the walls are initially at a common fixed temperature. Consequently, the deviation in the temperature at the steady state is on order of a few kelvins. To resolve fine structures in the flow, with differences on $O(1)$ kelvin, finer meshes are needed. We note minor $\sim5\%$ deviation between DSMC and DGFS for the x-velocity profile in Fig.~\ref{fig_lidCavity_Ux_hor}, and $\sim1\%$ for temperature in Fig.~\ref{fig_lidCavity_T_hor} at $y/H=0.8$ due to relatively small velocity grid $[-5,\;5]$ discretized with $N^3=24^3$ points. With a finer velocity grid $[-6,\;6]^3$ discretized using $N^3=48^3$ points and $N_r=12$, the deviation between DSMC and DGFS reduces to below $\sim0.5\%$. Taking the complexity of collision solver into account, the end-user has two choices: a) $N^3=24^3$ with $\sim5\%$ difference, or b) $N^3=48^3$ with $<1\%$ difference.
}

\begin{figure*}[tbp]
\centering
\begin{subfigure}{.5\textwidth}
  \centering
  \includegraphics[width=75mm,trim={0cm 0cm 0cm 0cm},clip]{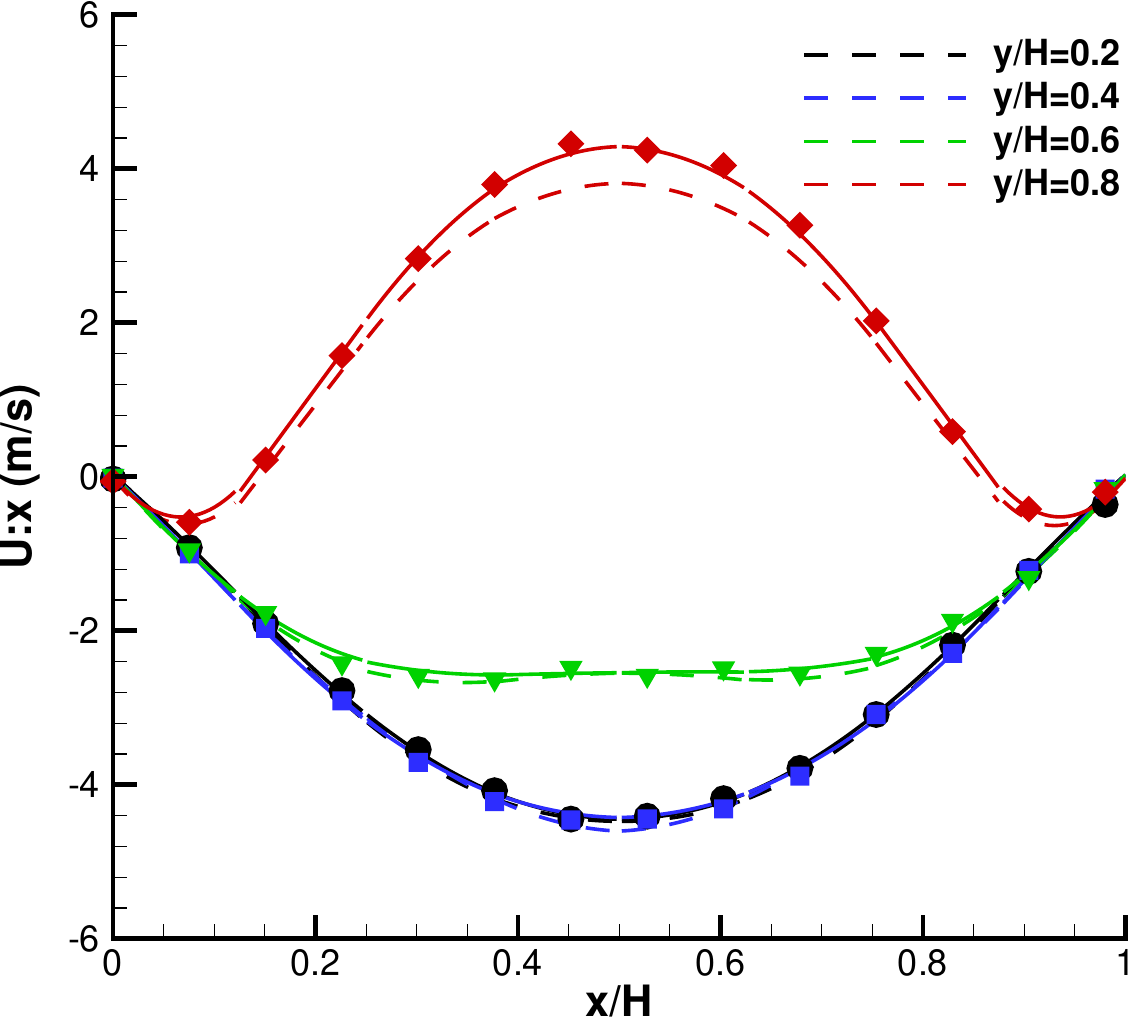}
  \caption{$x$-component of velocity (on horizontal lines)}
  \label{fig_lidCavity_Ux_hor}
\end{subfigure}%
\begin{subfigure}{.5\textwidth}
  \centering
  \includegraphics[width=75mm,trim={0cm 0cm 0cm 0cm},clip]{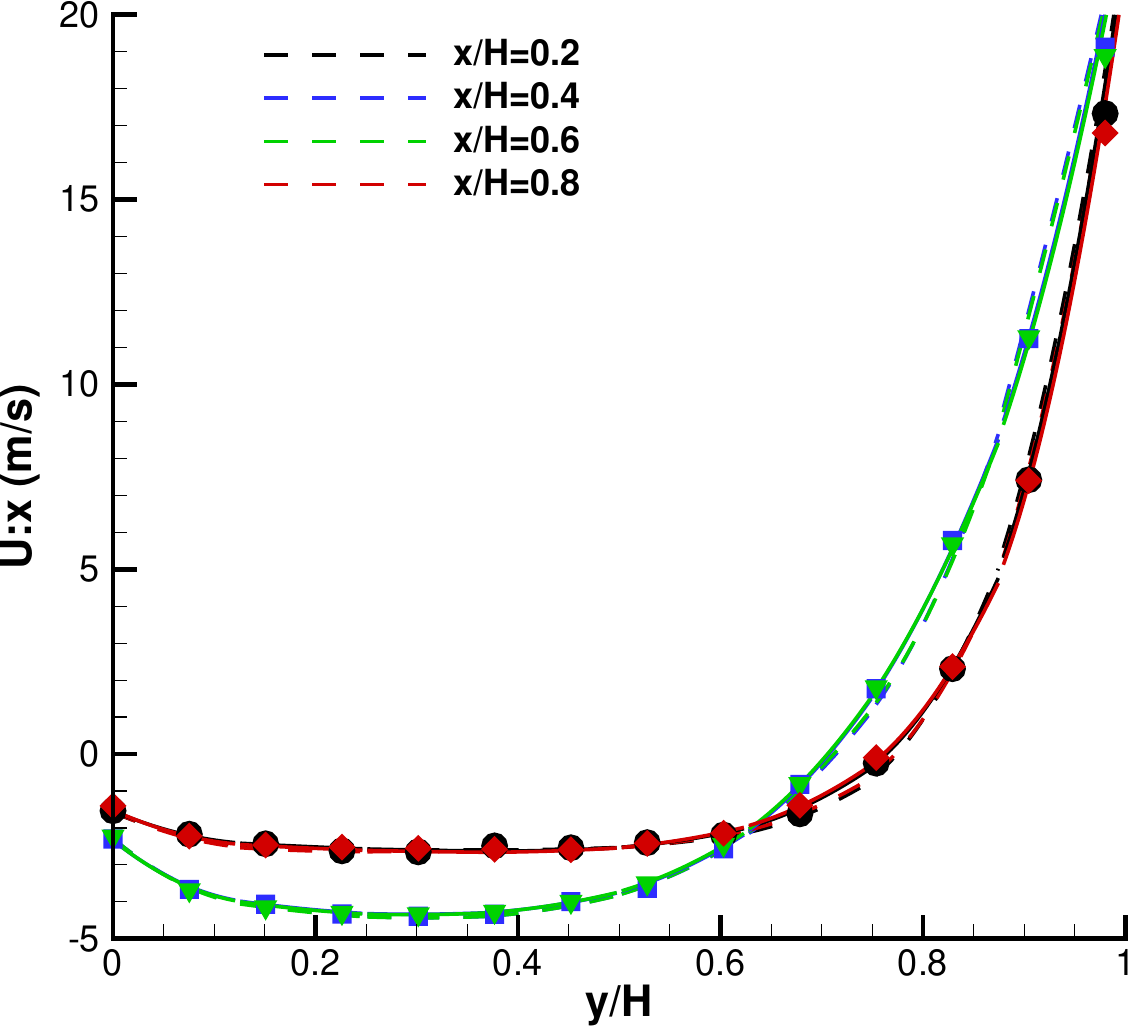}
  \caption{$x$-component of velocity (on vertical lines)}
  \label{fig_lidCavity_Ux_ver}
\end{subfigure}

\begin{subfigure}{.5\textwidth}
  \centering
  \includegraphics[width=75mm,trim={0cm 0cm 0cm 0cm},clip]{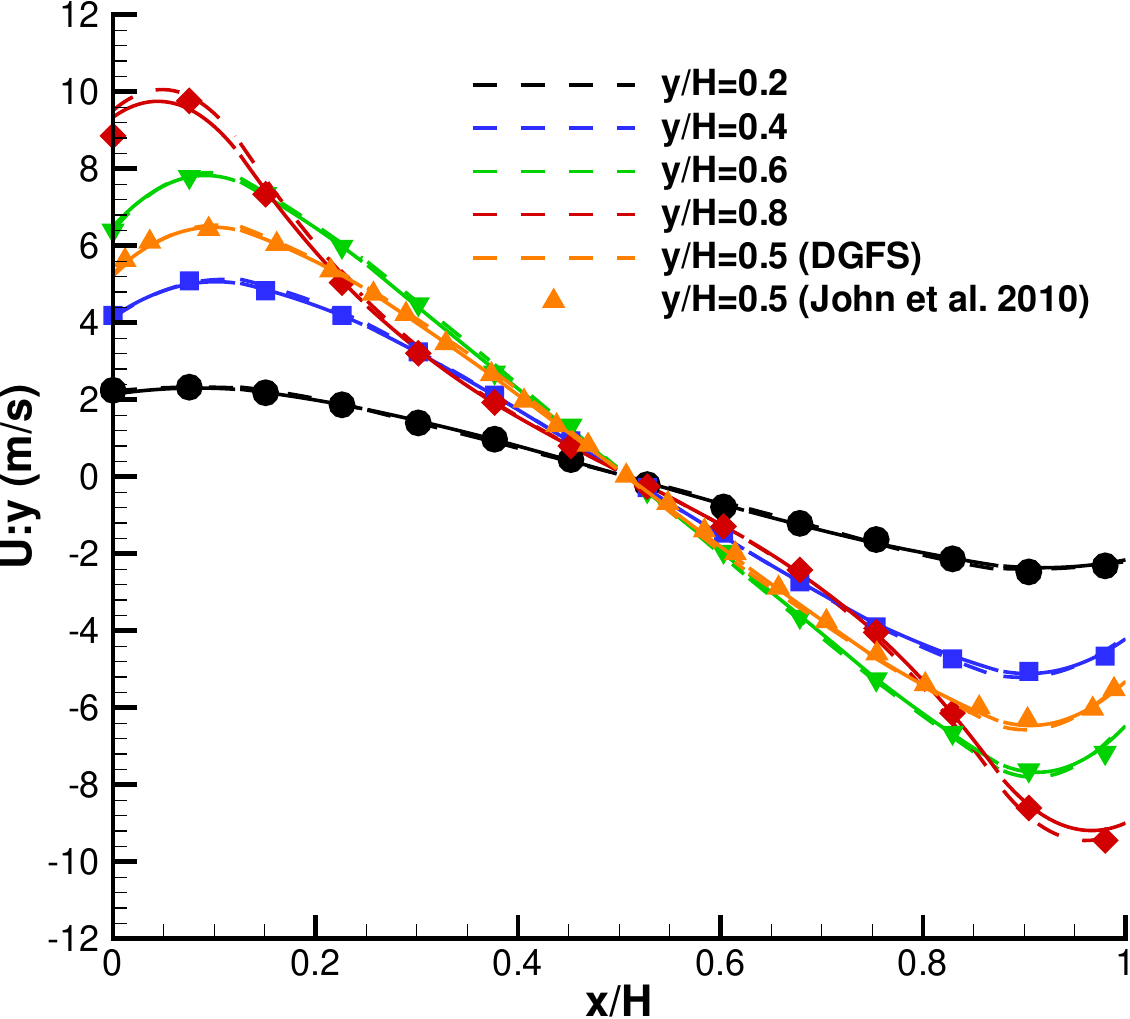}
  \caption{$y$-component of velocity (on horizontal lines)}
  \label{fig_lidCavity_Uy_hor}
\end{subfigure}%
\begin{subfigure}{.5\textwidth}
  \centering
  \includegraphics[width=75mm,trim={0cm 0cm 0cm 0cm},clip]{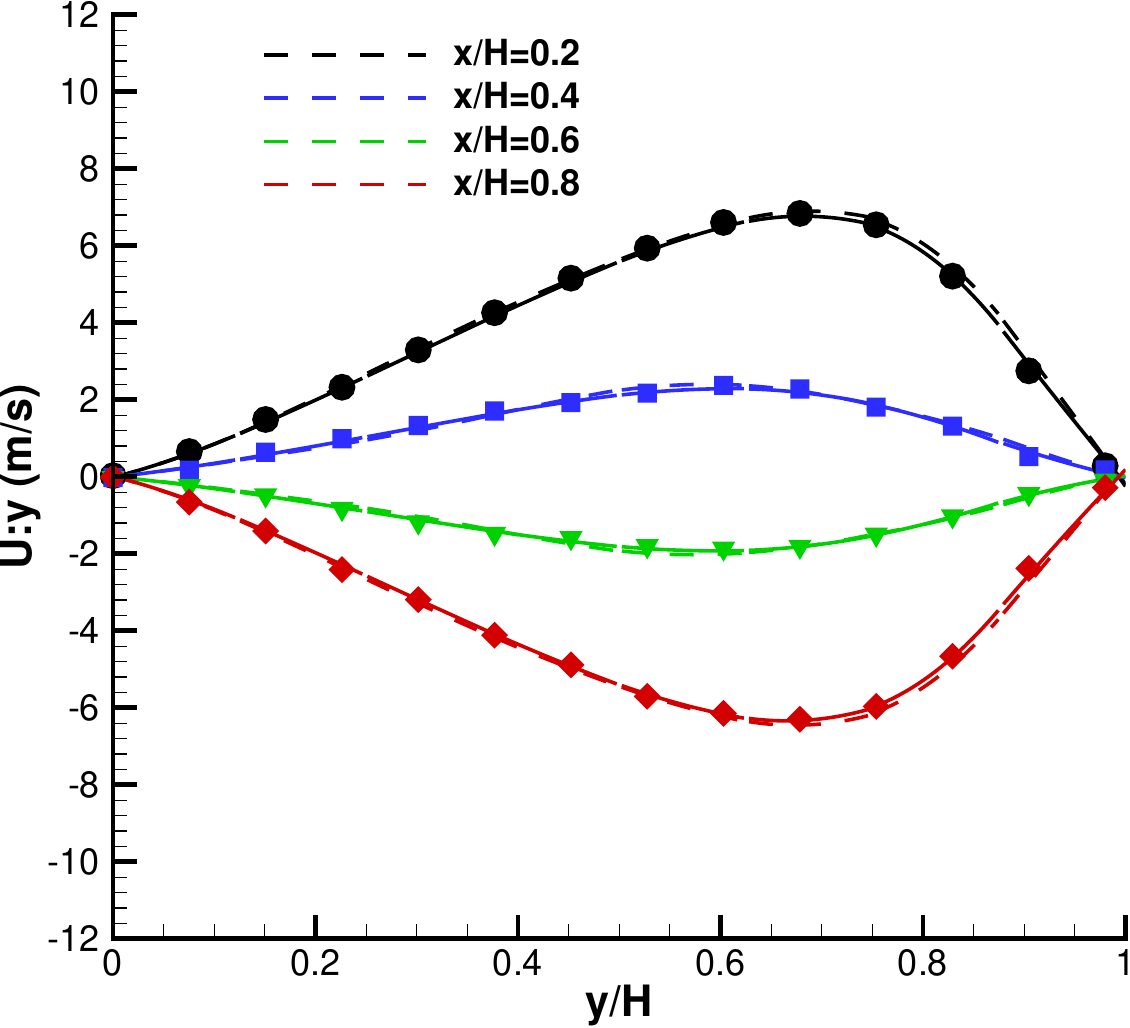}
  \caption{$y$-component of velocity (on vertical lines)}
  \label{fig_lidCavity_Uy_ver}
\end{subfigure}

\begin{subfigure}{.5\textwidth}
  \centering
  \includegraphics[width=75mm,trim={0cm 0cm 0cm 0cm},clip]{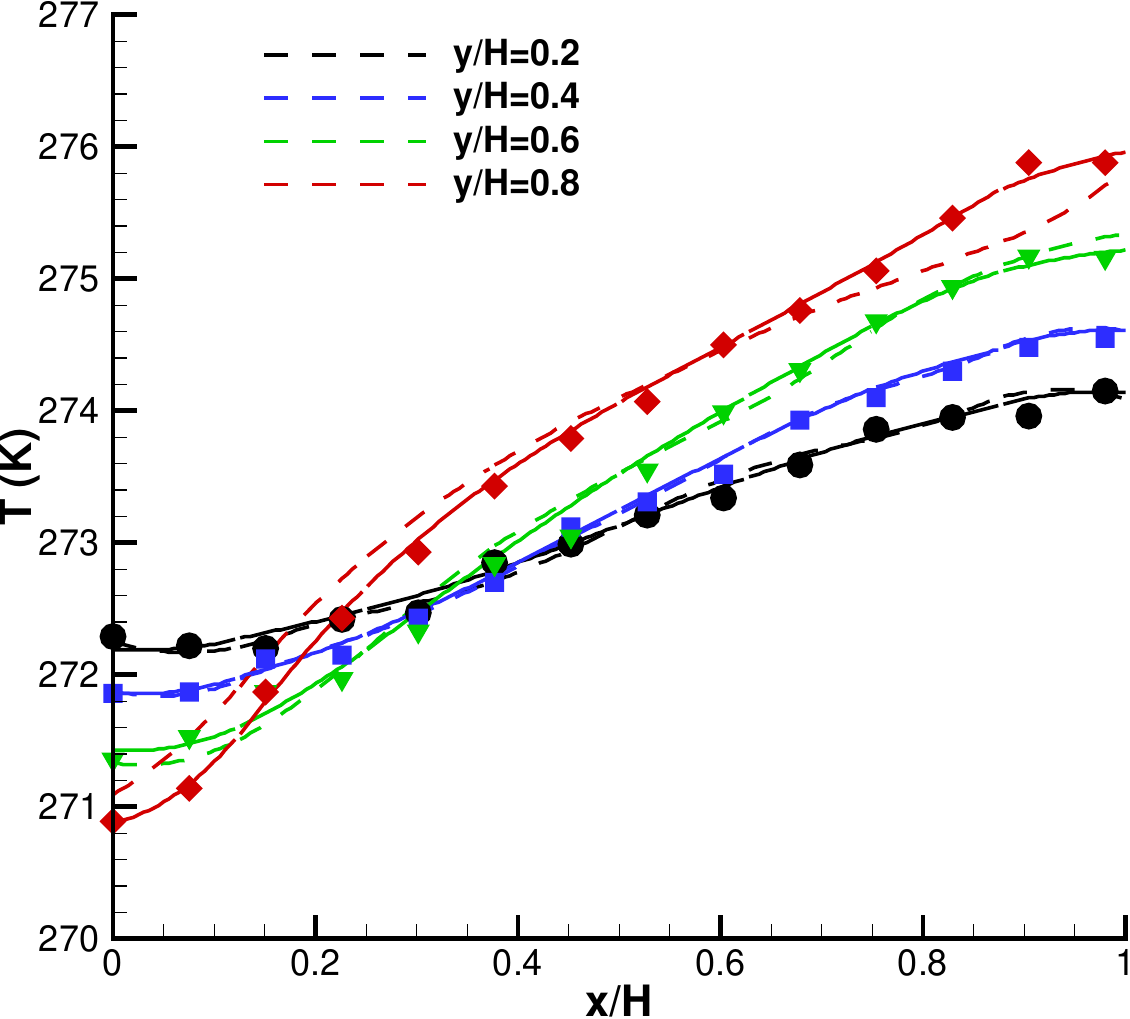}
  \caption{temperature (on horizontal lines)}
  \label{fig_lidCavity_T_hor}
\end{subfigure}%
\begin{subfigure}{.5\textwidth}
  \centering
  \includegraphics[width=75mm,trim={0cm 0cm 0cm 0cm},clip]{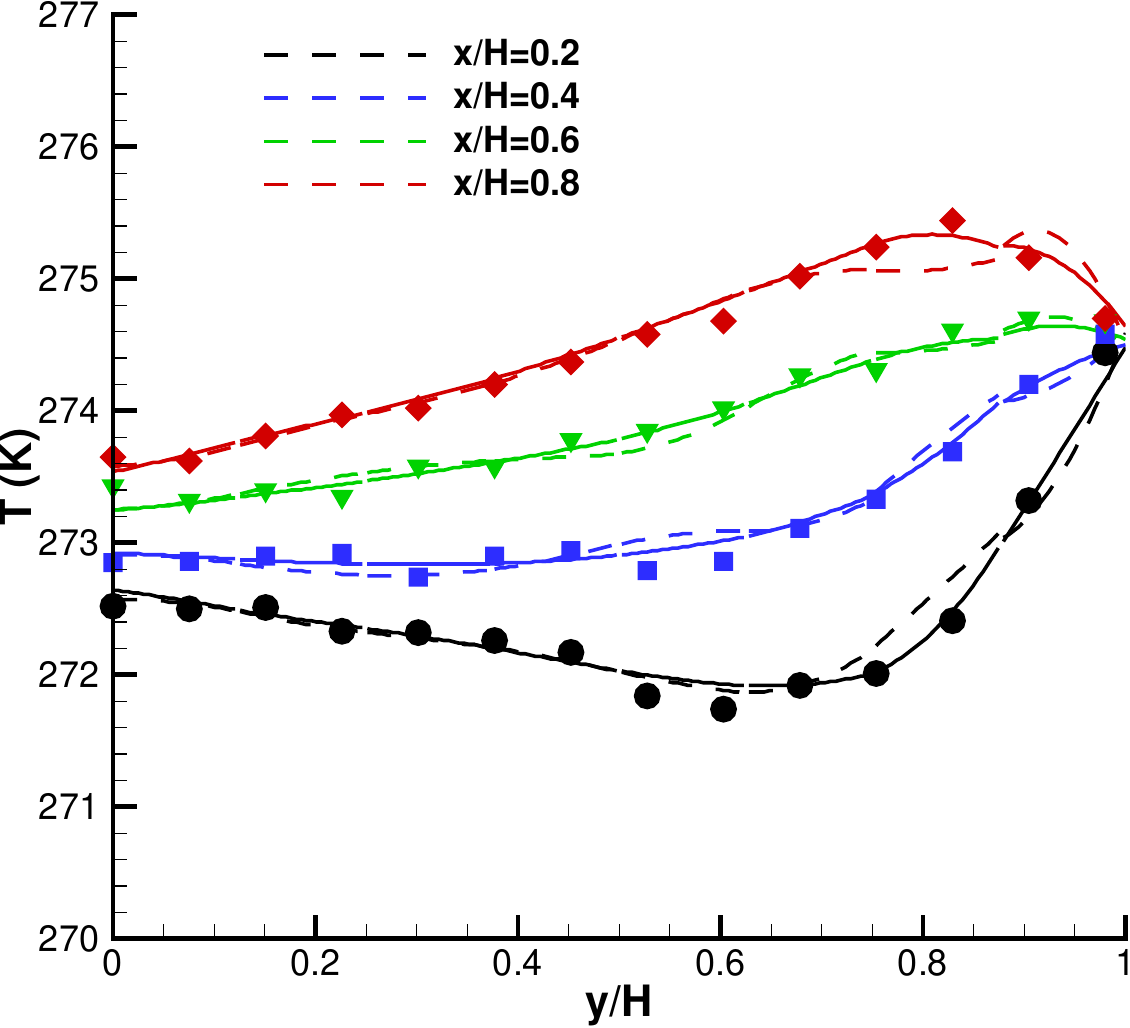}
  \caption{temperature (on vertical lines)}
  \label{fig_lidCavity_T_ver}
\end{subfigure}

\caption{{Variation of flow properties on horizontal and vertical lines for lid-driven cavity flow at $\Kn=1$. Symbols denote DSMC results, dashed lines denote DGFS solutions obtained using velocity space $[-5,\,5]^3$ discretized with $N^3=24^3$ points, and solid lines denote DGFS solutions obtained using velocity space $[-6,\,6]^3$ discretized with $N^3=48^3$ points and $N_r=12$. For DGFS, the physical space is discretized using $8 \times 8$ cells and DG order of 3. $M=6$ is used on the half sphere in all cases.}}
\label{fig_lidCavity_UxUyT}
\end{figure*}

\begin{figure*}[tbp]
\centering
\begin{subfigure}{.5\textwidth}
  \centering
  \includegraphics[width=75mm,trim={0cm 0cm 0cm 0cm},clip]{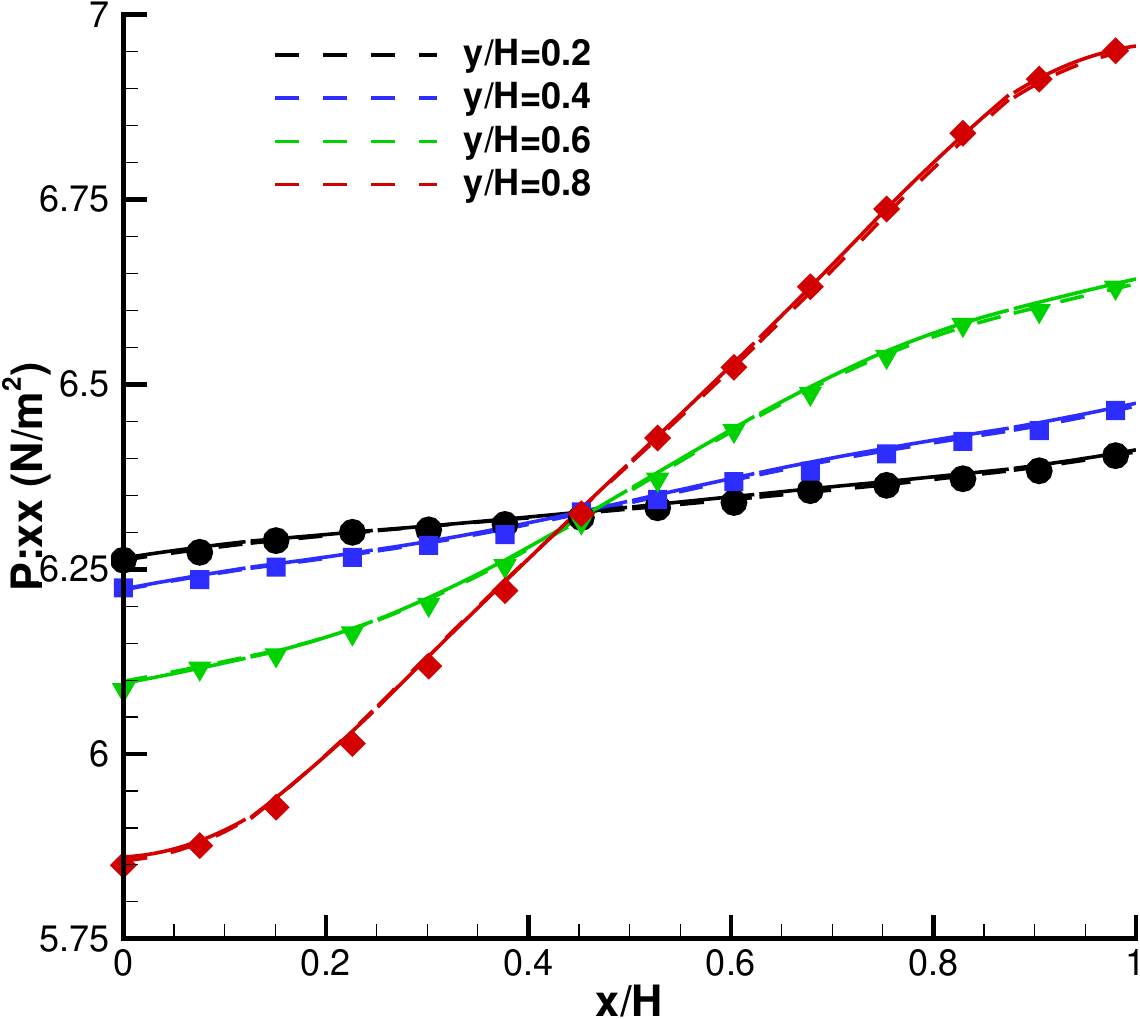}
  \caption{$xx$-component of stress (on horizontal lines)}
  \label{fig_lidCavity_Pxx_hor}
\end{subfigure}%
\begin{subfigure}{.5\textwidth}
  \centering
  \includegraphics[width=75mm,trim={0cm 0cm 0cm 0cm},clip]{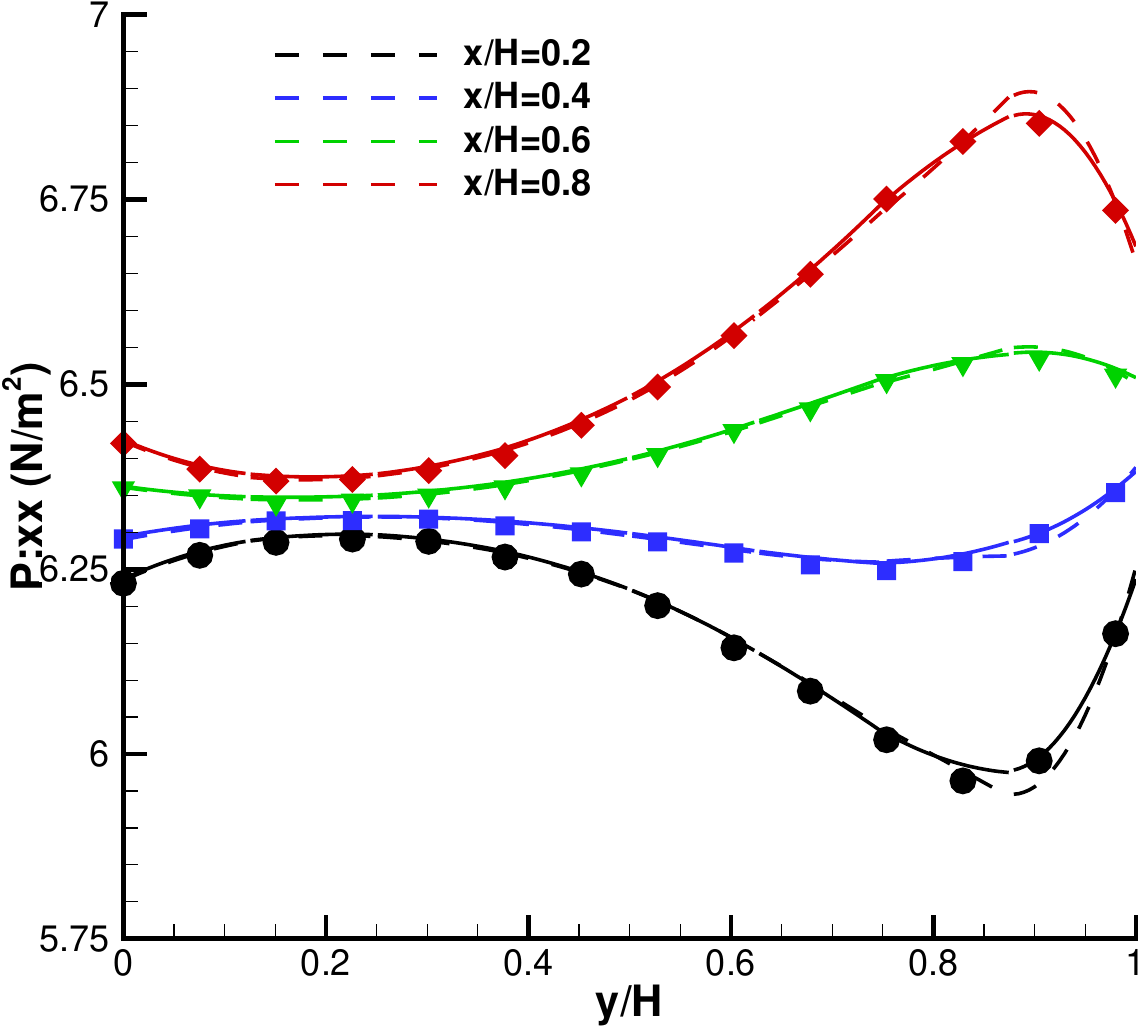}
  \caption{$xx$-component of stress (on vertical lines)}
  \label{fig_lidCavity_Pxx_ver}
\end{subfigure}

\begin{subfigure}{.5\textwidth}
  \centering
  \includegraphics[width=75mm,trim={0cm 0cm 0cm 0cm},clip]{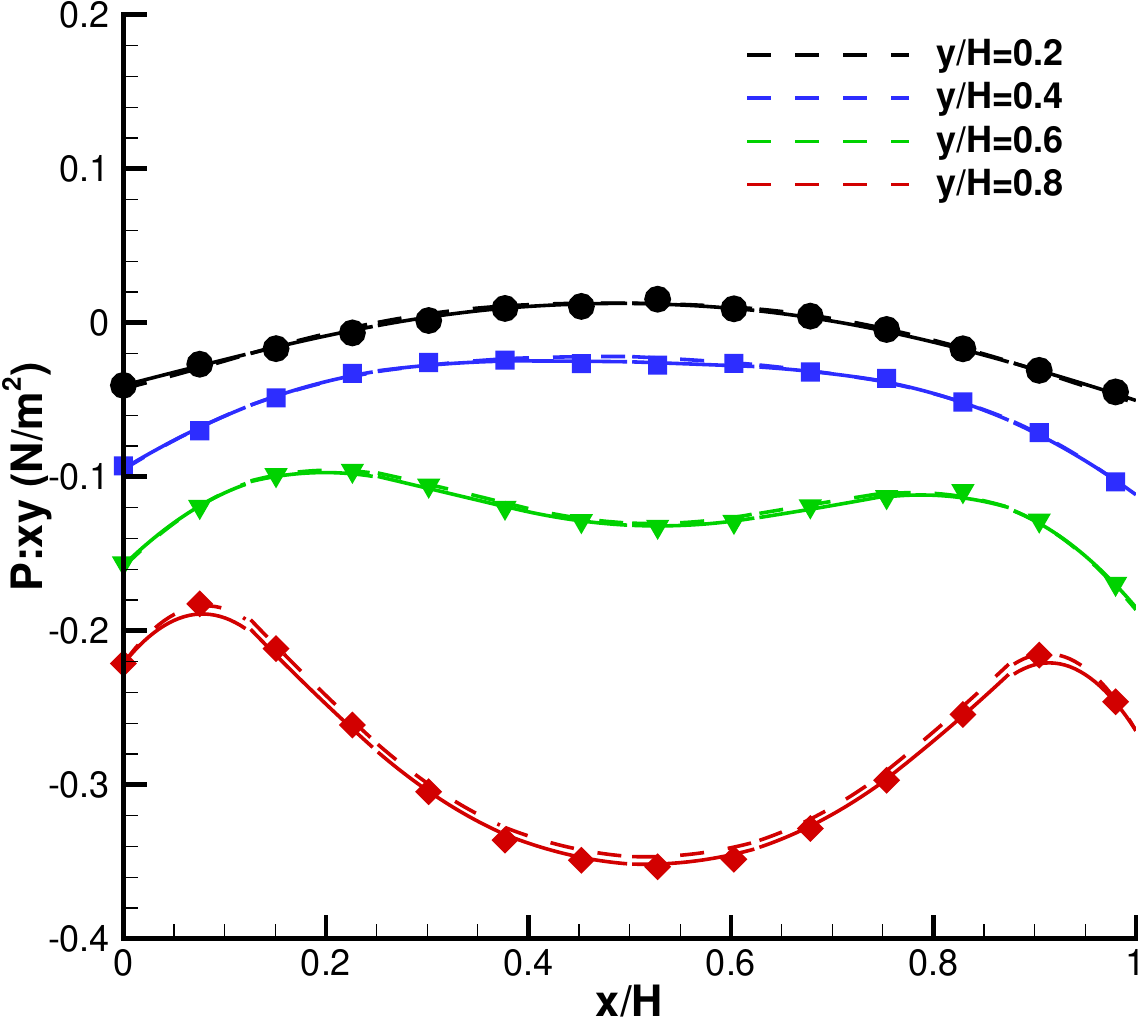}
  \caption{$xy$-component of stress (on horizontal lines)}
  \label{fig_lidCavity_Pxy_hor}
\end{subfigure}%
\begin{subfigure}{.5\textwidth}
  \centering
  \includegraphics[width=75mm,trim={0cm 0cm 0cm 0cm},clip]{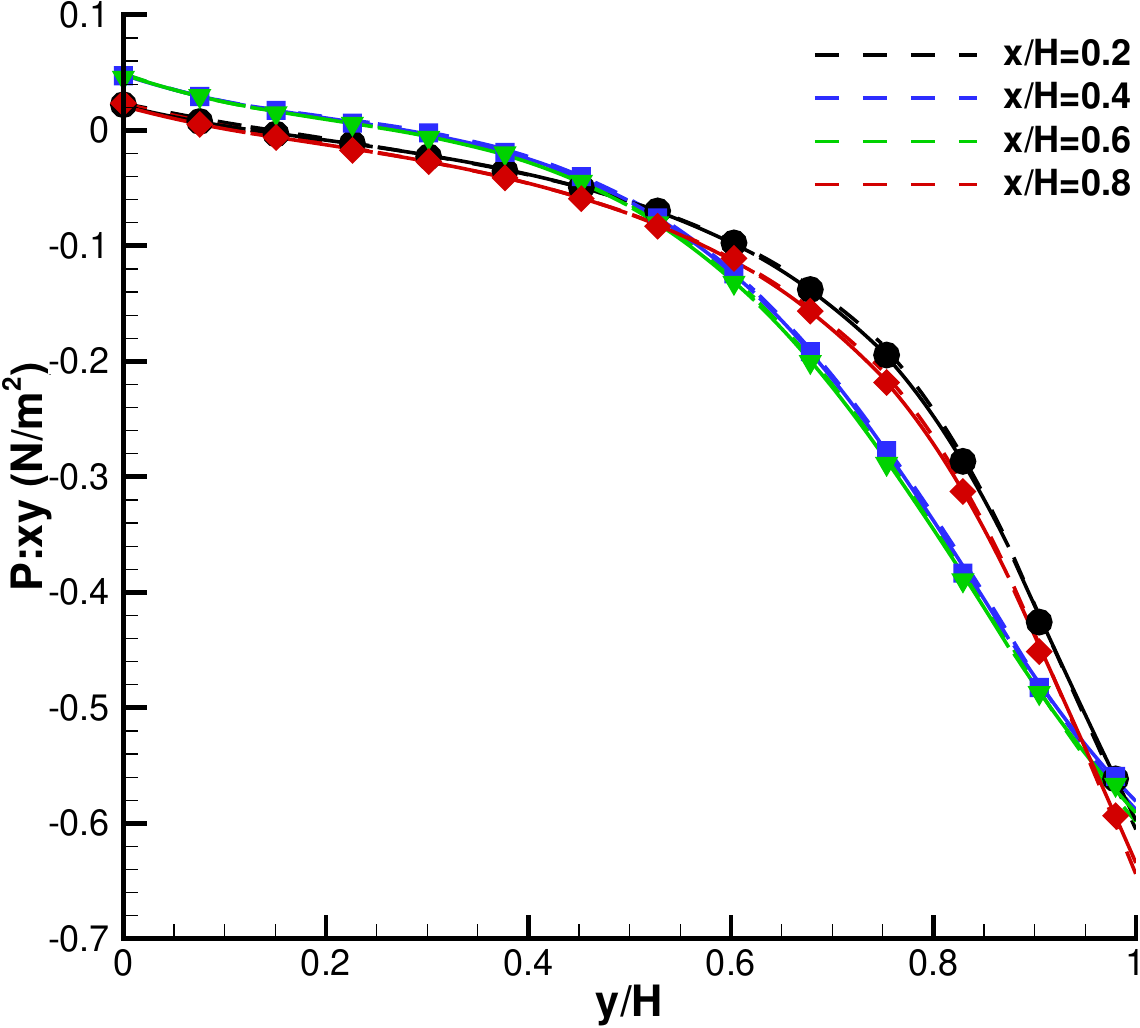}
  \caption{$xy$-component of stress (on vertical lines)}
  \label{fig_lidCavity_Pxy_ver}
\end{subfigure}

\begin{subfigure}{.5\textwidth}
  \centering
  \includegraphics[width=75mm,trim={0cm 0cm 0cm 0cm},clip]{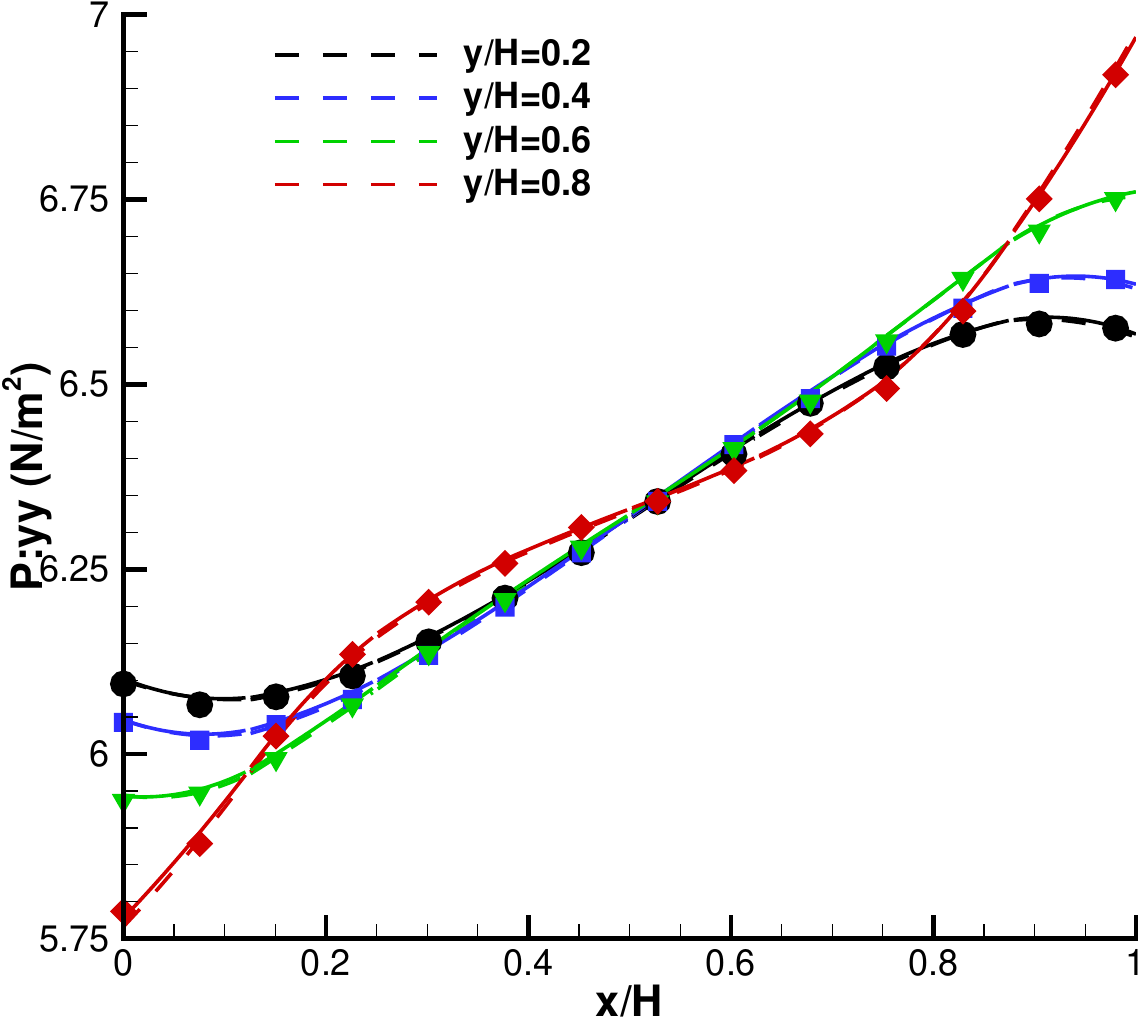}
  \caption{$yy$-component of stress (on horizontal lines)}
  \label{fig_lidCavity_Pyy_hor}
\end{subfigure}%
\begin{subfigure}{.5\textwidth}
  \centering
  \includegraphics[width=75mm,trim={0cm 0cm 0cm 0cm},clip]{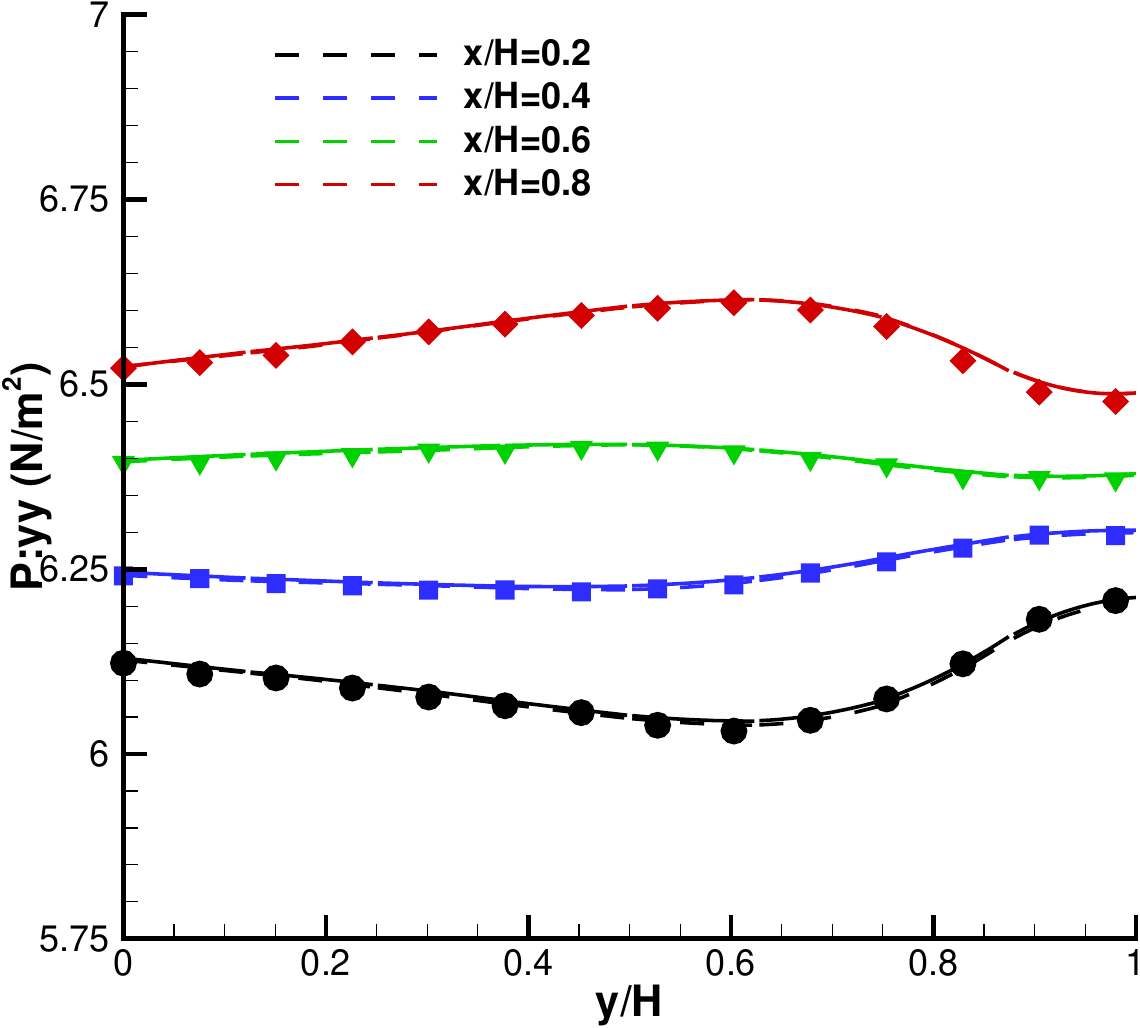}
  \caption{$yy$-component of stress (on vertical lines)}
  \label{fig_lidCavity_Pyy_ver}
\end{subfigure}

\caption{{Continuation of Fig.~\ref{fig_lidCavity_UxUyT}.}}
\label{fig_lidCavity_PxxPxyPyy}
\end{figure*}

\subsection{2D case: thermally driven cavity flow}
\label{subsec_thermalCavity}
We now consider the effect of flow induced due to thermal gradients. We consider a square box of length $H=1\times10^{-3}$ meters. The left and right walls are fixed at $T_c=263K$. At the top and bottom walls, we introduce a linearly increasing temperature (from $T_c$ to $T_h=283K$) in left half of domain, and a linearly decreasing temperature (from $T_h$ to $T_c$) in the right half. {The velocity space $[-6,\,6]^3$ is discretized using $N^3=32^3$ points. $M=6$ is used on the half sphere for all the cases.} The setup of the problem is given in Figure \ref{fig_thermalCavitySchematic}.
                                                                                                                               
\subsubsection{Boundary condition}

At the top and bottom walls, given $T_c$, $T_h$, and the position vector of end-points $\mathbf{r}_c$ and $\mathbf{r}_h$:
\begin{enumerate}
	\item \textbf{DGFS:}  Using the Lagrangian polynomial, we interpolate the temperature values at the known set of DG solution/quadrature points on the surface. Once the temperature $T_i$ is known at a given quadrature point, we then define a Maxwellian wall distribution around $T_i$ for that particular solution point. 
	\item \textbf{DSMC:} Given a particle on boundary with some position vector $\mathbf{r}_i$, we interpolate the temperature linearly using three-dimensional equation of line. And, then we emit the particle with the Maxwellian defined around $T_i$ (interpolated temperature for particle with position vector $\mathbf{r}_i$). 
\end{enumerate}

\begin{figure}[ht]
	\centering
\begin{tikzpicture}[scale=0.9]		
		\def\W{6};
		\def\H{4};
		\def\pW{0.15};	%
		\def\aW{1};
		\def\aH{0.75};
	
		\draw (0,0) -- (\W,0) -- (\W,\H) -- (0,\H) -- (0,0);		%

		\fill[pattern=north west lines, pattern color=gray, line width = 0.1mm, thin] (0,0) rectangle ({-\pW},\H);
		\fill[pattern=north east lines, pattern color=gray, line width = 0.1mm, thin] (\W,0) rectangle ({\W+\pW},\H);
		
		\fill[pattern=north west lines, pattern color=gray, line width = 0.1mm, thin] (0,0) rectangle (\W,{-\pW});
		\fill[pattern=north east lines, pattern color=gray, line width = 0.1mm, thin] (0,\H) rectangle ({\W},{\H+\pW});
		
		\draw[thick,-latex, line width=0.40mm] ({\W*0.75},{\H*0.5}) -- ({\W*0.75+\aW},{\H*0.5});	%
		\draw[thick,-latex, line width=0.40mm] ({\W*0.25},{\H*0.5}) -- ({\W*0.25-\aW},{\H*0.5});	%
		\node[] at ({\W*0.5},{\H*0.5}) {$T_c$};					%
	
		\draw[domain=0:23mm, samples=100, black, xshift={\W*0.6 cm}, yshift={\H cm}, -latex, thick] plot ({-2*cos(\x)},{-sin(\x)}) node[anchor=west] {$T_f$};
		
		\draw[domain=0:23mm, samples=100, black, xshift={\W*0.6 cm}, yshift={0 cm}, -latex, thick] plot ({-2*cos(\x)},{sin(\x)}) node[anchor=west] {$T_f$};

		\draw[thick,-latex, line width=0.45mm, black] (0,0) -- (0,1);				%
		\draw[thick,-latex, line width=0.45mm, black] (0,0) -- (1,0);				%
		
		\node[] at (0.5,-0.4) {$x$};
		\node[] at (-0.4,0.4) {$y$};

		\def\pr{1.22};
		\foreach \y in {1,...,5}
			\foreach \x in {1,...,9} {
				\draw[dotted] ({-\pr + (\pr)^\x},{-\pr + (\pr)^\y}) -- ({-\pr + (\pr)^(\x+1)},{-\pr + (\pr)^(\y)});
			}

		\foreach \y in {1,...,5}
			\foreach \x in {1,...,9} {
				\draw[dotted] ({-\pr + (\pr)^\x},{4 + \pr - (\pr)^\y}) -- ({-\pr + (\pr)^(\x+1)},{4 + \pr - (\pr)^(\y)});
			}

		\foreach \y in {1,...,5}
			\foreach \x in {1,...,6} {
				\draw[dotted] ({-\pr + (\pr)^\x},{-\pr + (\pr)^\y}) -- ({-\pr + (\pr)^(\x)},{-\pr + (\pr)^(\y+1)});
			}

		\foreach \y in {1,...,5}
			\foreach \x in {1,...,6} {
				\draw[dotted] ({-\pr + (\pr)^\x},{4 + \pr - (\pr)^\y}) -- ({-\pr + (\pr)^(\x)},{4 + \pr - (\pr)^(\y+1)});
			}

		\foreach \y in {1,...,5}
			\foreach \x in {1,...,6} {
				\draw[dotted] ({6 + \pr - (\pr)^\x},{-\pr + (\pr)^\y}) -- ({6 + \pr - (\pr)^(\x)},{-\pr + (\pr)^(\y+1)});
			}

		\foreach \y in {1,...,5}
			\foreach \x in {1,...,6} {
				\draw[dotted] ({6 + \pr - (\pr)^\x},{4 + \pr - (\pr)^\y}) -- ({6 + \pr - (\pr)^(\x)},{4 + \pr - (\pr)^(\y+1)});
			}

		\def\Off{0.75};
		
		\draw[dashed](0,-\Off) -- (0,-{\H+\Off});
		\draw[dashed] (0,-{\H+\Off}) node[anchor=east] {$T_c$} -- (\W,-{\H+\Off}) node[anchor=north east] {\footnotesize Temperature profile : $T_f$};
		\draw[dashed](\W,-\Off) -- (\W,-{\H+\Off});
		\draw(0,-{\H+\Off}) -- (\W/2,-\Off);
		\draw (\W/2,-\Off) -- (\W,-{\H+\Off});		
		\draw[dotted] (0,-\Off) node[anchor=east] {$T_h$} -- (\W/2,-\Off);
	\end{tikzpicture} 	\caption{Numerical setup for thermally driven cavity flow. The representative linearly-graded mesh is shown with dotted lines. Due to symmetry of the problem, part of the domain, denoted by thick dashed red line, is used in simulation.}
	\label{fig_thermalCavitySchematic}
\end{figure}
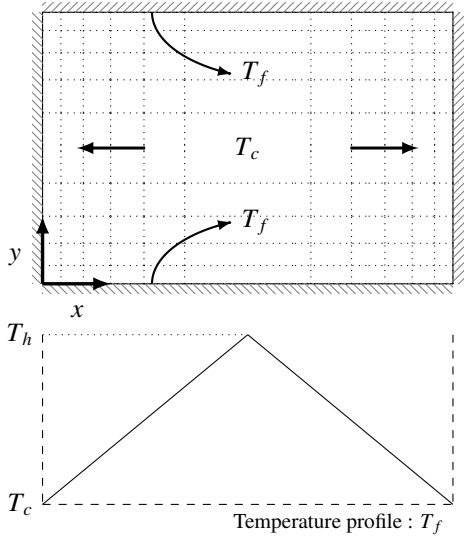

{Considering the symmetry of the problem, we simulate the $[0,\,H/2]^2$ region of the spatial domain, denoted by thick dashed red line in the Fig.~(\ref{fig_thermalCavitySchematic}). Consequently, at the top and the right boundaries, a symmetry boundary condition is imposed:}
\begin{equation}
	f_{sym}(t,\bx,\bc^j) =  f(t,\bx,\bc^{r}) ,
\end{equation}
{where $f$ is the interior domain solution adjacent to the boundary, $\bc^{r}=\bc^j - 2(\bc^j\cdot{\bf\hat{n}})\,{\bf\hat{n}}$ is the reflected velocity, and $r$ is the index associated with the discrete velocity which is computed using the minimum of $|\bc^{r}-\bc^t|,\ \ t=1,\dots,N^3$.}

\subsubsection{Flow properties}
\begin{figure*}[tbp]
\centering
\begin{subfigure}{.5\textwidth}
  \centering
  \includegraphics[width=75mm,trim={0cm 0cm 0cm 0cm},clip]{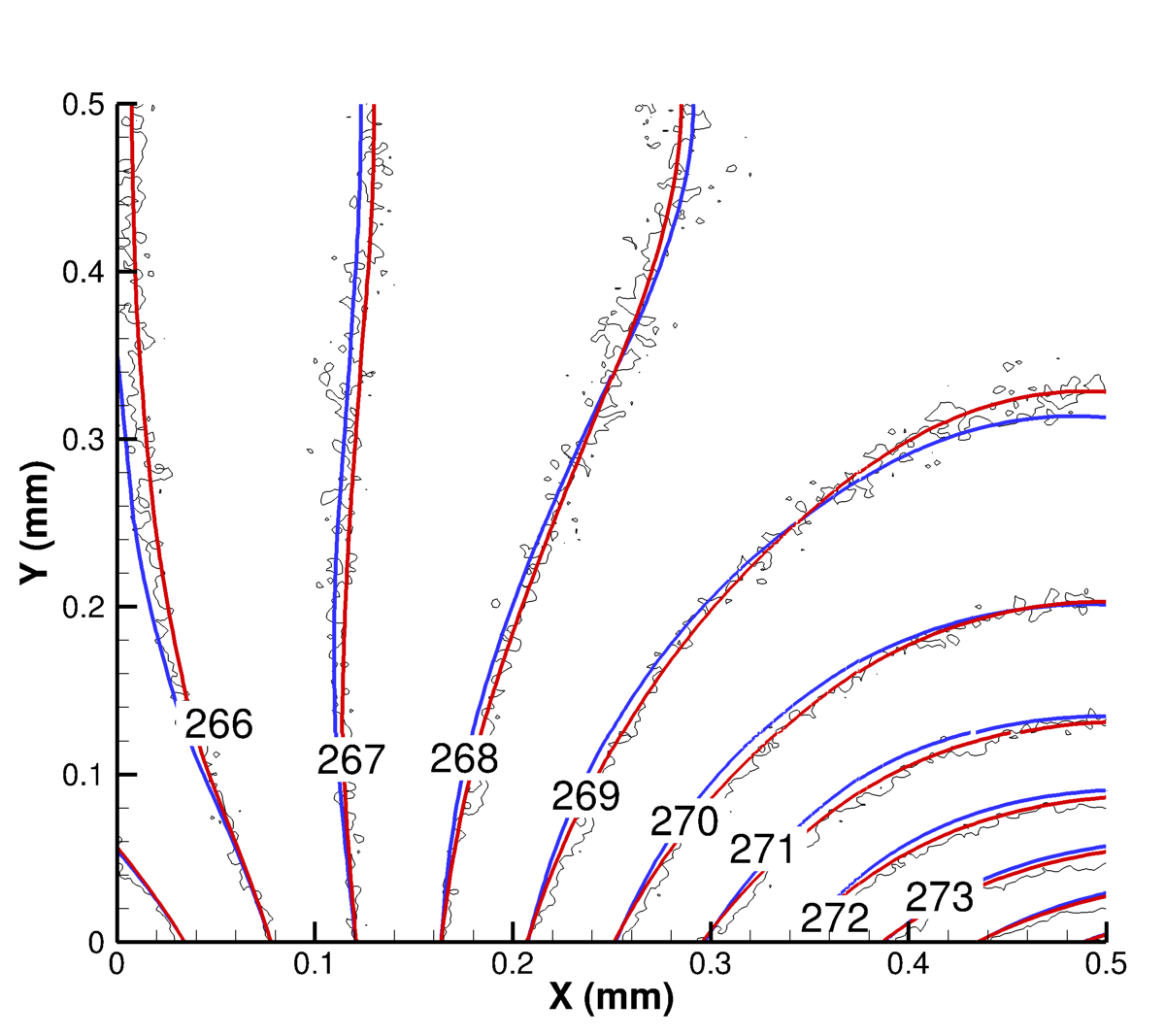}
  \caption{temperature}
  \label{fig_cavity_T_contour}
\end{subfigure}%
\begin{subfigure}{.5\textwidth}
  \centering
  \includegraphics[width=75mm,trim={0cm 0cm 0cm 0cm},clip]{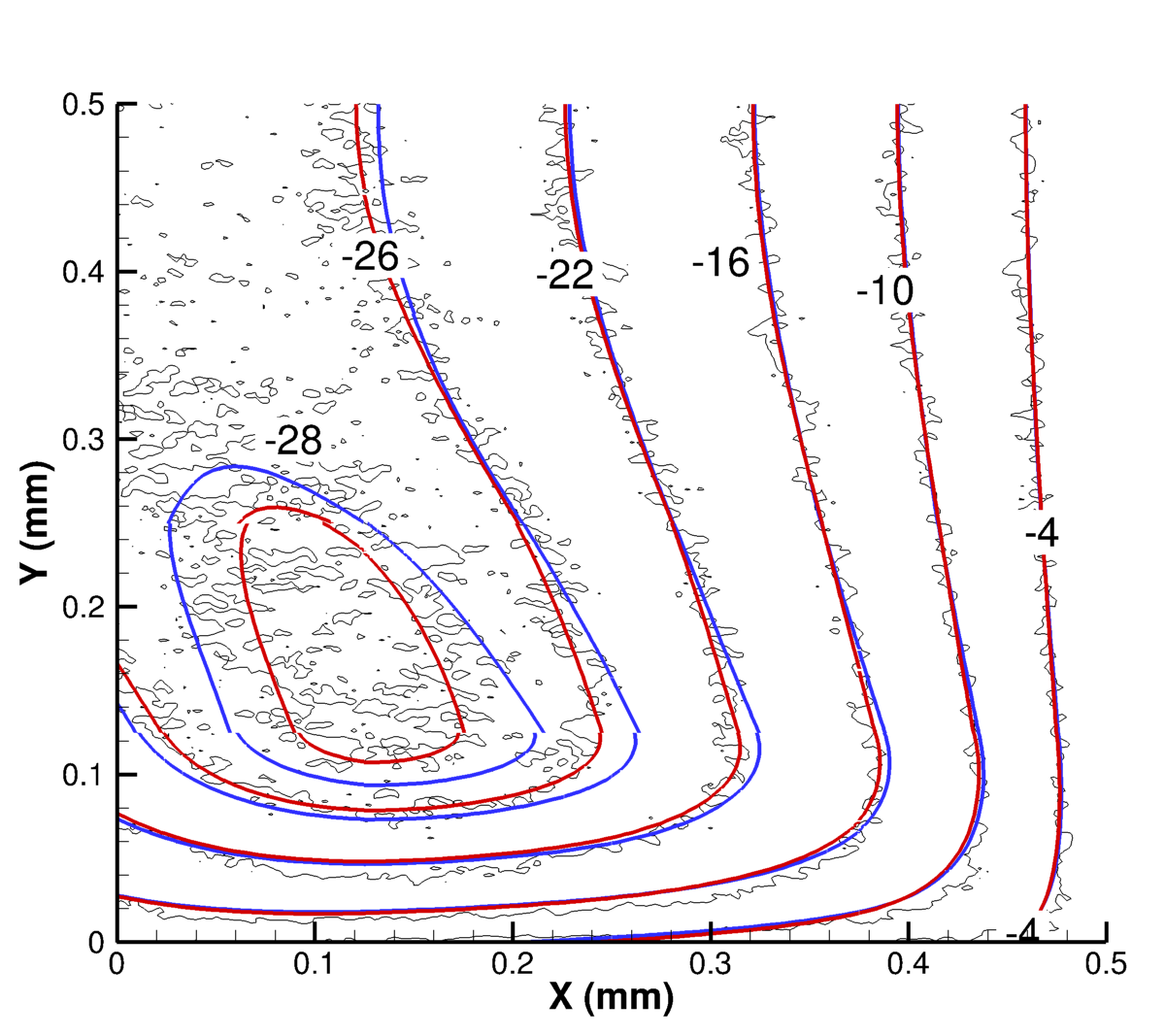}
  \caption{$x$-component of heat-flux}
  \label{fig_cavity_Qx_contour}
\end{subfigure}

\begin{subfigure}{.5\textwidth}
  \centering
  \includegraphics[width=75mm,trim={0cm 0cm 0cm 0cm},clip]{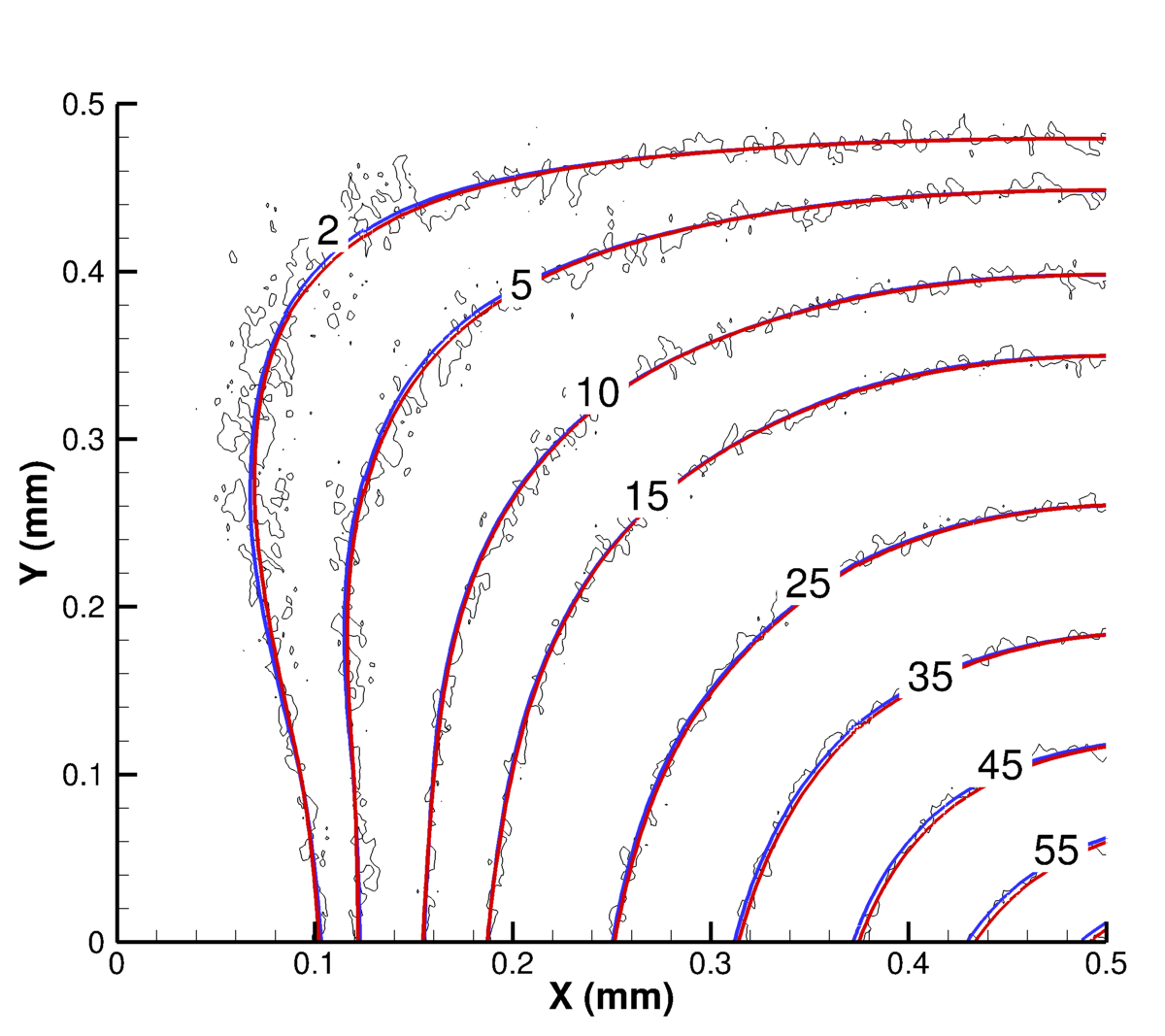}
  \caption{$y$-component of heat-flux}
  \label{fig_cavity_Qy_contour}
\end{subfigure}%
\begin{subfigure}{.5\textwidth}
  \centering
  \includegraphics[width=75mm,trim={0cm 0cm 0cm 0cm},clip]{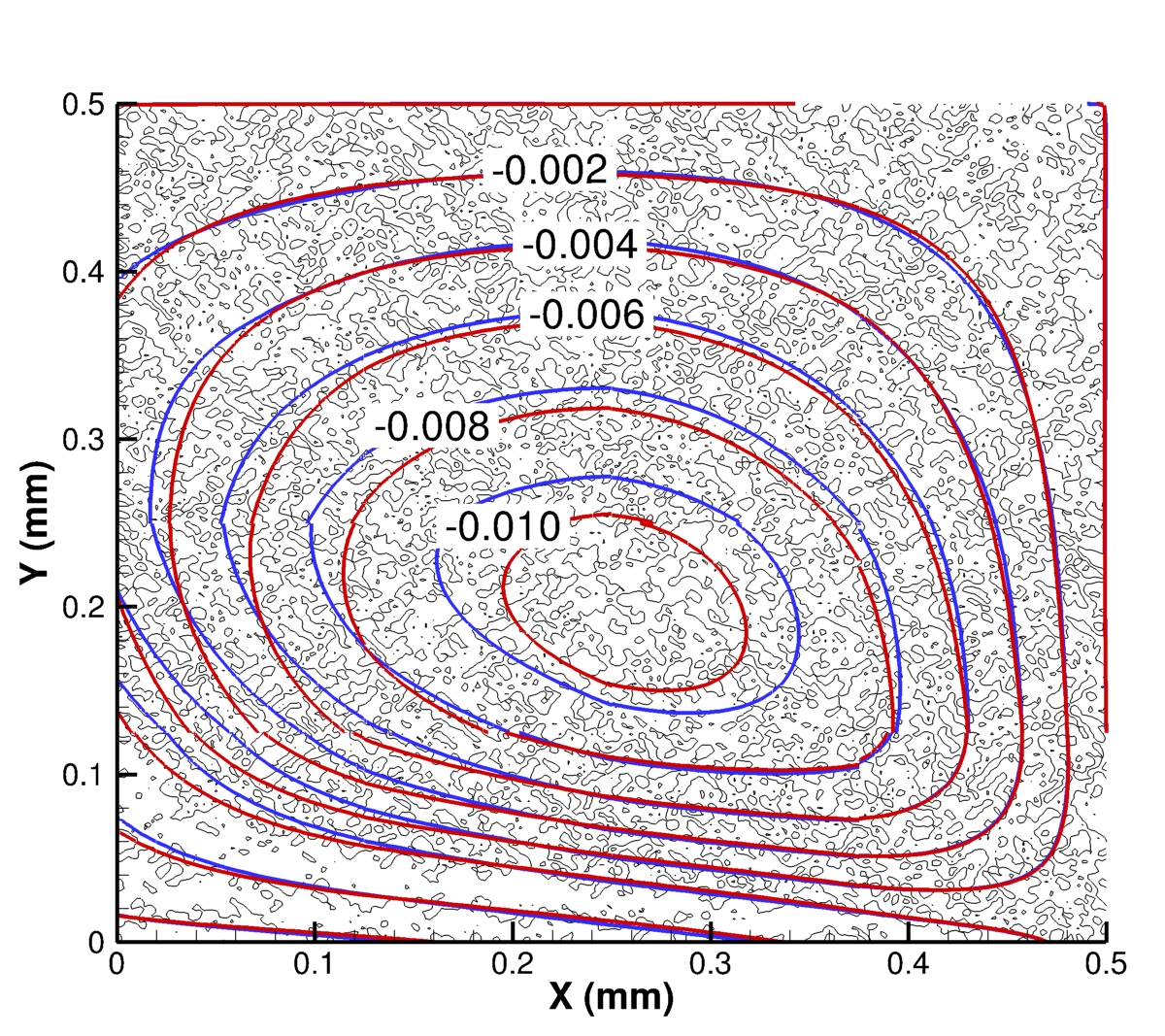}
  \caption{$xy$-component of stress}
  \label{fig_cavity_Pxy_contour}
\end{subfigure}

\begin{subfigure}{.5\textwidth}
  \centering
  \includegraphics[width=75mm,trim={0cm 0cm 0cm 0cm},clip]{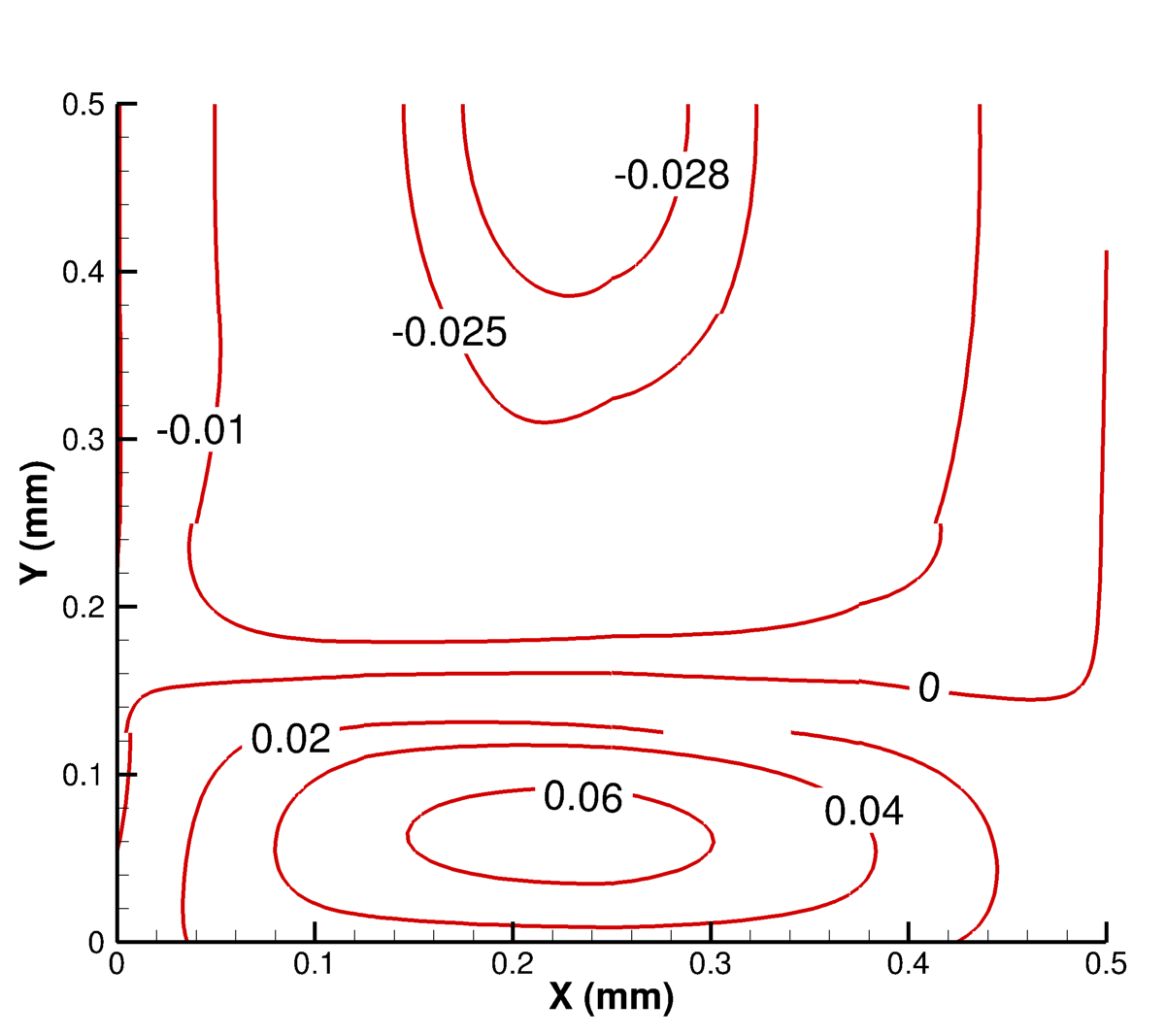}
  \caption{$x$-component of velocity}
  \label{fig_cavity_Ux_contour}
\end{subfigure}%
\begin{subfigure}{.5\textwidth}
  \centering
  \includegraphics[width=75mm,trim={0cm 0cm 0cm 0cm},clip]{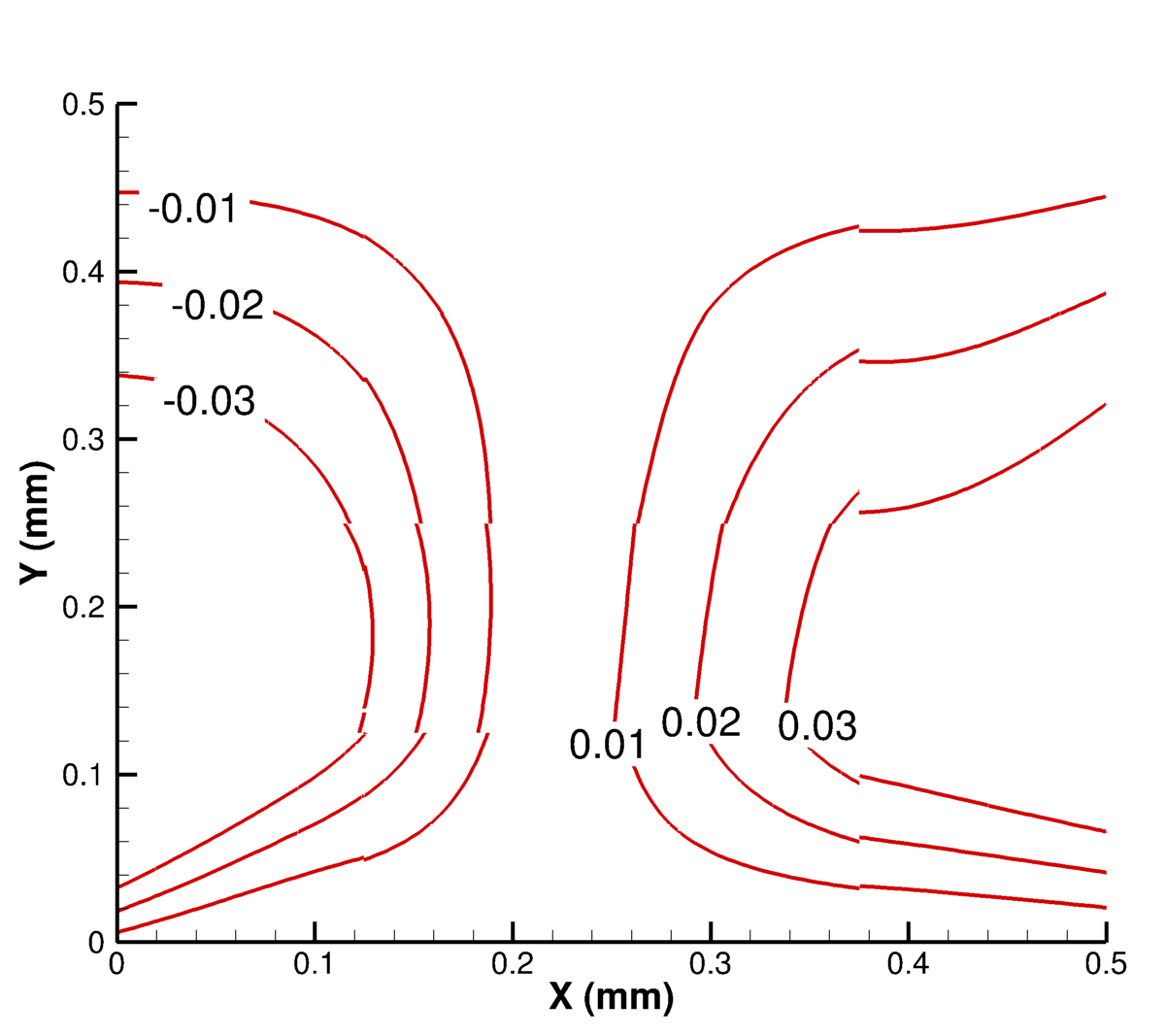}
  \caption{$y$-component of velocity}
  \label{fig_cavity_Uy_contour}
\end{subfigure}

\caption{{Contours of various flow properties for thermal-driven flow at $\Kn=1$ obtained with DSMC (thin black lines), DGFS employing velocity space $[-5,\,5]^3$ discretized with $N^3=24^3$ points (solid blue lines), and DGFS employing velocity space $[-6,\,6]^3$ discretized with $N^3=48^3$ points and $N_r=12$ (solid red lines). For DGFS, the physical space is discretized using $4 \times 4$ cells and DG order of 3. $M=6$ is used on the half sphere in all cases. In the present case, due to the presence of temperature gradients, a very low-velocity gas motion is induced \cite{kogan1976stresses}. Due to high statistical noise, DSMC results for velocity have been removed from Figs.~\ref{fig_cavity_Ux_contour},~\ref{fig_cavity_Uy_contour}.}}
\label{fig_thermalCavity_contour}
\end{figure*}

{Figure \ref{fig_thermalCavity_contour} shows the contour plot of various flow properties. Figures~\ref{fig_cavity_TQxQy},~\ref{fig_cavity_PxyUxUy} illustrate the comparison of flow properties on vertical and horizontal lines along and across the domain. We observe a fair agreement between DSMC and DGFS results except for velocity. Due to the presence of temperature gradients, a very low-velocity gas motion is induced \cite{kogan1976stresses}. DSMC finds it difficult to reproduce the \textit{slow} gas-motion due to the statistical noise. Note that the DSMC simulations for the present case employed 100 billion samples in an attempt to reproduce a meaningful average. Increasing the sample size in DSMC should further resolve the fluctuations in the shear-stress and velocity components. However, the same remains elusive from a computational viewpoint. }%

{From a computation viewpoint, DSMC-SPARTA simulations at $Kn=1$, with $500\times 500$ cells, 30 particles per cell, a time-step of 2e-9 sec, 200,000 unsteady time-steps, and 1200000 steady time-steps, on 32 CPU processors took 109155.55 sec. On the other hand, DGFS simulations on a single GPU at $Kn=1$, with $4 \times 4$ elements, 3rd order DGFS, $N^3=24^3$, $M=6$ took $\sim$56020.99 sec to achieve $(\|f^{n+1}-f^{n}\|/\|f^{n}\|_{L_2})/(\|f^{2}-f^{1}\|/\|f^{1}\|_{L_2}) < 3\times10^{-5}$. Note that these are representative simulation times for indicating the computational efforts required in DGFS and DSMC for 2-D simulations. Our experience shows that even heavily tuned codes can be further improved. A detailed comparison between CPU and GPU performance is subject of future study.}

\begin{figure*}[tbp]
\centering
\begin{subfigure}{.5\textwidth}
  \centering
  \includegraphics[width=70mm,trim={0cm 0cm 0cm 0cm},clip]{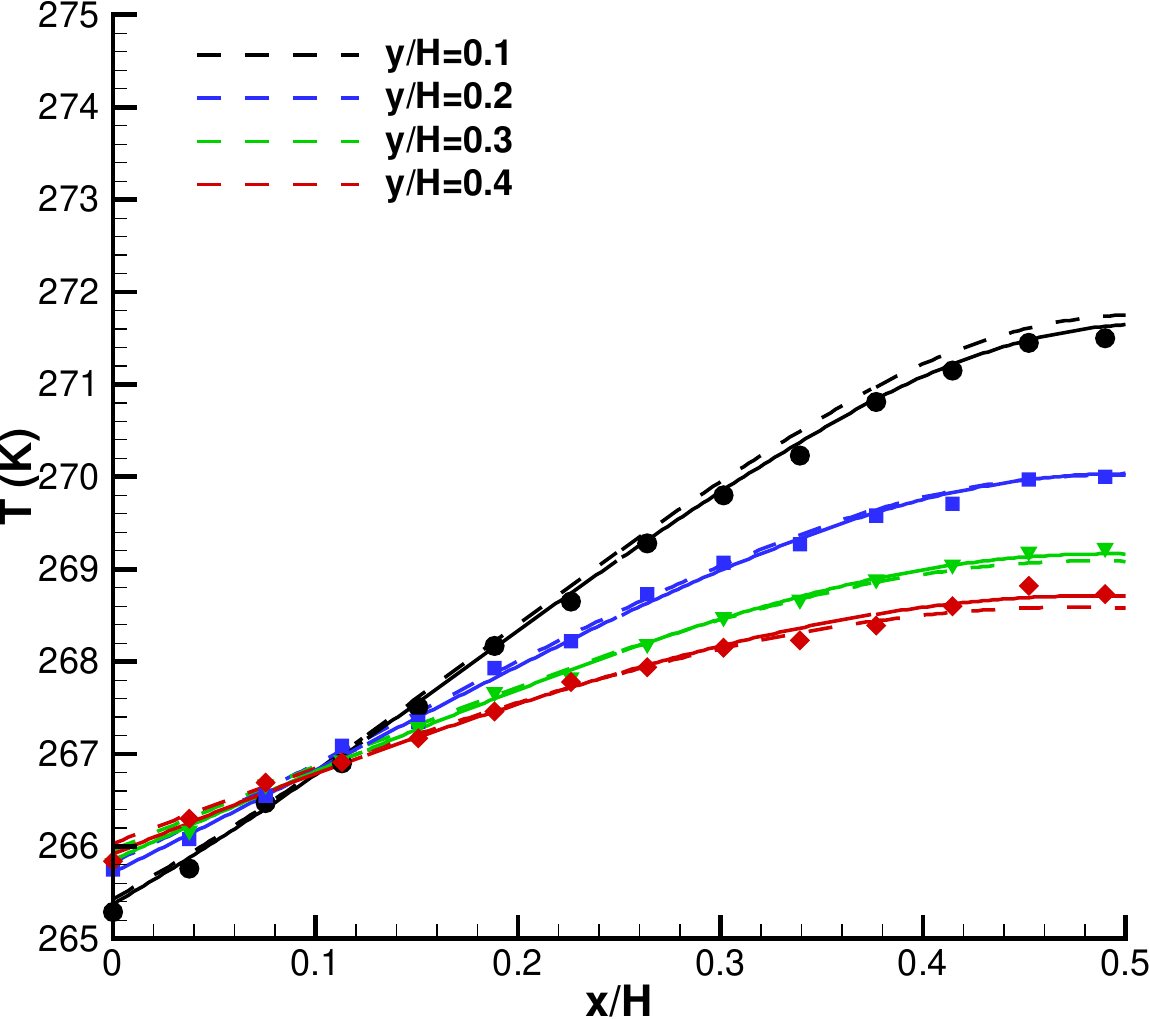}
  \caption{temperature (on horizontal lines)}
  \label{fig_cavity_T_hor}
\end{subfigure}%
\begin{subfigure}{.5\textwidth}
  \centering
  \includegraphics[width=70mm,trim={0cm 0cm 0cm 0cm},clip]{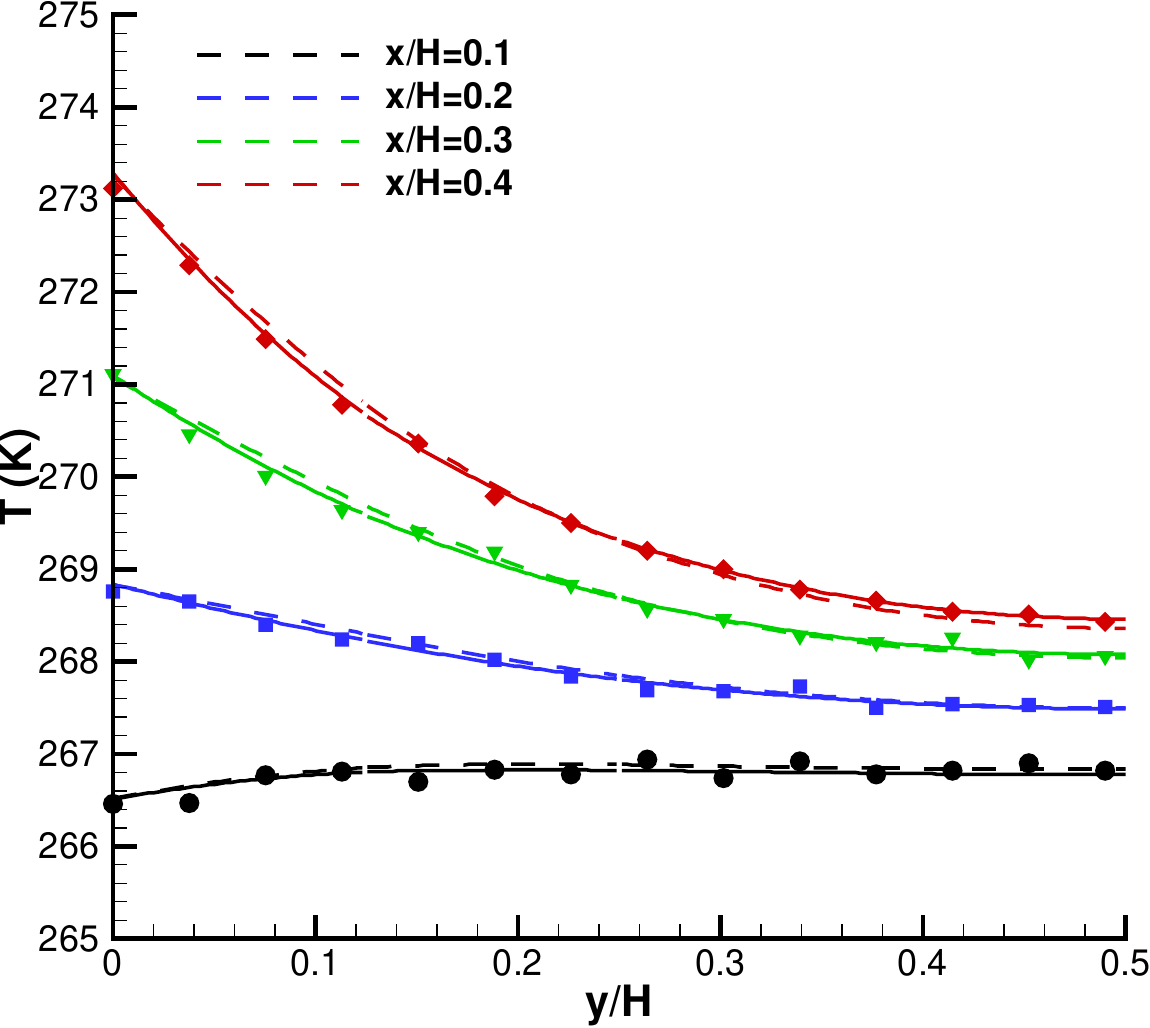}
  \caption{temperature (on vertical lines)}
  \label{fig_cavity_T_ver}
\end{subfigure}

\begin{subfigure}{.5\textwidth}
  \centering
  \includegraphics[width=70mm,trim={0cm 0cm 0cm 0cm},clip]{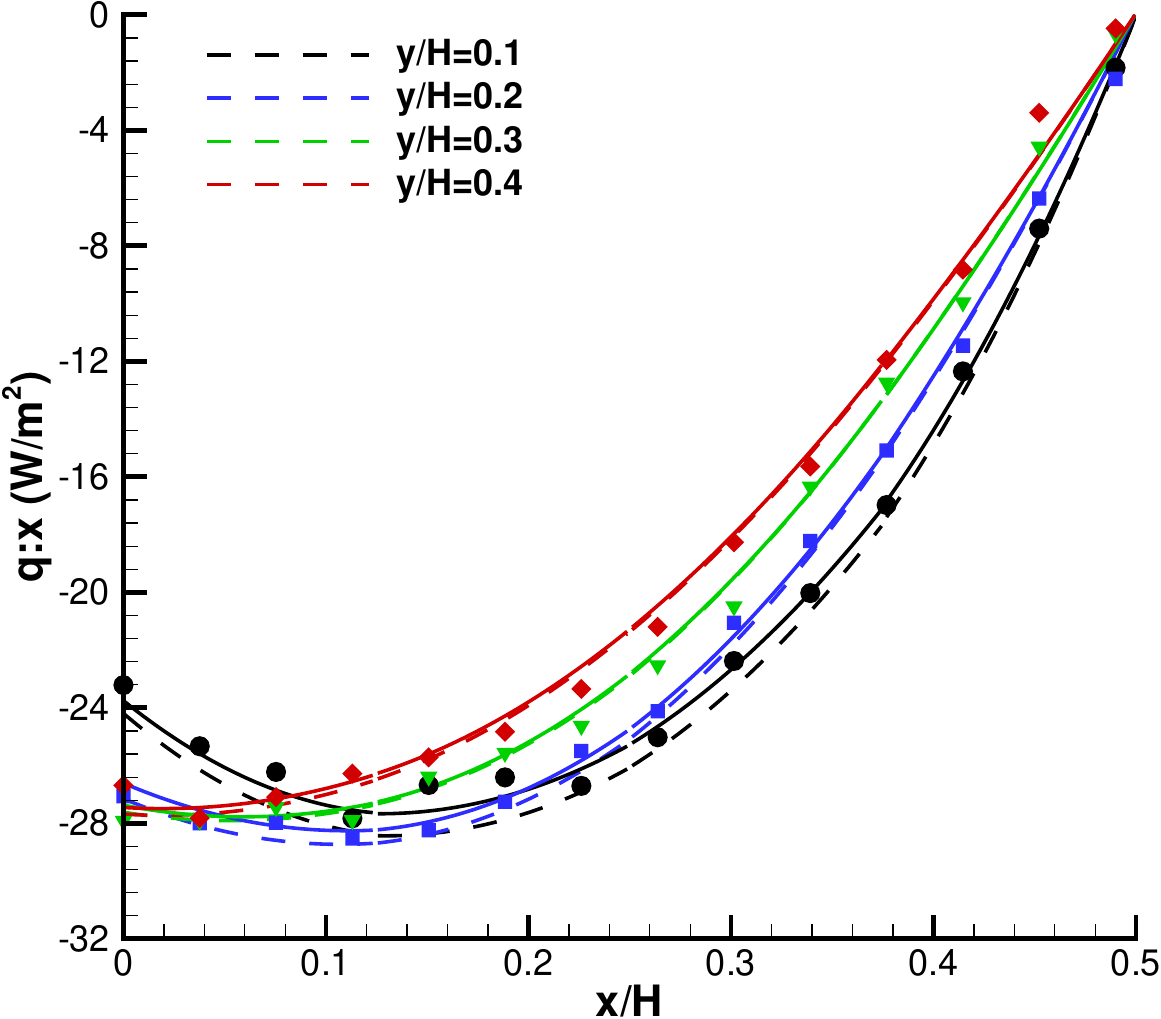}
  \caption{$x$-component of heat-flux (on horizontal lines)}
  \label{fig_cavity_Qx_hor}
\end{subfigure}%
\begin{subfigure}{.5\textwidth}
  \centering
  \includegraphics[width=70mm,trim={0cm 0cm 0cm 0cm},clip]{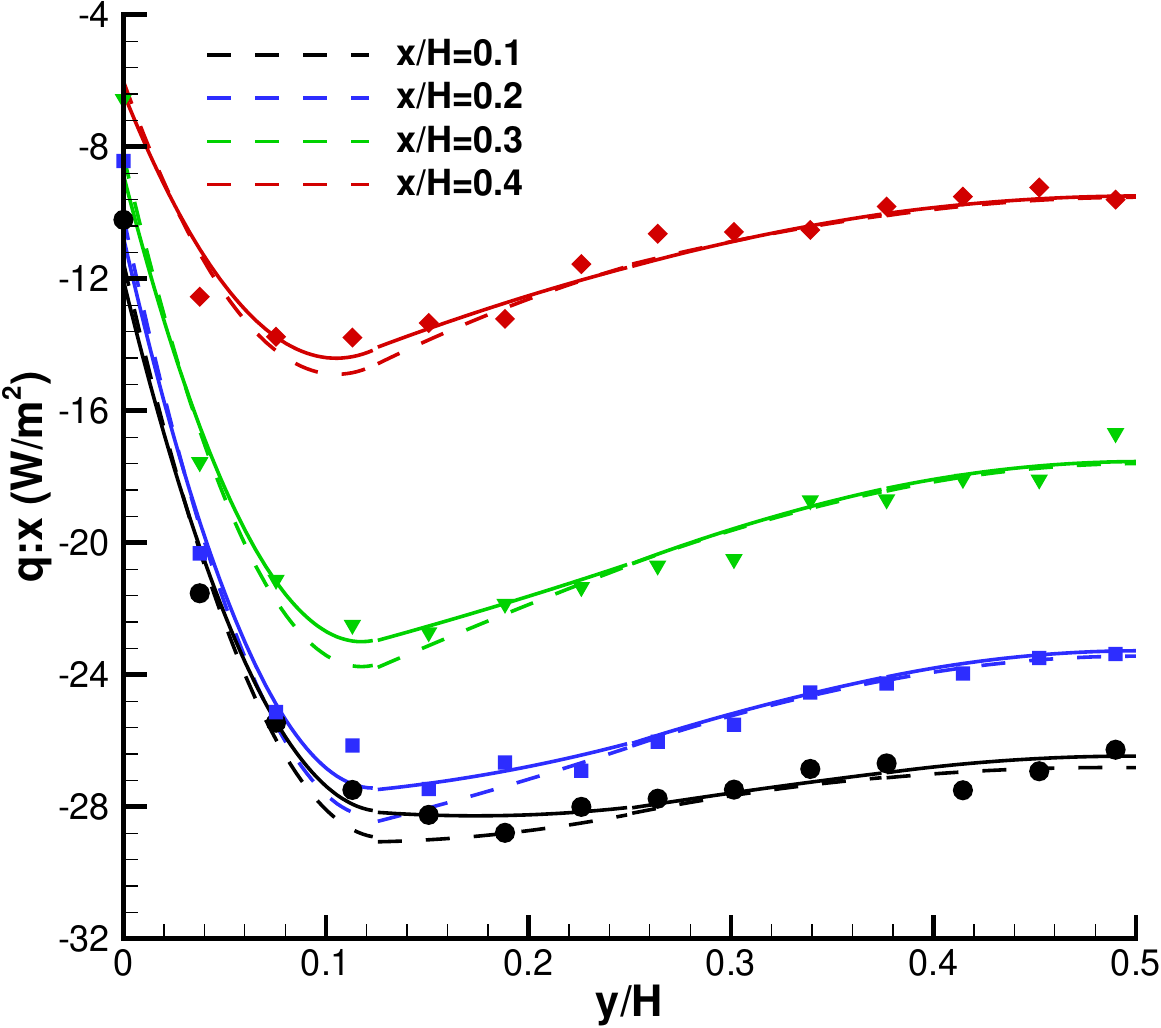}
  \caption{$x$-component of heat-flux (on vertical lines)}
  \label{fig_cavity_Qx_ver}
\end{subfigure}

\begin{subfigure}{.5\textwidth}
  \centering
  \includegraphics[width=70mm,trim={0cm 0cm 0cm 0cm},clip]{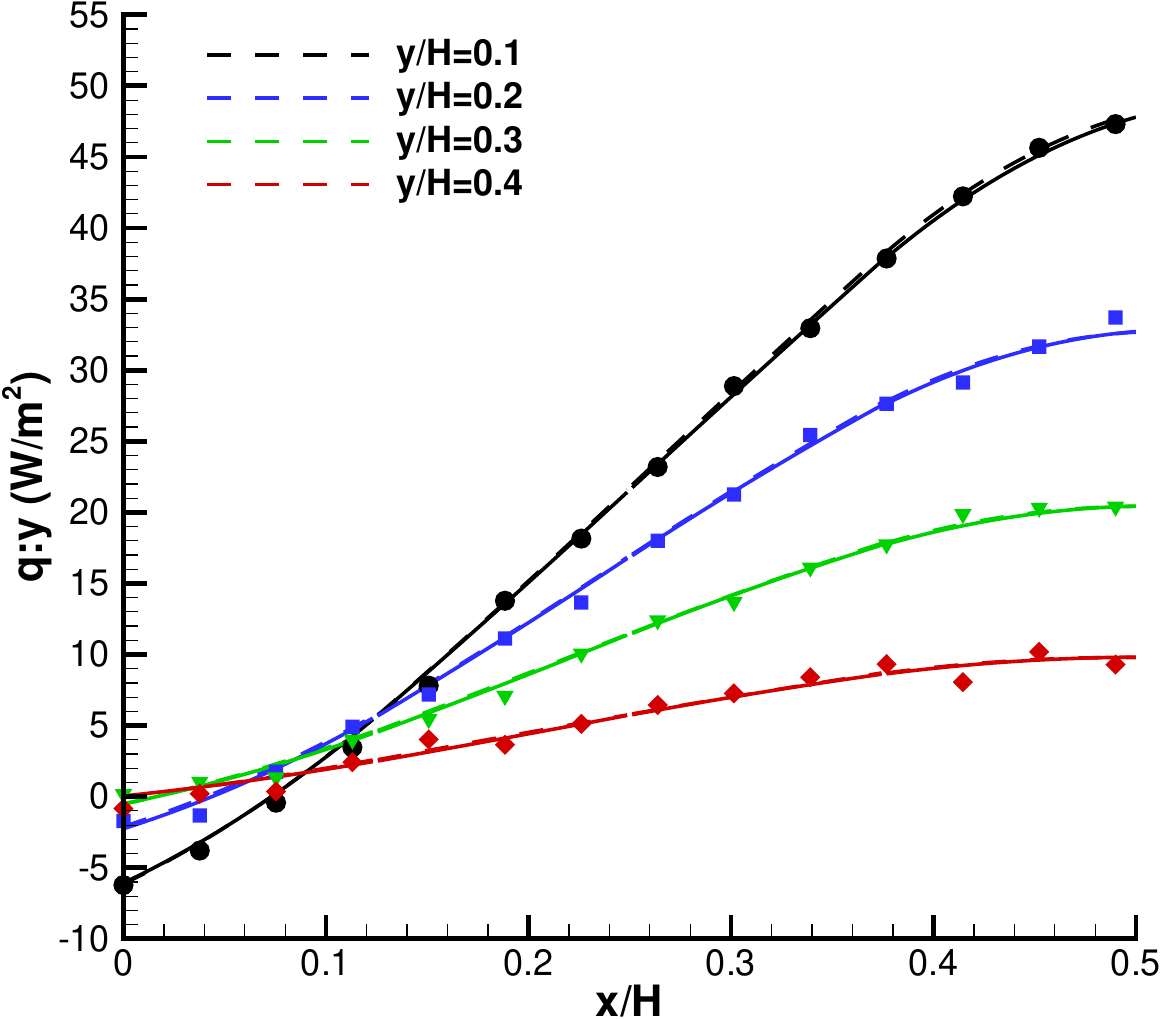}
  \caption{$y$-component of heat-flux (on horizontal lines)}
  \label{fig_cavity_Qy_hor}
\end{subfigure}%
\begin{subfigure}{.5\textwidth}
  \centering
  \includegraphics[width=70mm,trim={0cm 0cm 0cm 0cm},clip]{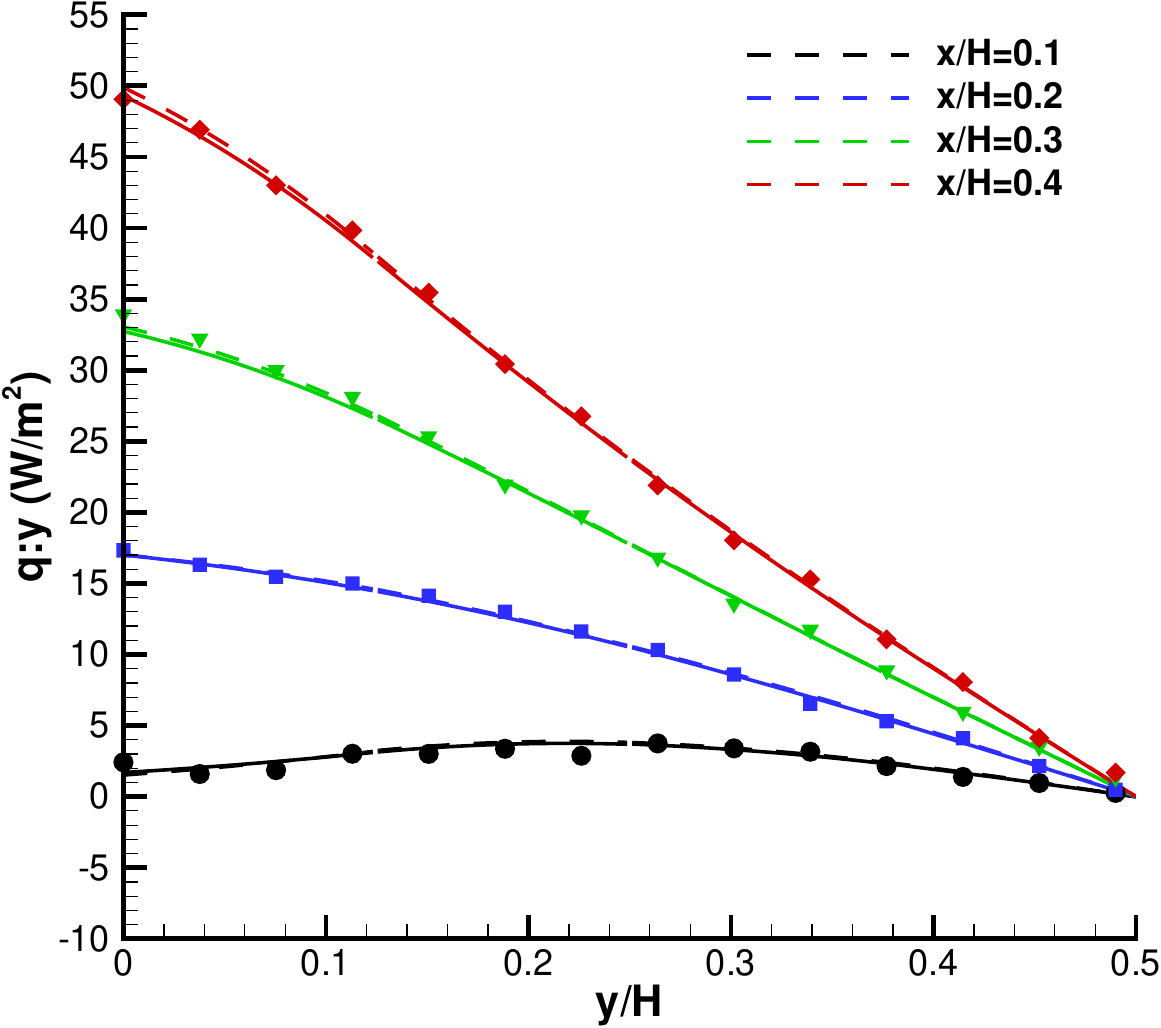}
  \caption{$y$-component of heat-flux (on vertical lines)}
  \label{fig_cavity_Qy_ver}
\end{subfigure}

\caption{{Variation of flow properties on horizontal and vertical lines for thermal-driven flow at $\Kn=1$. Symbols denote DSMC results, dashed lines denote DGFS solutions obtained using velocity space $[-5,\,5]^3$ discretized with $N^3=24^3$ points, and solid lines denote DGFS solutions obtained using velocity space $[-6,\,6]^3$ discretized with $N^3=48^3$ points and $N_r=12$. For DGFS, the physical space is discretized using $4 \times 4$ cells and DG order of 3. $M=6$ is used on the half sphere in all cases.}}
\label{fig_cavity_TQxQy}
\end{figure*}

\begin{figure*}[tbp]
\centering
\begin{subfigure}{.5\textwidth}
  \centering
  \includegraphics[width=70mm,trim={0cm 0cm 0cm 0cm},clip]{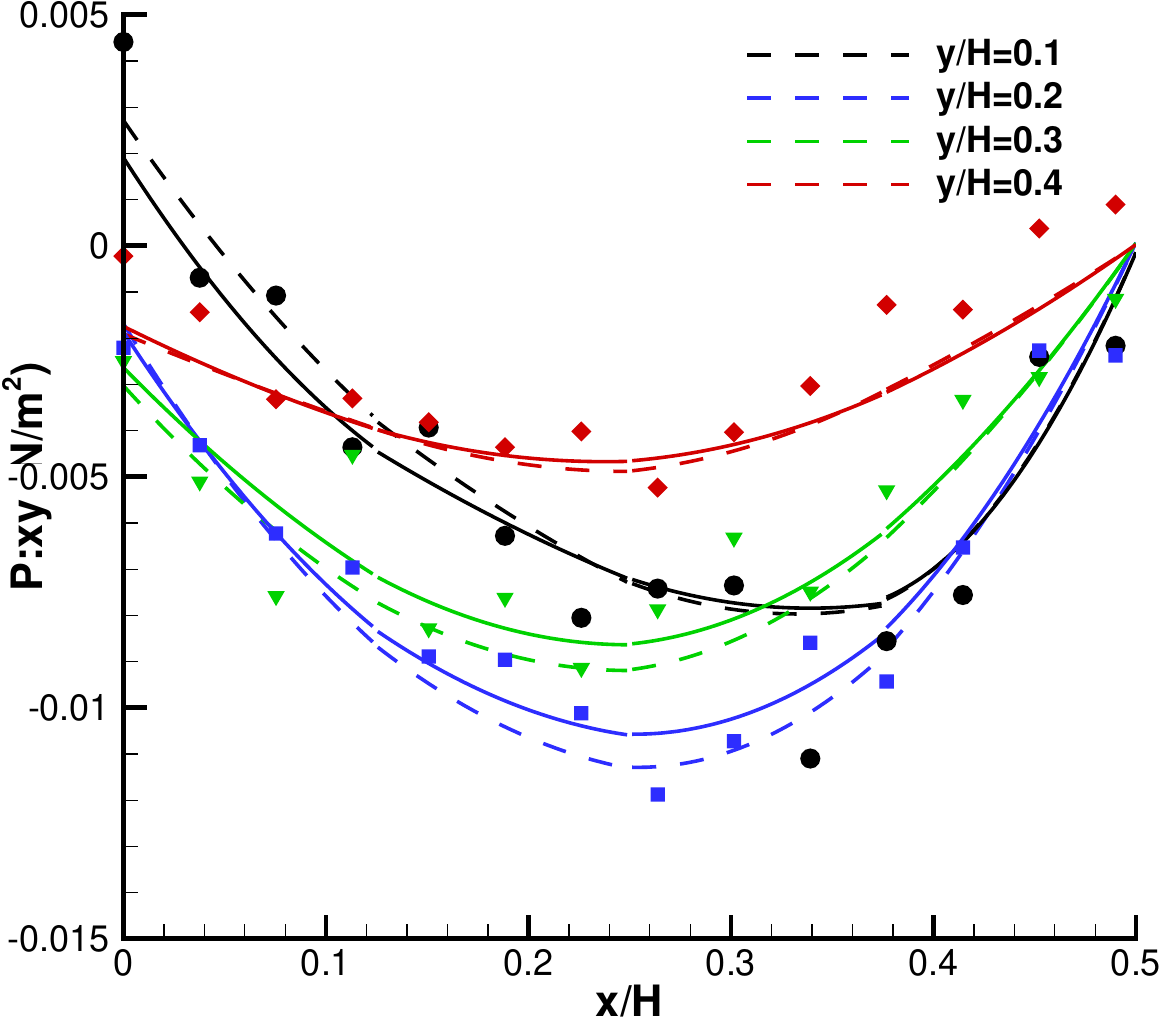}
  \caption{$xy$-component of stress (on horizontal lines)}
  \label{fig_cavity_Pxy_hor}
\end{subfigure}%
\begin{subfigure}{.5\textwidth}
  \centering
  \includegraphics[width=70mm,trim={0cm 0cm 0cm 0cm},clip]{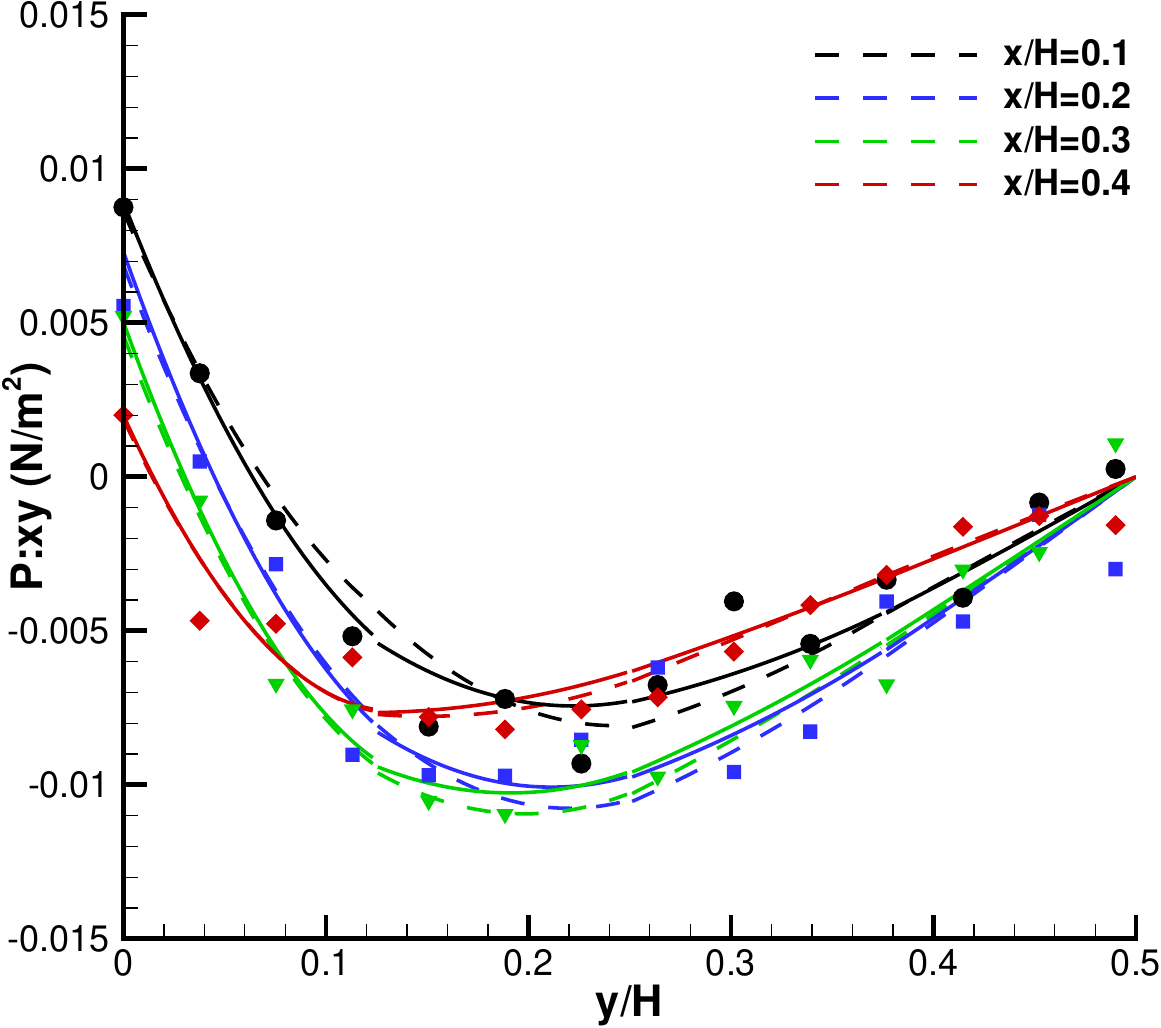}
  \caption{$xy$-component of stress (on vertical lines)}
  \label{fig_cavity_Pxy_ver}
\end{subfigure}

\begin{subfigure}{.5\textwidth}
  \centering
  \includegraphics[width=70mm,trim={0cm 0cm 0cm 0cm},clip]{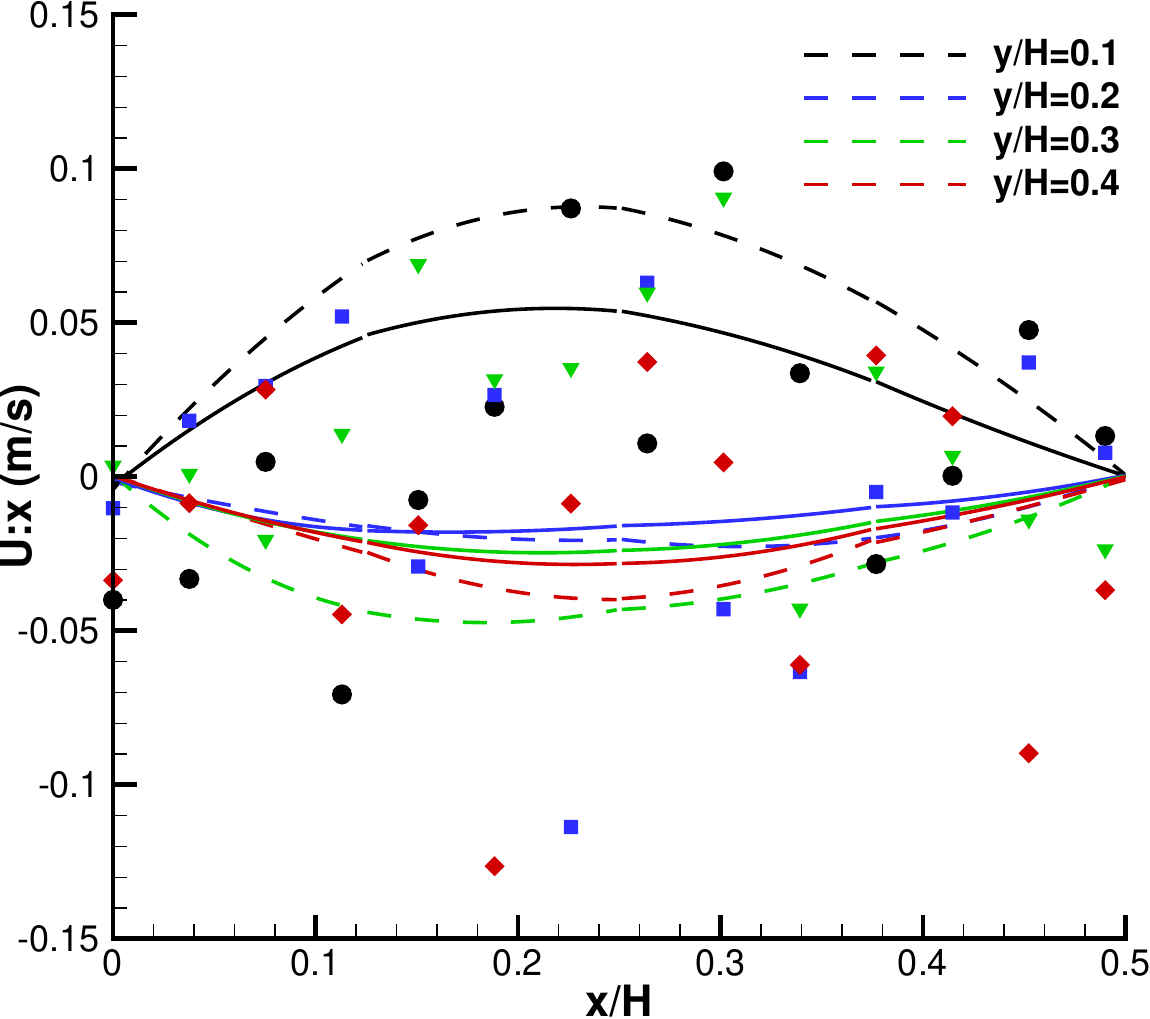}
  \caption{$x$-component of velocity (on horizontal lines)}
  \label{fig_cavity_Ux_hor}
\end{subfigure}%
\begin{subfigure}{.5\textwidth}
  \centering
  \includegraphics[width=70mm,trim={0cm 0cm 0cm 0cm},clip]{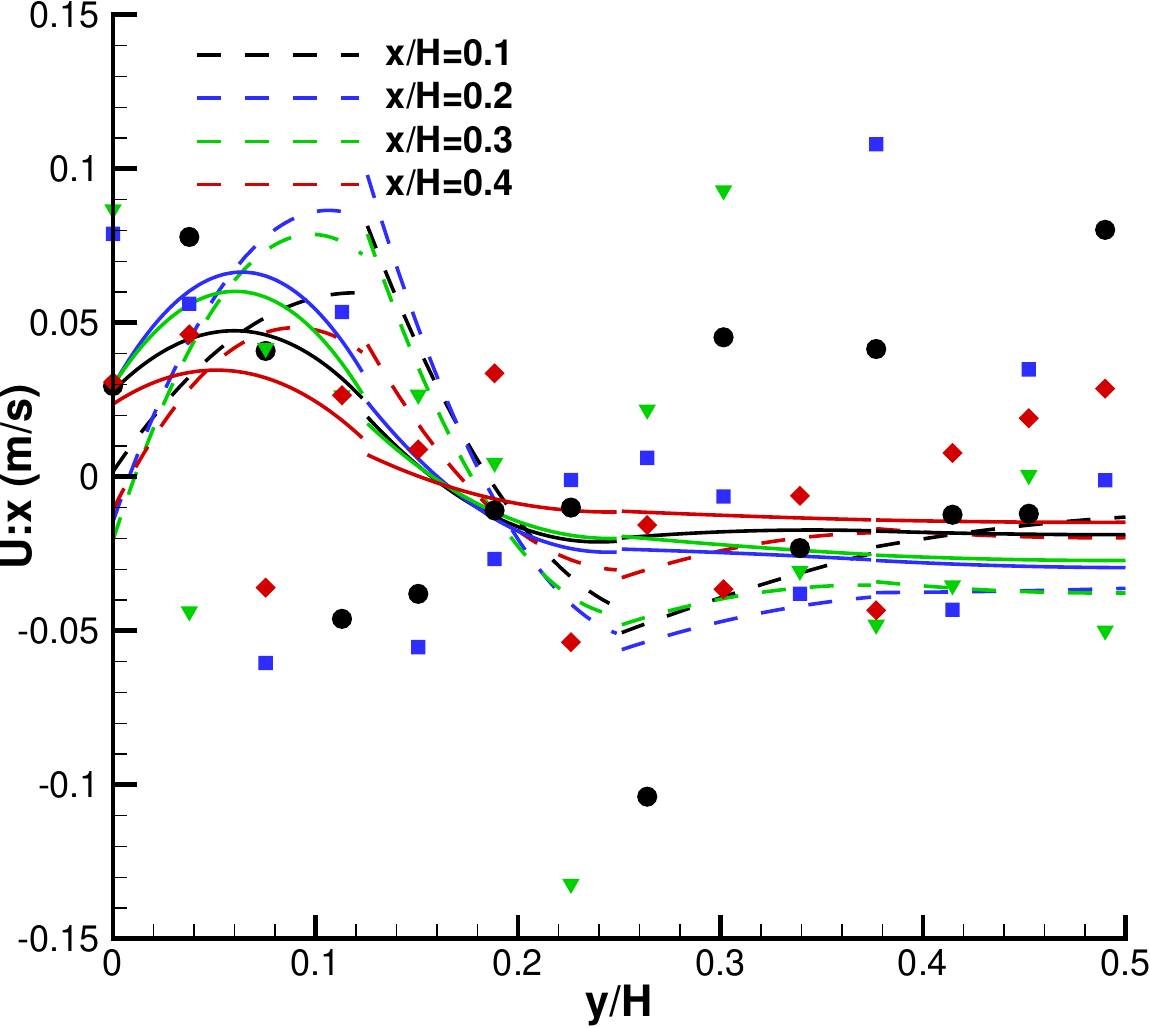}
  \caption{$x$-component of velocity (on vertical lines)}
  \label{fig_cavity_Ux_ver}
\end{subfigure}

\begin{subfigure}{.5\textwidth}
  \centering
  \includegraphics[width=70mm,trim={0cm 0cm 0cm 0cm},clip]{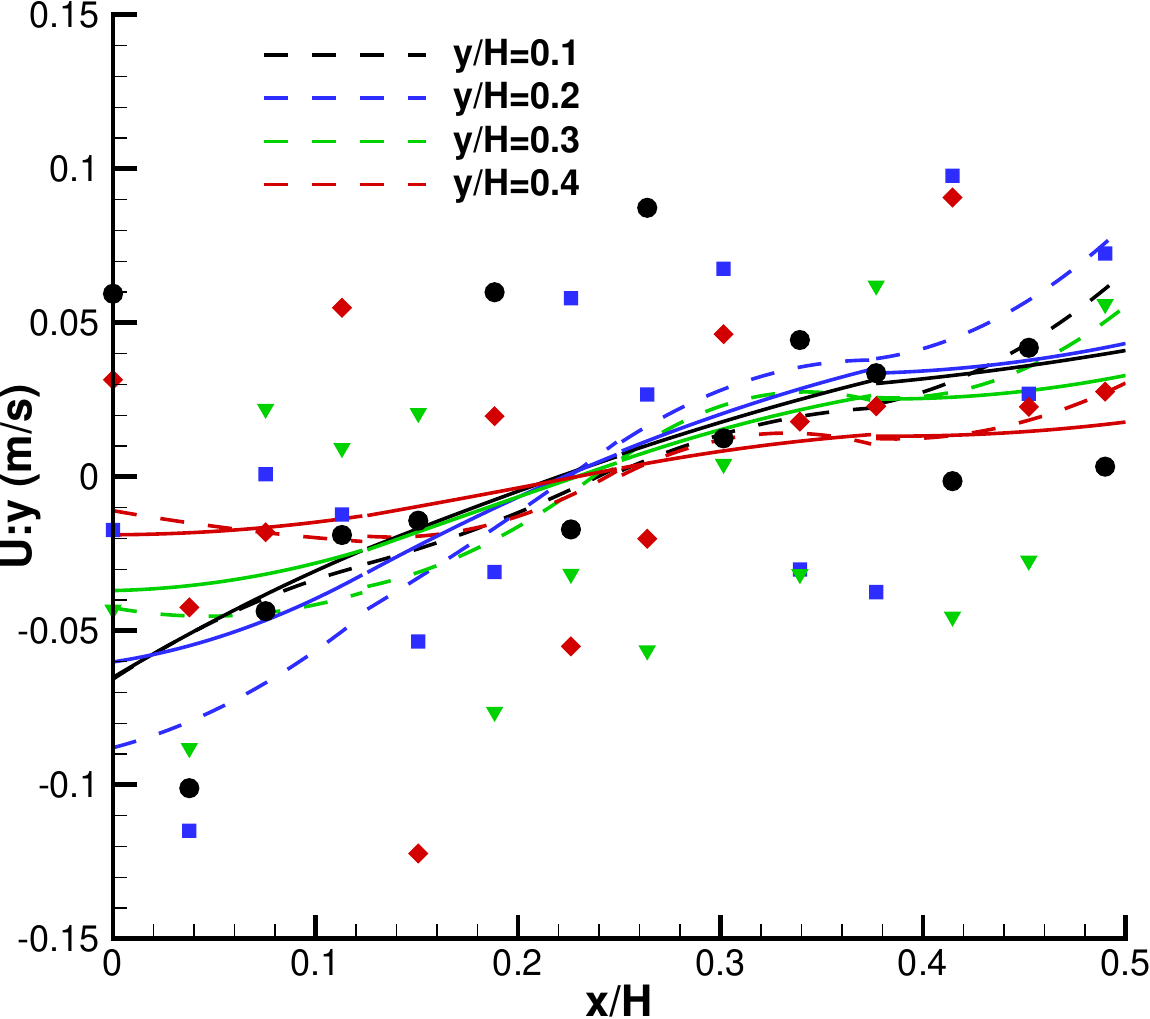}
  \caption{$y$-component of velocity (on horizontal lines)}
  \label{fig_cavity_Uy_hor}
\end{subfigure}%
\begin{subfigure}{.5\textwidth}
  \centering
  \includegraphics[width=70mm,trim={0cm 0cm 0cm 0cm},clip]{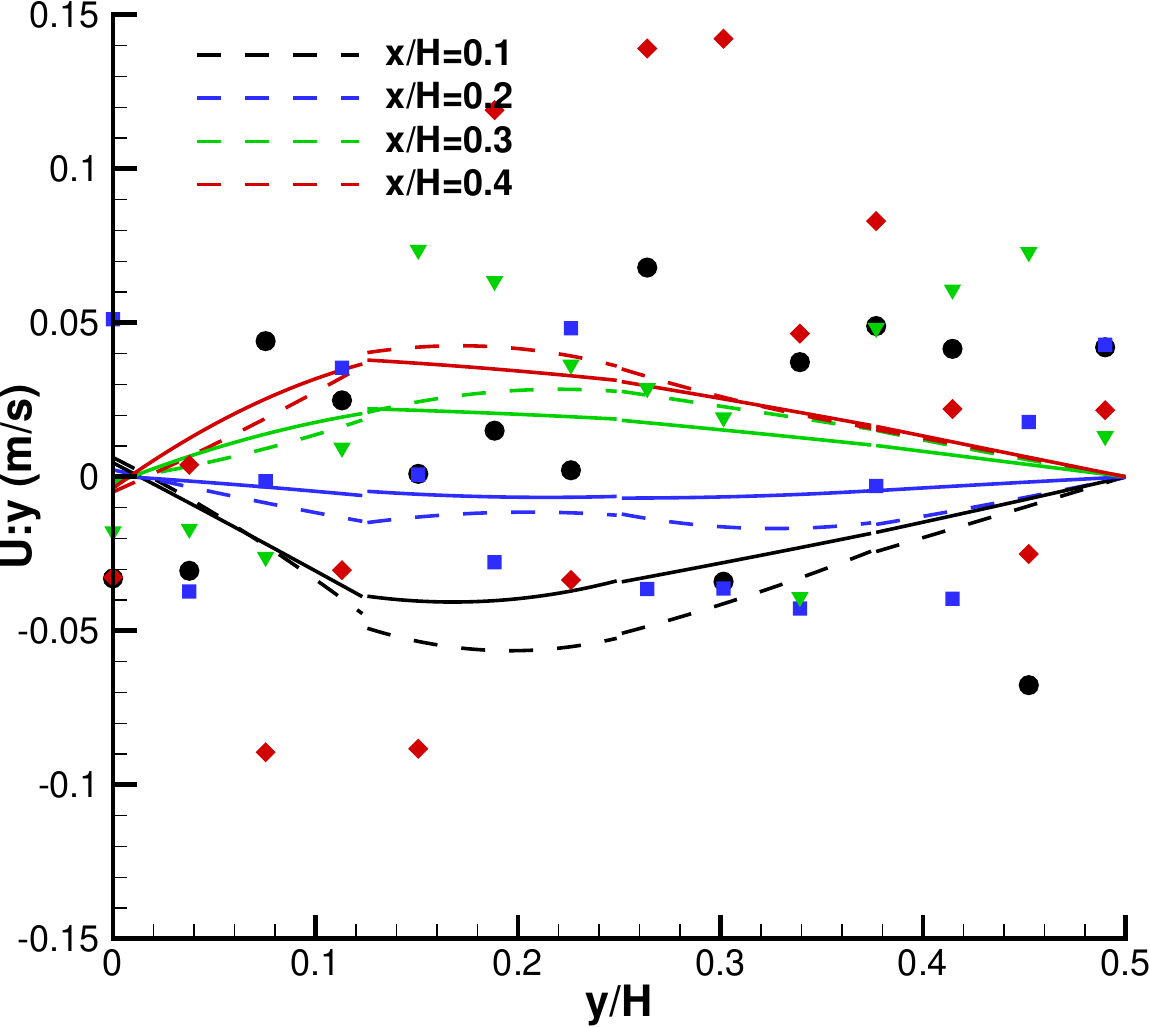}
  \caption{$y$-component of velocity (on vertical lines)}
  \label{fig_cavity_Uy_ver}
\end{subfigure}

\caption{{Continuation of Figure~\ref{fig_cavity_TQxQy}.}}
\label{fig_cavity_PxyUxUy}
\end{figure*}

\section{Conclusions}

We have presented a deterministic numerical method for the full Boltzmann equation. The method combines the discontinuous Galerkin discretization in the physical space and the fast Fourier spectral method in the velocity space to yield highly accurate numerical solutions. The DG-type formulation employed in the present work has advantage of having high order accuracy at the element-level, and its element-local compact nature (and that of our collision algorithm) enables effective parallelization on massively parallel architectures. Our fast spectral method for evaluating the Boltzmann collision operator does not rely on any assumption or parameter fitting of the collision kernel in contrast to the previously proposed methods in literature. Further, we have proposed a novel SVD based collision algorithm to further reduce the cost in evaluating the collision operator resulting from the DG formulation.

To verify the proposed DGFS method, we carried out rarefied gas flow simulations for spatially homogeneous, Fourier, Couette, oscillatory Couette, normal shock, lid-driven, and thermally driven cavity flows at different Knudsen numbers. Each of these cases have been run with different collision kernel to highlight the general nature of our collision algorithm. We conclude that the results obtained with our deterministic solver and DSMC are inextricable ignoring the statistical noise and the errors therein. The deterministic solution of the Boltzmann equation by the DGFS method, in particular, is suitable for studying low-speed and unsteady flows.

\section*{Appendix}

In this appendix, we give a brief description of the fast Fourier spectral method proposed in \cite{GHHH17}. Our implementation here differs from \cite{GHHH17} in mainly two aspects: 1) the symmetrized version of the collision kernel is used which allows the integration to be performed on the half sphere rather than whole sphere; 2) a different spherical quadrature is adopted which shows better numerical performance.

First of all, from equations (\ref{eq_dim_Q_full}), (\ref{cc}) and (\ref{eq_dim_B}), it is easy to see that one can replace the collision kernel by its symmetrized version:
\begin{equation}
B_{\text{sym}}(|\bc-\bc_*|,\cos \chi) =\frac{B(|\bc-\bc_*|,\cos \chi)+B(|\bc-\bc_*|,-\cos \chi)}{2}.
\end{equation}
Second, from the discussion in Section~\ref{subsec:collision}, all we need is to evaluate an operator of the form
\begin{equation} \label{eq_anpdx_Q1}
\begin{split}
\mathcal{Q}(f,g)(\bc) = \int_{\mathbb{R}^3} \int_{\mathcal{S}^2} &B_{\text{sym}}( |\bc -  \bc_*|, \cos\chi)[  f( \bc')  g( \bc'_*)\\& -  f( \bc) g( \bc_*) ] \rd{\sigma} \rd{\bc_*}.
\end{split}
\end{equation}

The main steps of the Fourier spectral approximation of (\ref{eq_anpdx_Q1}) can be summarized as follows:
\begin{itemize}
\item Change the variable $\bc_*$ to the relative velocity $\bc_r = \bc-\bc_*$:
\begin{equation}
\begin{split}
    \mathcal{Q}(f,g)(\bc) = \int_{\mathbb{R}^3} \int_{\mathcal{S}^2} &B_{\text{sym}} ( c_r, \sigma\cdot \hat{\bc}_r) [  f( \bc')  g( \bc'_*) \\&-  f( \bc) g(\bc-\bc_r) ] \rd{\sigma} \rd{\bc_r},
    \end{split}
\label{eq_apndx_Q}
\end{equation}
where $c_r$ is the magnitude of $\bc_r$, $\hat{\bc}_r$ is the unit vector along $\bc_r$, and
\begin{equation}
\bc'=\bc-\frac{\bc_r}{2}+\frac{c_r}{2}\sigma, \quad \bc_*'=\bc-\frac{\bc_r}{2}-\frac{c_r}{2}\sigma.
\end{equation}
\item Determine the computational domain $D_L=[-L,L]^3$ as described in Section \ref{subsec:vel}, and periodically extend $f$, $g$ to $\mathbb{R}^3$.
\item Truncate the integral in $\bc_r$ to a ball $B_R$ with $R=\frac{4}{3+\sqrt{2}}L$ (criterion based on \cite{PR00}).
\item Approximate $f$, $g$ by truncated Fourier series
\begin{equation} \label{summation}
f^N(\bc)=\sum_{k=-N/2}^{N/2-1}\hat{f}_k e^{i\frac{\pi}{L}{k}\cdot \bc}, \quad g^N(\bc)=\sum_{k=-N/2}^{N/2-1}\hat{g}_{k} e^{i\frac{\pi}{L}k\cdot \bc}.
\end{equation}
Note here $k=(k_1,k_2,k_3)$ is a 3D index, and the summation in (\ref{summation}) is understood to be over the lattice $\{k\in \mathbb{Z}^3: \, -N/2\leq k_1,k_2,k_3\leq N/2-1\}$.
\item Substitute $f^N$, $g^N$ into (\ref{eq_apndx_Q}), and perform the standard Galerkin projection
\begin{equation}
\begin{split}
\hat{\mathcal{Q}}_{k}:&=\frac{1}{(2L)^3}\int_{D_L} \mathcal{Q}(f^N,g^N)(\bc)e^{-i\frac{\pi}{L}k\cdot \bc}\rd{\bc}\\
&=\sum_{\substack{l,m=-N/2\\l+m=k}}^{N/2-1}[G(l,m)-G(m,m)]\hat{f}_l\,\hat{g}_m,
\end{split}
\end{equation}
where $k=-N/2,\dots,N/2-1$, and the kernel mode $G$ is given by
\begin{equation}
\label{GG}
G(l,m) = \int_{B_R} \int_{\mathcal{S}^2} B_{\text{sym}}(c_r, \sigma \cdot \hat{\bc}_r) \; e^{-i\frac{\pi}{L} \frac{l+m}{2} \cdot \bc_r + i\frac{\pi}{L}c_r \frac{l-m}{2} \cdot \sigma} \rd{\sigma}\rd{\bc_r}.
\end{equation}
\end{itemize}

It is clear that a direct evaluation of $\hat{\mathcal{Q}}_k$ (for all $k$) would require $O(N^6)$ complexity. But if we can find a low-rank, separated expansion of $G(l,m)$ as
\begin{equation} \label{lowrank}
G(l,m) \approx \sum_{r=1}^R \alpha_r (l+m) \; \beta_r(l) \; \gamma_r(m),
\end{equation}
then the gain term (positive part) of $\hat{\mathcal{Q}}_k$ can be rearranged as
\begin{equation}
\hat{\mathcal{Q}}^+_k = \sum_{r=1}^R \alpha_r (k) \sum_{\substack{l,\;m=-N/2 \\ l+m=k}}^{N/2-1} \; \left(\beta_r(l) \hat{f}_l \right) \; \left(\gamma_r(m) \hat{g}_m \right),
\label{eq_apndx_fs_Glm_Qk}
\end{equation}
which is a convolution of two functions $\beta_r(l) \hat{f}_l$ and $\gamma_r(m) \hat{g}_m$, hence can be computed via FFT in $O(RN^3 \log N)$ operations. Note that the loss term (negative part) of $\hat{\mathcal{Q}}_k$ is readily a convolution and can be computed via FFT in $O(N^3\log N)$ operations.

In order to find the decomposition as in (\ref{lowrank}), we simplify (\ref{GG}) as (using the symmetry of the kernel)
\begin{equation} \label{realGG}
G(l,m)=2\int_0^R \int_{\mathcal{S}^{2+}}F(l+m,c_r,\sigma)\cos\left(\frac{\pi}{L}c_r\frac{l-m}{2}\cdot \sigma\right)\rd{\sigma}\rd{c_r},
\end{equation}
where $\mathcal{S}^{2+}$ denotes the half sphere, and
\begin{equation}
F(l+m,c_r,\sigma):=2c_r^2\int_{\mathcal{S}^{2+}}B_{\text{sym}}(c_r,\sigma\cdot \hat{\bc}_r)\cos\left(\frac{\pi}{L}c_r\frac{l+m}{2}\cdot \hat{\bc}_r\right)\rd{\hat{\bc}_r}.
\end{equation}
Now using the fact that $\cos(\alpha-\beta)=\cos\alpha \cos \beta+\sin\alpha \sin \beta$, if we approximate the integral in (\ref{realGG}) by a quadrature, we obtain
\begin{equation} \label{lowrank1}
\begin{split}
&G(l,m)\approx  2 \sum_{c_r,\sigma} w_{c_r}w_{\sigma} F(l+m,c_r,\sigma) \\ &\cdot \left [\cos\left(\frac{\pi}{L}c_r\frac{l}{2}\cdot \sigma\right)  \cos\left(\frac{\pi}{L}c_r\frac{m}{2}\cdot \sigma\right)  +\sin\left(\frac{\pi}{L}c_r\frac{l}{2}\cdot \sigma\right)\sin\left(\frac{\pi}{L}c_r\frac{m}{2}\cdot \sigma\right)\right],
\end{split}
\end{equation}
where $(c_r, w_{c_r})$ and $(\sigma,w_{\sigma})$ are the quadrature (points,weights) for the line integral and the spherical integral. (\ref{lowrank1}) is exactly in the desired form (\ref{lowrank}).

In the implementation, we use the Gauss-Legendre quadrature for $c_r$. As the integrand oscillates on the scale of $O(N)$, the total number of quadrature points $N_r$ needed for $c_r$ should be $O(N)$. For the integration on the half sphere, we choose to use the {\it spherical design} (SD) \cite{Womersley}, which is the near optimal quadrature on the sphere \cite{Beentjes15}. Other quadratures are possible, for example, the Lebedev quadrature as used in \cite{GHHH17}. Through numerical tests, we found that SD usually yields better results than Lebedev, probably due to the fact that the quadrature points are more uniformly distributed in SD. Let $M$ denote the number of quadrature points used on the half sphere (in practice $M\ll N^2$), the total number of terms in the expansion (\ref{lowrank}) is thus $R=O(MN)$. Therefore, the final computational cost of evaluating $\hat{\mathcal{Q}}_k$ (for all $k$) is reduced from $O(N^6)$ to $O(MN^4\log N)$.

\section*{Acknowledgments}

J. Hu's research was supported by NSF grant DMS-1620250 and NSF CAREER grant DMS-1654152. Support from DMS-1107291: RNMS KI-Net is also gratefully acknowledged.

\bibliographystyle{elsarticle-num-names}
\bibliography{elsarticle_bib}

\end{document}